\documentclass[12pt]{article}
\usepackage{latexsym}

\usepackage[totalheight = 25cm, totalwidth = 17cm]{geometry}

\newcommand{\ud}{\mathrm{d}}

\begin{document}

\title{Intersecting brane solutions in string and M-theory}
\author{Douglas J. Smith\footnote{email: Douglas.Smith@durham.ac.uk} }

\maketitle

\begin{center}

{\em
Department of Mathematical Sciences \\
University of Durham \\
Durham \\
DH1 3LE \\
UK
}

\end{center}

\vspace{0.5cm}

\begin{abstract}
We review various aspects of configurations of intersecting branes, including
the conditions for preservation of supersymmetry. In particular, we discuss the
projection conditions on the Killing spinors for given brane configurations
and the relation to calibrations. This
highlights the close connection between intersecting branes and branes wrapping
supersymmetric cycles as well as special holonomy manifolds. We also explain how these conditions can be used to find supergravity
solutions without directly solving the Einstein equations. The description of intersecting branes is considered both in terms of
the brane worldvolume theories and as supergravity solutions.
There are well-known simple procedures (harmonic function rules) for writing down the supergravity solutions
for supersymmetric configurations of orthogonally intersecting branes. However, such solutions involve smeared or delocalised branes. We describe several
methods of constructing solutions with less smearing, including some fully localised solutions.
Some applications of these supergravity
solutions are also considered -- in particular the study of black holes and gauge theories.

\end{abstract}

\vspace{-18.0cm}
\begin{flushright}
DTP/02/67 \\
hep-th/0210157
\end{flushright}

\thispagestyle{empty}

\newpage

\tableofcontents

\newpage

\setcounter{page}{1}

\section{Outline}
\label{outline}

The aim of this review is to describe intersecting branes in ten and eleven
dimensions. We restrict to configurations of branes which preserve
supersymmetry. We begin in section~\ref{HalfBPS} by reviewing the properties
of the individual branes. There are, of course many other more detailed
reviews and lectures notes covering this material such as
\cite{Duff:1995an, Polchinski:1996fm, Polchinski:1996na, Duff:1996zn, Stelle:1996tz, Lu:1998hb, Townsend:1997wg, Stelle:1998xg, Bachas:1998rg, Johnson:2000ch}
and several books covering string theory such as \cite{Green:1987sp, Green:1987mn} and more recently
also branes such as \cite{Polchinski:1998rq, Polchinski:1998rr} and particularly \cite{Johnson:2003gi}.
We first present the supergravity solutions
for parallel branes and discuss the supersymmetry of these solutions. In
particular we show how we can understand which supersymmetries should be
preserved and how this enables us to derive the supergravity solutions
without explicitly solving the Einstein equations. We also discuss the use of
brane probes and the relation between branes (particularly D-branes) and gauge
theories, including the AdS/CFT correspondence. The remaining sections all
build on these concepts for more complicated brane configurations. There will
be some overlap with Gauntlett's review of intersecting branes
\cite{Gauntlett:1997cv}. Throughout we will try to explain the main ideas by
giving the simplest non-trivial examples, referring to the literature for
more general (and usually technically more complicated) cases.

In section~\ref{IntandEnd} we will consider the conditions for (both orthogonal
and non-orthogonal) intersections of
branes to preserve supersymmetry. We will see that there are close relations
between intersecting branes and branes wrapping smooth cycles in special
holonomy manifolds. In particular we will review the relation between the
existence of Killing spinors (required for supersymmetry) and calibrations
which is a very useful method for understanding the conditions for branes to
wrap or intersect supersymmetrically. We will also briefly review the
description of such configurations in terms of the brane worldvolume field
theory.

We introduce the ``harmonic function rules'', which are a prescription for
writing down supergravity solutions for intersecting branes, in
section~\ref{Smeared}. We will see that the draw-back of this construction is
that the branes must be smeared or delocalised in some directions -- i.e.\ the
supergravity solutions have isometries in directions transverse to the branes.
We will present an important application of such solutions, as the ten- or eleven-
dimensional description of lower dimensional black holes. The smearing is not
problematic in this context since we compactify in those directions anyway. So
these solutions provide an interpretation of the black holes in terms of branes,
from which the microscopic entropy can be calculated.

We move on, in section~\ref{semilocalised}, to consider solutions where the
branes are more localised, using a more general set of harmonic function
rules. The draw-back is that it is now typically not possible to find explicit
solutions and there is still some smearing in most cases. We present some
explicit examples and show how some cases involving D6-branes have a very
simple eleven-dimensional geometrical interpretation. We then discuss a more
general construction of various (generically non-orthogonal) intersecting
branes, starting from eleven-dimensional solutions which may not contain any
branes.

Finally, in section~\ref{HananyWitten} we review how many gauge theories can
be described using configurations of intersecting branes. In particular we
focus on the case of four-dimensional $\mathcal{N} = 2$ theories. These cases
involve intersecting D4- and NS5-branes in type IIA which are related to a
single M5-brane wrapping a non-compact Riemann surface (identified with the
Seiberg-Witten curve) in eleven dimensions. We then describe how the
conditions for preserving supersymmetry can be used to find the localised
supergravity solution, at least in the limit appropriate to the AdS/CFT
correspondence.

\section{Review of one half BPS branes}
\label{HalfBPS}

Before trying to describe complicated configurations of branes it is useful to
consider the description of parallel branes in supergravity and string theory.
As well as fixing notation and conventions, we will see that there are several
ways of deriving the supergravity solutions. By describing the simplest cases
here in detail we can understand many of the essential features of the
intersecting brane solutions we will consider later, but without many of the
technical complications. In particular we will explain which supersymmetries
are preserved by a given brane and how this can be used to derive the
corresponding supergravity solution. We will also briefly describe the method
of using branes as probes of supergravity backgrounds, the relation between
D-branes and gauge theories, and the AdS/CFT correspondence.

\subsection{Brane solutions in supergravity}
\label{BraneSolns}

In supergravity theories, electric $p$-branes are solitonic objects (solutions
of the equations of motion with or without source terms)
charged with respect to a $(p+2)$-form U(1) field
strength $F_{(p+2)}$.
If the brane carries no other charges then it is a
solution of a subsector of the supergravity theory with action, in the Einstein
frame,
\begin{equation}
S = \frac{1}{2\kappa_D^2} \int \ud^Dx \sqrt{-g} \left( R - \frac{1}{2(p+2)!} F_{(p+2)}^2 \right)
\end{equation}
where $\kappa_D$ is related to the $D$-dimensional Newton's constant, $G_D$,
and Planck length, $l_P$ by
\begin{equation}
2\kappa_D^2 = 16\pi G_D = (2\pi)^{D-3}l_P^{D-2}
\end{equation}
For example with $D=11$ and $p=2$, this, apart from a Chern-Simons term
\begin{equation}
\frac{1}{2\kappa_D^2} \int \frac{1}{6} F_{(4)} \wedge F_{(4)} \wedge A_{(3)}
\end{equation}
is the bosonic part of the eleven-dimensional supergravity action, where
$F_{(4)} = \ud A_{(3)}$. In ten-dimensional supergravity, branes typically couple
to the dilaton $\phi$. We will consider this extra complication later.

It can easily be checked that a solution to the equations of motion is
\begin{eqnarray}
\ud s^2 & = & H^{-\frac{D-p-3}{D-2}} \ud x_{(1,p)}^2 + H^{\frac{p+1}{D-2}} \ud x_{(D-p-1)}^2 \label{branemet} \\
F_{(p+2)} & = & \pm \ud (H^{-1}) \wedge \epsilon_{1,p} \label{braneF} \\
H & = & 1 + \frac{c_pN}{r^{D-p-3}} \label{braneharm}
\end{eqnarray}
where $\ud x_{(1,p)}^2$ is the $(p+1)$-dimensional Minkowski metric with volume
form $\epsilon_{1,p}$ and
$$ \ud x_{(D-p-1)}^2 = \ud r^2 + r^2 \ud\Omega_{D-p-2}^2 $$
is the $(D-p-1)$-dimensional Euclidean metric. This solution is interpreted as
$N$ coincident branes with a $(p+1)$-dimensional Minkowski worldvolume located at $r=0$.
Branes and anti-branes differ in the sign of $F_{(p+2)}$ or equivalently in
the orientation of their worldvolume. The
equations of motion reduce to the condition that $H$ satisfies the Laplace
equation in the transverse space so we see that this is a solution with a
source term at $r=0$. Nevertheless, the question of whether we need a source term
is rather subtle since it may be possible to analytically continue the solution
through $r=0$, i.e.\ view $r=0$ as simply a coordinate singularity. This is
indeed possible in many cases. However, even in such cases the curvature will
become very large for some solutions or choices of parameters as we approach $r=0$ and so we should not trust the
supergravity description of the system. For our purposes we will consider
branes to be solutions of the supergravity equations of motion with appropriate
source terms.

These source terms can be understood as arising from the coupling of a $p$-brane
action to the supergravity action
\begin{equation}
S \sim S_{Supergravity} + S_{Brane}
\end{equation}
This leads to a relation between the
(dimensionful) constant $c_p$ appearing in the Harmonic function $H$ and the
$p$-brane tension $T_p$
\begin{equation}
\label{c_p}
c_p = \frac{2\kappa_D^2 T_p}{(D-p-3)V(S^{D-p-2})}
\end{equation}
where $V(S^{D-p-2})$ is the volume of a $(D-p-2)$-sphere of unit radius. See
\cite{Duff:1991xz} for the case of the 2-brane in eleven dimensions.

There are also magnetic $(D-p-4)$-brane solutions which are magnetically
charged under $F_{(p+2)}$. At the level of classical equations of motion we can
equally consider $F_{(p+2)}$ or its Hodge dual $F_{(D-p-2)} \equiv \ast F_{(p+2)}$
to be the fundamental field strength. So, at least at the level of classical
solutions of supergravity, there is no distinction between electric and
magnetic solutions except that conventionally $F_{(p+2)}$ with $p+2 \le D/2$
is assumed to be the fundamental field strength.

As with electric particles and magnetic monopoles in four dimensions, there is
a Dirac quantisation condition relating the charges (and tensions) of electric
and magnetic branes charged under the same field strength. In terms of the
brane tensions this takes the form \cite{Nepomechie:1985wu, Teitelboim:1986yc}
\begin{equation}
2\kappa_D^2 T_p T_{D-p-4} = 2\pi n \; , \; n \in Z
\end{equation}
This is satisfied by the branes we will consider with $n=1$.

In 11-dimensional supergravity the only field strength is $F_{(4)}$ and so the
only brane solutions are (electric) 2-branes and (magnetic) 5-branes, usually
labelled M2- and M5-branes. So equations (\ref{branemet}), (\ref{braneF}) and
(\ref{braneharm}) with $D=11$ give the supergravity solution for M2-branes
\cite{Duff:1991xz} (with $p=2$) and M5-branes \cite{Gueven:1992hh} (with $p=5$.)
The tensions of these branes are related by the
Dirac quantisation condition. Furthermore, the tension of say the M2-brane is
fixed in terms of the eleven-dimensional Planck length due to the presence
of the Chern-Simons term in the supergravity action. This results in
\cite{Duff:1995wd}
\begin{equation}
\label{TM2}
T_{M2} = \frac{1}{4\pi^2 l_P^3}
\end{equation}

In the following two sections we will briefly review the branes in
ten-dimensional type IIA and type IIB theories. More details from the
supergravity perspective can be found in e.g.\ \cite{Stelle:1998xg} while
\cite{Johnson:2000ch} provides a detailed review of D-branes in terms of string
theory.

\subsubsection{Branes in type IIA supergravity}

In type IIA supergravity there are Ramond-Ramond (RR) field strengths $F_{(2)}$
giving rise to D0- and D6-branes, and $F_{(4)}$ with corresponding D2- and
D4-branes. There are also D8-branes which are domain walls in ten dimensions.
These are solutions of massive IIA supergravity \cite{Bergshoeff:1995vh}. They
are predicted to exist in string theory by T-duality from other D-branes but,
unlike the other branes in type IIA string theory, it is not clear
how they are related to an eleven-dimensional theory.

From the string theory
point of view D$p$-branes in type IIA and type IIB are $(p+1)$-dimensional submanifolds on which
open strings can end. As we will discuss later, this leads to the result that
the low energy dynamics of D$p$-branes is described by a $(p+1)$-dimensional
gauge theory. The tension of a D$p$-brane can be calculated from a 1-loop
open string amplitude \cite{Polchinski:1995mt}. The result is
\begin{equation}
\label{TDp}
T_{Dp} = \frac{1}{(2\pi)^p g_s l_s^{p+1}}
\end{equation}

There are also non-dynamical RR-charged objects known as orientifold $p$-planes.
These are the fixed planes of a $\mathbf{Z}_2$ action which consists of a
reflection of the $9-p$ transverse coordinates together with a reversal of the
orientation of the string worldsheet. The charge and tension of these
orientifold $p$-planes is given in terms of the D$p$-brane tension by
\begin{equation}
\label{TOp}
T_{Op} = \pm 2^{p-5} T_{Dp}
\end{equation}

Finally in both type IIA and type IIB there is also a NS-NS (for Neveu-Schwarz) three-form
field strength $H_{(3)}$ and so there are 1- and 5-brane solutions. These
1-branes correspond to the fundamental strings and are sometimes referred to
as F1-branes or NS1-branes. The 5-branes are called NS5-branes. Note that these
objects are not D-branes (they are not endpoints for fundamental open strings.)

The tension of the fundamental string
\begin{equation}
\label{TF1}
T_{F1} = \frac{1}{2\pi l_s^2}
\end{equation}
defines the string length $l_s = \sqrt{\alpha'}$. This string length is in turn
related to the ten-dimensional Newton's constant by
\begin{equation}
2\kappa_{10}^2 = 16\pi G_{10} = (2\pi)^7g_s^2 l_s^8
\end{equation}
where the string coupling constant $g_s \equiv e^{\phi_{\infty}}$ is related to
the asymptotic value of the dilaton $\phi \rightarrow \phi_{\infty}$. We can
of course shift $\phi$ so that it vanishes at infinity, and include factors
of $g_s$ explicitly. This is the convention we use throughout this paper.
In the string frame metric the supergravity action takes the form
\begin{equation}
S_{IIA} = \frac{1}{2\kappa_{10}^2} \int \ud^{10}x \sqrt{-g} \left[ e^{-2\phi}
	\left( R + 4 |\nabla \phi |^2 - \frac{1}{2.3!}H_{(3)}^2 \right) -
	\frac{1}{2} \sum_n \frac{1}{n!} F_{(n)}^2 \right]
\label{IIAStrAction}
\end{equation}
where the summation is over $n = 2,4$, together with a Chern-Simons term
\begin{equation}
\frac{1}{2\kappa_{10}^2} \int \frac{1}{2} \ud C_{(3)} \wedge \ud C_{(3)} \wedge B_{(2)}
\end{equation}
where
\begin{equation}
H_{(3)} = dB_{(2)} \; , \; F_{(2)} = \ud C_{(1)} \; , \; F_{(4)} = \ud C_{(3)} + C_{(1)} \wedge H_{(3)}
\end{equation}
We can write this action in the Einstein
frame where the gravitational part takes the standard Einstein-Hilbert form
by rescaling the string-frame metric $g_{MN}$. In terms of the Einstein-frame
metric
\begin{equation}
g^{(E)}_{MN} \equiv e^{-\phi/2} g_{MN}
\end{equation}
the string action takes the form
\begin{equation}
S_{IIA} = \frac{1}{2\kappa_{10}^2} \int \ud^{10}x \sqrt{-g^{(E)}} \left[
        R -\frac{1}{2} |\nabla \phi |^2 - \frac{1}{2.3!}e^{-\phi}H_{(3)}^2
        - \frac{1}{2} \sum_n \frac{1}{n!} e^{\frac{5-n}{2}\phi}F_{(n)}^2 \right]
\label{IIAEinstAction}
\end{equation}
where all quantities (e.g.\ $R$ and $F^2$) are calculated using the
Einstein-frame metric.

As is well-known, type IIA supergravity arises from the Kaluza-Klein reduction
of eleven-dimensional supergravity on a circle of radius $R_{11}$, with the
fundamental string being interpreted as an M2-brane wrapping the circle. This
relates the type IIA string coupling, $g_s$, and string
length, $l_s$, to the eleven-dimensional Planck length, $l_P$, and $R_{11}$
\begin{equation}
\label{IIA11dconst}
R_{11} = g_s l_s \;\; , \;\; l_P = g_s^{\frac{1}{3}}l_s
\end{equation}
The explicit relation between the type IIA string-frame metric $\ud s_{(1,9)}^2$,
dilaton $\phi$, RR one-form potential $C_{\mu}$, RR four-form field strength
$F_{(4)}$ and NS-NS three-form field strength $H_{(3)}$,
and the eleven-dimensional metric $\ud s_{(1,10)}^2$ and four-form field strength
$\tilde{F}_{(4)}$ is
\begin{eqnarray}
\label{IIA11dmetric}
\ud s_{(1,10)}^2 & = & e^{-\frac{2\phi}{3}}\ud s_{(1,9)}^2 + e^{\frac{4\phi}{3}} \left(
	R_{11} d\psi + C_{\mu}\ud x^{\mu} \right)^2 \\
\tilde{F}_{(4)} & = & F_{(4)} + H_{(3)} \wedge \ud x^{10}
\end{eqnarray}
where $\psi$ has period $2\pi$ and $x^{10} = R_{11}\psi$.

In this way all type IIA branes (except the D8-brane) can be understood in terms
of dimensional reduction from eleven dimensions \cite{Townsend:1995kk}. Since
$\tilde{F}_{(4)}$ reduces to $H_{(3)}$ and $F_{(4)}$,
fundamental strings and D2-branes are simply M2-branes
wrapped or not wrapped on the eleventh dimension. Similarly D4- and NS5-branes
both correspond to M5-branes in eleven dimensions. The field strength $F_{(2)}$
is just the usual Kaluza-Klein gauge fieldstrength and so the D0-branes are
Kaluza-Klein particles while the D6-branes are Kaluza-Klein monopoles.
In particular the eleven-dimensional supergravity solution which reduces to a
D6-brane in ten dimensions is given by a geometrical background which is a
product of $(6+1)$-dimensional Minkowski spacetime with the Taub-NUT space,
which we will refer to as a KK6-brane.

\subsubsection{Branes in type IIB supergravity}

In type IIB supergravity there are RR field strengths $F_{(1)} = \ud C_{(0)}$ (D(-1)-branes
and D7-branes), $F_{(3)} = \ud C_{(2)} - C_{(0)} \wedge H_{(3)}$ (D1-branes and D5-branes), and a self-dual five-form
$F_{(5)} = \ud C_{(4)} - \frac{1}{2}C_{(2)} \wedge H_{(3)} + \frac{1}{2}B_{(2)} \wedge H_{(3)}$ so there are D3-branes, but not electric and magnetic
versions since $F_{(5)}$ is self-dual. Formally we have the same action as
for type IIA, equation~(\ref{IIAStrAction}) with the summation now over
$n = 1, 3, 5$. However, we should use this action with the understanding that
the self-duality condition is to be imposed on the five-form when solving the
equations of motion and consequently include an extra factor $\frac{1}{2}$ in the kinetic term for $F_{(5)}$. There are also D9-branes which fill all of spacetime. The
only consistent way of including them is to have 16 D9-branes with an
orientifold 9-plane of the opposite charge. This background of type IIB is equivalent to type I string
theory. The tension of the D$p$-branes is
given by the same formula as in type IIA, equation~(\ref{TDp}). Again there is
a NS-NS three-form
field strength $H_{(3)}$ so there are fundamental strings and NS5-branes, with
the fundamental string tension also given by equation~(\ref{TF1}). Some
comments are required on the branes in type IIB. First of all, the
D(-1)-brane or D-instanton is a solution localised at a point in spacetime.
Secondly there is an SL(2, $\mathbf{Z}$) symmetry\footnote{There is apparently an SL(2, $\mathbf{R}$) symmetry in type IIB supergravity but Dirac charge
quantisation breaks this to SL(2, $\mathbf{Z}$).} which acts on the doublet
$(F_{(3)}, H_{(3)})$ as well as the complex coupling constant
$\tau = C_{(0)} + ie^{-\phi}$ and so there are
dyonic $(p,q)$-strings and
$(p,q)5$-branes where $p$ and $q$ are relatively prime integers. These branes
can also be interpreted as one half BPS bound states of $p$ fundamental
strings (NS5-branes) with $q$ D1-branes (D5-branes.) This symmetry can be
directly related to the SL(2, $\mathbf{Z}$) duality of $\mathcal{N} = 4$
3+1-dimensional SYM. Essentially this is the low energy dynamics of the
D3-branes and the complex gauge coupling is simply $\tau$ while the electric
particles correspond to fundamental strings ending on the D3-branes. The
D3-branes are invariant under SL(2, $\mathbf{Z}$) transformations but the
fundamental strings are transformed into dyonic $(p,q)$-strings.

The type IIB branes are related to the type IIA branes by T-duality. T-dualising
in a direction parallel or perpendicular to a D$p$-brane in type IIA/B produces
a D$(p-1)$-brane or D$(p+1)$-brane respectively in type IIB/A. Hence all the
D-branes are related to each other by T-duality and we can see the necessity of
including D8- and D9-branes. The NS5-branes in type IIA and type IIB are
related to each other under T-duality in a direction parallel to the branes.
A T-duality transformation perpendicular to a NS5-brane relates it to a
Kaluza-Klein 5-brane, i.e.\ a geometry consisting of the factors
$(5+1)$-dimensional Minkowski spacetime and Taub-NUT \cite{Ooguri:1996wj} (see
also \cite{Gregory:1997te} for a detailed discussion of this T-duality.) At the
 level of
supergravity solutions, T-duality can only be performed along an isometry
direction and so is of limited use in generating new solutions. For example
T-duality parallel to a D4-brane solution would produce a D3-brane solution
where the D3-brane(s) were smeared over this `transverse' direction. See
\cite{Buscher:1987sk, Buscher:1988qj, Bergshoeff:1995as} for the explicit form
of T-duality transformations of supergravity solutions.

The supergravity solution for $N$ coincident D$p$-branes (in the string-frame)
is
\begin{eqnarray}
\ud s^2 & = & H^{-1/2}\ud x_{(1,p)}^2 + H^{1/2}\ud x_{(9-p)}^2 \\
F_{(p+2)} & = & -\ud(H^{-1}) \wedge \epsilon_{1,p} \\
e^{\phi} & = & H^{\frac{3-p}{4}} \\
H & = & 1 + \frac{c_p N}{r^{7-p}}
\end{eqnarray}
There are also explicit solutions describing NS5-branes \cite{Dabholkar:1990yf}
\begin{eqnarray}
\ud s^2 & = & \ud x_{(1,5)}^2 + H\ud x_{(4)}^2 \\
H_{(3)} & = & \ast \left( \ud(\ln H) \wedge \epsilon_{1,5} \right) \\
e^{\phi} & = & H^{\frac{1}{2}} \\
H & = & 1 + \frac{l_s^2 N}{r^{2}}
\end{eqnarray}
and fundamental strings \cite{Callan:1991dj}
\begin{eqnarray}
\ud s^2 & = & H^{-1}\ud x_{(1,1)}^2 + \ud x_{(8)}^2 \\
H_{(3)} & = & -\ud(H^{-1}) \wedge \epsilon_{1,1} \\
e^{\phi} & = & H^{-\frac{1}{2}} \\
H & = & 1 + \frac{2^5 \pi^2 g_s^2 l_s^6 N}{r^{6}}
\end{eqnarray}
In each case $r$ is the radial coordinate in the directions transverse to the
branes. Note that in all cases the brane solution is determined by a harmonic
function $H$. Solutions describing separated parallel branes are the same as
above, with $H$ replaced by a multi-centred harmonic function.

\subsubsection{Supersymmetry of brane solutions}
\label{KappaSym}

An important property of the brane solutions we have presented
is that they preserve half
the supersymmetry of the supergravity theory. To see this explicitly we need
to consider the form of the supersymmetry transformations in the appropriate supergravity
theory.
In all cases, since we are considering
purely bosonic solutions, the supersymmetry transformations of all bosonic fields
vanish. So we only consider the supersymmetry transformations of the fermionic
fields. The subset of all allowed transformations which vanish for the given
solution are those which are preserved by the solution.
For example, if we consider 
11-dimensional supergravity \cite{Cremmer:1978km} with a field content consisting of the metric
$g_{MN}$, four-form field strength $F_{(4)}$ and Rarita-Schwinger fermion
$\psi_{M\alpha}$, the supersymmetry transformations (in a bosonic background) are given in terms of
a 32-component Majorana spinor $\epsilon_{\alpha}$ by
\begin{equation}
\delta_{\epsilon} \psi_M = D_M\epsilon +
   \frac{1}{288}{\Gamma_M}^{NPQR}F_{NPQR}
\epsilon
-\frac{1}{36}\Gamma^{PQR}F_{MPQR}\epsilon
\equiv \tilde{D}_M\epsilon
\label{11dSUSY}
\end{equation}
where
\begin{equation}
D_{\mu} = \partial_M + \frac{1}{4}\omega^n_{~pM}\hat{\Gamma}_n^{~p}
\end{equation}
is the usual covariant derivative acting on spinors. We use the notation
$\Gamma_M$ for the spacetime Dirac gamma-matrices and $\hat{\Gamma}_m$
for the tangent-space gamma-matrices. The are related by the vielbein
$e_M^m$. I.e.\
\begin{equation}
g_{MN} = e_M^me_N^n\eta_{mn} \;\; , \;\;
\Gamma_M = e_M^m \hat{\Gamma}_m \;\; , \;\;
\left\{ \Gamma_M, \Gamma_N \right\} = 2g_{MN} \;\; , \;\;
\left\{ \hat{\Gamma}_m, \hat{\Gamma}_n \right\} = 2\eta_{mn}
\end{equation}
and antisymmetric combinations are denoted
\begin{equation}
\Gamma_{M_1 \cdots M_p} \equiv \frac{1}{p!} \left( \Gamma_{M_1}\Gamma_{M_2} \cdots \Gamma_{M_p} - \Gamma_{M_2}\Gamma_{M_1} \cdots \Gamma_{M_p} + \cdots \right)
\end{equation}

It can relatively easily be checked that for the M2- and M5-brane solutions
presented in the previous section, these supersymmetry variations vanish with
an arbitrary choice of half of the components of the spinors $\epsilon$. More
precisely, in each case these supersymmetry variations vanish when $\epsilon$
is some specific function multiplying a constant spinor $\epsilon_0$ which
satisfies a projection condition
\begin{equation}
\hat{\Gamma}_{012} \epsilon_0 = \hat{\Gamma}_0 \hat{\Gamma}_1 \hat{\Gamma}_2 \epsilon_0 = \epsilon_0
\end{equation}
for M2-branes with worldvolume directions 012, or
\begin{equation}
\hat{\Gamma}_0 \cdots \hat{\Gamma}_5 \epsilon_0 = \epsilon_0
\end{equation}
for M5-branes with worldvolume directions 012345. Similar results hold for the
brane solutions of type IIA and type IIB supergravity. In the following sections
we will explain why such projection conditions arise and also show how the
brane solutions (taking the M2-brane solution as an example) can be derived
from the requirement of preserving precisely those supersymmetries.

\subsection{Supersymmetry and $\kappa$-symmetry}
We have remarked that $p$-brane solutions preserve supersymmetry. We will now
consider the situation from the brane worldvolume point of view. The idea is
to understand which spacetime supersymmetries should be preserved without
using a specific supergravity solution. We will then
use the fact that the corresponding supersymmetry transformations of the
fermionic supergravity fields must vanish to derive relations between the
bosonic fields. We will then see that this enables us to find the supergravity
solution by solving first-order differential equations rather than the full
set of second-order equations of motion.

One way to understand the relation between spacetime and worldvolume
supersymmetry is through the notion of brane probes. A brane probe is
essentially a brane placed into a fixed background as a `test brane' -- i.e.\
the backreaction of the brane on the background is neglected. We can consider
which supersymmetries are preserved by a probe brane in a given background. The
point is that if the background is generated by the same type (and orientation)
of branes then we expect that the probe brane does not break any of the
supersymmetries preserved by the background. The fact that we are ignoring the
backreaction should not matter since we can consider related backgrounds,
preserving exactly the same supersymmetries, where the backreaction is as small
as we want.

The simplest case is $N+1$ parallel branes where we can consider one to be a
probe brane. The backreaction here is an effect of order $1/N$ but the
supersymmetries preserved do not depend on $N$ and so it is reasonable to
expect that the backreaction does not affect the determination of which
supersymmetries are preserved.

The supersymmetric worldvolume brane action can be derived by (super-)embedding
the brane worldvolume in superspace. In the cases we consider this superspace
has 10 or 11 bosonic spacetime coordinates $X^{\mu}$ and 32 fermionic
coordinates $\Theta_{\alpha}$. In order to have worldvolume supersymmetry the
number of on-shell bosonic and fermionic degrees of freedom must match. This
requires a symmetry of the worldvolume action called $\kappa$-symmetry which
projects out half the components of $\Theta$ on the brane worldvolume. The
$\kappa$-symmetry transformations are very similar to supersymmetry
transformations and can be understood in this way from superembedding the
brane worldvolume superspace into target superspace (see \cite{Sorokin:1999jx}
for a comprehensive review.) In all cases the $\kappa$-symmetry transformation
takes the form
\begin{equation}
\delta_{\kappa} \Theta = \frac{1}{2}(1+\Gamma)\kappa
\end{equation}
while the supersymmetry transformation is
\begin{equation}
\delta_{\epsilon} \Theta = \epsilon
\end{equation}
with the bosonic worldvolume fields not transforming when we have set the
fermionic fields to zero. The form of $\Gamma$ depends on the type of brane
but in all cases $\Gamma^2 = 1$ so that
$\mathcal{P}_{\pm} \equiv \frac{1}{2}(1 \pm \Gamma)$ are projection
operators. Also $\Gamma$ is traceless so that each projection operator
projects out precisely half the components of an arbitrary spinor. Hence decomposing $\Theta$ under
these projections we see that
\begin{eqnarray}
\delta_{\kappa} (\mathcal{P}_{-} \Theta) & = & 0 \\
\delta_{\kappa} (\mathcal{P}_{+} \Theta) & = & \mathcal{P}_{+}\kappa
\end{eqnarray}
Similarly the supersymmetry transformation becomes
\begin{eqnarray}
\delta_{\epsilon} (\mathcal{P}_{-} \Theta) & = & \mathcal{P}_{-}\epsilon \\
\delta_{\epsilon} (\mathcal{P}_{+} \Theta) & = & \mathcal{P}_{+}\epsilon
\end{eqnarray}
So we can consistently set $(1 + \Gamma) \Theta = 0$, fixing $\kappa$-symmetry
and leaving $\mathcal{P}_{-} \Theta$ as the worldvolume fermionic degrees of
freedom. I.e.\ we use $\kappa$-symmetry to set $(1 + \Gamma) \Theta = 0$ and
then we preserve this gauge choice by compensating a supersymmetry
transformation with a $\kappa$-symmetry transformation with parameter
$\kappa = -\epsilon$. This effectively removes $\mathcal{P}_{+} \Theta$,
leaving only $\mathcal{P}_{-} \Theta$ with the supersymmetry transformations
\begin{equation}
\delta_{\epsilon} (\mathcal{P}_{-} \Theta) = \mathcal{P}_{-}\epsilon
\end{equation}
The condition for preservation of worldvolume supersymmetry is then
$\delta_{\epsilon} (\mathcal{P}_{-} \Theta) = 0$, i.e.\
\begin{equation}
\Gamma \epsilon = \epsilon
\end{equation}
Since $\Gamma$ depends on the worldvolume fields we see that the brane locally
preserves half (i.e.\ 16 out of 32) of the background global supersymmetries.
So the brane preserves at most 16 supersymmetries but in general there will
be different projection conditions at different parts of the worldvolume so
that typically all supersymmetry will be broken.

For completeness we give the form of the projector $\Gamma$ for each type of
brane in type IIA, type IIB and M-theory. In each case $\{ \sigma^{\mu}\}$
are worldvolume coordinates on the brane and $\{ X^{M}(\mathbf{\sigma})\}$
describe the bosonic embedding of the brane in spacetime. The induced metric on the
brane worldvolume, $G_{\mu\nu}$, is the pullback of the spacetime metric
$g_{MN}$
\begin{equation}
G_{\mu\nu} = \partial_{\sigma^{\mu}}X^{M}\partial_{\sigma^{\nu}}X^{N}g_{MN}
\end{equation}
and $G = \det(G_{\mu\nu})$. The worldvolume gamma-matrices
$\{ \gamma_{\mu}\}$ are the pullback of the spacetime gamma-matrices
\begin{equation}
\gamma_{\mu} = \partial_{\sigma^{\mu}}X^{M}\Gamma_{M}
\end{equation}
and so
\begin{equation}
\left\{ \gamma_{\mu}, \gamma_{\nu} \right\} = 2G_{\mu\nu}
\end{equation}
So for a $p$-brane we can define (using the conventions
$\epsilon_{01\ldots p} = +1$ so that $\epsilon^{01\ldots p} = G^{-1}$)
\begin{equation}
\gamma_{(p+1)} = \frac{\sqrt{|G|}}{(p+1)!} \epsilon^{\mu_0\mu_1\ldots\mu_p} \gamma_{\mu_0} \gamma_{\mu_1} \ldots \gamma_{\mu_p}
\end{equation}
It is easy to check that
\begin{equation}
\gamma_{(p+1)}^2 = (-1)^{\frac{p(p+1)}{2} + 1}
\end{equation}

In all cases we consider zero worldvolume fieldstrengths -- see e.g.\
\cite{Bergshoeff:1997kr} for the case of D-branes with non-zero worldvolume
fieldstrengths.

\subsubsection{M-branes}

We have M2- and M5-branes in M-theory and eleven-dimensional supergravity. The
projector takes the same simple form in each case
\begin{equation}
\Gamma = \pm \gamma_{(p+1)}
\end{equation}
for M2-branes \cite{Bergshoeff:1987cm} and M5-branes \cite{Bandos:1997ui, Aganagic:1997zq}
with $p=2$ and $p=5$ respectively. In this case $\epsilon$
is a 32-component real spinor. Note also that
\begin{equation}
\hat{\Gamma}_0 \ldots \hat{\Gamma}_{(10)} = 1
\end{equation}

\subsubsection{Type IIA branes}
For D$(2p)$-branes we have
\cite{Aganagic:1997pe, Cederwall:1997ri, Bergshoeff:1997tu}
\begin{equation}
\Gamma = \pm \gamma_{(2p+1)} \Gamma_{(11)}^{p+1}
\end{equation}
where $\Gamma_{(11)} = \hat{\Gamma}_0 \cdots \hat{\Gamma}_9$ when we use
irreducible 16-component real spinors $\epsilon_L$ and $\epsilon_R$ of
opposite chirality. Alternatively we can use a 32-component real spinor
$\epsilon$ as in eleven dimensions and $\Gamma_{(11)}$ is identified with
$\hat{\Gamma}_{(10)}$. This makes the eleven dimensional origin of the type IIA
branes obvious. The projection conditions in each case are
\begin{equation}
\Gamma \epsilon_L = \epsilon_R \;\; \mathrm{or} \;\;
\Gamma \epsilon = \epsilon
\end{equation}

For the NS5-brane we have
\begin{equation}
\Gamma = \pm \gamma_{(6)}
\end{equation}
while for the fundamental string we have
\begin{equation}
\Gamma = \pm \gamma_{(2)} \Gamma_{(11)}
\end{equation}
and in both cases the projection conditions are
\begin{equation}
\Gamma \epsilon_L = \epsilon_L \;\; \mathrm{and} \;\;
\Gamma \epsilon_R = \epsilon_R \;\; \mathrm{or} \;\;
\Gamma \epsilon = \epsilon
\end{equation}

\subsubsection{Type IIB branes}
For D$(2p-1)$-branes we have
\cite{Cederwall:1997pv, Aganagic:1997pe, Cederwall:1997ri, Bergshoeff:1997tu}
\begin{equation}
\Gamma = \pm i\sigma_3^{p}\sigma_2\otimes\gamma_{(2p)}
\end{equation}
where
\begin{equation}
\sigma_2 = \left( \begin{array}{cc} 0 & -i \\ i & 0 \end{array} \right) \; \;
\mathrm{and} \; \;
\sigma_3 = \left( \begin{array}{cc} 1 & 0 \\ 0 & -1 \end{array} \right)
\end{equation}
are the usual Pauli $\sigma$-matrices, acting
on a 32-component spinor which can be decomposed into two 16-component
spinors of the same chirality
\begin{equation}
\epsilon = \left( \begin{array}{c} 1 \\ 0 \end{array} \right) \otimes \epsilon_L
 + \left( \begin{array}{c} 0 \\ 1 \end{array} \right) \otimes \epsilon_R
\end{equation}
For the NS5-brane we have
\begin{equation}
\Gamma = \pm \sigma_3 \otimes \gamma_{(6)}
\end{equation}
and similarly for the fundamental string
\begin{equation}
\Gamma = \pm \sigma_3 \otimes \gamma_{(2)}
\end{equation}
In all cases the projection condition is
\begin{equation}
\Gamma \epsilon = \epsilon
\end{equation}

In the case of $(p,q)$-strings (5-branes)
the projection condition is a linear combination of the projection conditions
for the component branes \cite{Sen:1998xi} (see also \cite{Ortin:1995su})
\begin{equation}
\label{SUSYpqstring}
\Gamma = \pm \left( \begin{array}{cc} \cos \theta & \sin \theta \\
	\sin \theta & -\cos \theta \end{array} \right) \otimes \gamma_{(2)}
\end{equation}
for $(p,q)$-strings and
\begin{equation}
\label{SUSYpq5brane}
\Gamma = \pm \left( \begin{array}{cc} \cos \theta & \sin \theta \\
        \sin \theta & -\cos \theta \end{array} \right) \otimes \gamma_{(6)}
\end{equation}
for $(p,q)$5-branes. In both cases $\theta$ is the argument of $p + q \tau$.

\subsection{M2-brane example}

We will briefly consider the M2-brane solution \cite{Duff:1991xz} as an example of brane solutions,
to explain why the solutions are given in terms of harmonic functions and to
show explicitly how the Killing spinor equations restrict the form of the
solution.
Here, and throughout this paper, we will consider M-branes as simple examples
since typically brane solutions in ten dimensions share the same essential
features but have some extra technical complications, e.g. due to the dilaton.

The form of the solution can be deduced from symmetry considerations. We expect
to have $(2+1)$-dimensional Poincar{\'e} invariance of the M2-brane worldvolume
and, for coincident M2-branes,
SO(8) rotational invariance in the space transverse to the M2-brane. This
restricts the metric to be of the form
\begin{equation}
\ud s^2 = e^{2A} \eta_{\mu\nu} \ud x^{\mu} \ud x^{\nu} + e^{2B} \delta_{\alpha\beta} \ud x^{\alpha} \ud x^{\beta}
\end{equation}
where $\mu$, $\nu$ run over $\{ 0, 1, 2 \}$ and $\alpha$, $\beta$ over
$\{ 3, 4, \ldots , 10 \}$. We will refer to such a configuration as an M2-brane
(or a set of M2-branes) with worldvolume directions 012. The functions $A$ and
$B$ only depend on the transverse coordinates $x^{\alpha}$ while for coincident
M2-branes the dependence is only on the transverse radial coordinate
$r = \sqrt{\delta_{\alpha\beta} x^{\alpha} x^{\beta}}$.

The M2-branes
source the four-form field strength $F_{(4)} = \ud A_{(3)}$. Since the three-form potential,
$A_{(3)}$, couples directly to the M2-brane worldvolume we expect it to have
only non-zero components $A_{012}$ (again with no dependence on the $012$
coordinates) and so the field strength will have the form
\begin{equation}
F_{(4)} = \ud \left( e^C \right) \wedge \epsilon_{1,2}
\end{equation}
Clearly $F_{(4)}$ is an exact 4-form, provided $e^C$ is globally well-defined,
and so will obey the (source-free)
Bianchi identity
\begin{equation}
\ud F_{(4)} = 0
\end{equation}
as expected. We will see explicitly that the solution for $e^C$ is well-defined
everywhere.

Now in this case the $\kappa$-symmetry requirement imposes the projection
condition
\begin{equation}
\hat{\Gamma}_{012} \epsilon = \epsilon
\end{equation}
Using this together with the above metric and four-form we find
\begin{eqnarray}
\tilde{D}_{\mu}\epsilon & = & \partial_{\mu}\epsilon +
	\frac{1}{2} e^{-B} \partial_{\alpha}(e^A) \hat{\Gamma}_{\mu}
	\hat{\Gamma}^{\alpha} \epsilon +
   \frac{1}{6} e^{-2A} e^{-B} \partial_{\alpha}(e^C)
	\hat{\Gamma}_{\mu} \hat{\Gamma}^{\alpha} \epsilon \\
\tilde{D}_{\alpha}\epsilon & = & \partial_{\alpha}\epsilon +
   \frac{1}{2} (\partial_{\beta}B) \hat{\Gamma}_{\alpha}^{\,\; \beta}\epsilon -
   \frac{1}{12} e^{-3A} \partial_{\beta}(e^C)
	\hat{\Gamma}_{\alpha}^{\,\; \beta}\epsilon +
	\frac{1}{6} e^{-3A} \partial_{\alpha}(e^C) \epsilon
\end{eqnarray}
For a supersymmetric solution we require $\tilde{D}_{\mu}\epsilon = 0$ and
$\tilde{D}_{\alpha}\epsilon = 0$. To preserve half the supersymmetry we cannot
impose any additional projection conditions on $\epsilon$. So we can now simply
extract the coefficients of the linearly independent combinations of
$\hat{\Gamma}$-matrices acting on $\epsilon$. This gives the following set of
equations
\begin{eqnarray}
\partial_{\mu}\epsilon & = & 0 \\
\frac{1}{2} \partial_{\alpha}(e^A) & = & -\frac{1}{6} e^{-2A} \partial_{\alpha}(e^C) \\
\partial_{\alpha}\epsilon & = & -\frac{1}{6} e^{-3A} \partial_{\alpha}(e^C) \epsilon \\
\partial_{\beta}B & = & \frac{1}{6} e^{-3A} \partial_{\beta}(e^C)
\end{eqnarray}
We can fix the relevant integration constants by requiring that the metric is
asymptotically the standard eleven-dimensional Minkowski metric. This determines
the following relations
\begin{eqnarray}
e^A & = & e^{-2B} \\
e^C & = & -e^{3A} + \mathrm{constant}
\end{eqnarray}
and
\begin{equation}
\epsilon = e^{A/2}\epsilon_0
\end{equation}
for a constant spinor $\epsilon_0$ which must satisfy
\begin{equation}
\hat{\Gamma}_{012} \epsilon_0 = \epsilon_0
\end{equation}

So the supersymmetry preservation conditions leave us with only one unknown
function. We can easily determine this function by solving the equation of
motion for $F_{(4)}$. Noting that $F_{(4)} \wedge F_{(4)} = 0$ we want
$\ud \ast F_{(4)} = 0$ (or more precisely we should specify source terms in this
equation corresponding to the specific location of the M2-branes.) The only
non-zero components of $F_{(4)}$ are given by
\begin{equation}
F_{\alpha 012} = \partial_{\alpha} \left( e^C \right)
\end{equation}
and so the equation $\ud \ast F_{(4)} = 0$ becomes
\begin{eqnarray}
0 & = & \partial_{\alpha} \left( \sqrt{-g} F_{(4)}^{\alpha 012} \right) =
	\sum_{\alpha} \partial_{\alpha} \left( e^{3A+8B} e^{-2B} e^{-6A} \partial_{\alpha} \left( e^C \right) \right) \nonumber \\
 & = & -\sum_{\alpha} \partial_{\alpha} \left( e^{-6A} \partial_{\alpha} \left( e^{3A} \right) \right) =
	\sum_{\alpha} \partial_{\alpha} \partial_{\alpha} \left( e^{-3A} \right)
\end{eqnarray}
Hence we see that $e^{-3A}$ is a harmonic function of the transverse
coordinates $x^{\alpha}$. In terms of $H \equiv e^{-3A}$ the solution is
given by
\begin{eqnarray}
\ud s^2 & = & H^{-\frac{2}{3}} \eta_{\mu\nu} \ud x^{\mu} \ud x^{\nu} + H^{\frac{1}{3}} \delta_{\alpha\beta} \ud x^{\alpha} \ud x^{\beta} \label{M2metric} \\
F_{(4)} & = & -\ud \left( H^{-1} \right) \wedge \epsilon_{1,2} \label{M24form} \\
H & = & 1 + \sum_i \frac{K}{|x^{\alpha} - y_i^{\alpha}|^6} \label{M2harmonic}
\end{eqnarray}
where $y_i^{\alpha}$ are the locations of the M2-branes and the correctly
normalised source would give, from equations (\ref{c_p}) and (\ref{TM2})
\begin{equation}
K = 2^5 \pi^2 l_P^6
\end{equation}

Notice that because we have looked for a supersymmetric solution, we have only
had to solve first order differential equations coming from the Killing spinor
equations, in order to express the metric and four-form in terms of a single
unknown function, $H$. The equation of motion for $F_{(4)}$ then became a second
order PDE determining $H$. Fortunately this could be solved in general since
it reduced to the flat-space Laplace equation. Note in particular that we have not yet
considered the Einstein equations\footnote{Again we should more precisely
combine the M2-brane action with the supergravity action which would lead to
the correct source terms in the Einstein equations.}
\begin{equation}
R_{MN} - \frac{1}{2} g_{MN} R = \frac{1}{12} F_{MPQR}F_N^{\:\: PQR} -
	\frac{1}{96} g_{MN}F_{PQRS}F^{PQRS}
\end{equation}
However it turns out that these equations are automatically satisfied once we
have imposed the conditions for the existence of Killing spinors and satisfied
the equations of motion for $F_{(4)}$. Of course, this must be the case if such
supersymmetric solutions are to exist since we have fully determined the metric
and four-form without directly using the Einstein equations. So although in general it is
necessary to check the Einstein equations, the constraints from the
first-order Killing spinor equations will usually greatly simplify the problem.
Also, in some cases the existence of Killing spinors automatically ensures that
the Einstein equations are satisfied, provided the equations of motion (and
Bianchi identities) for the field strengths are satisfied. This happens in the
cases where all 32 supersymmetries are preserved or when the metric is
diagonal \cite{Kaya:1999mm, Kaya:2000zs}.

\subsection{Brane probes}
\label{ParBraneProbe}

Here we briefly discuss the idea of using branes to probe a supergravity
background \cite{Douglas:1998uz, Banks:1996nj, Sen:1997sk, Douglas:1997yp}. There are many applications of brane probes. The point of view
we will consider here is that a brane probing a supersymmetric background generated by
branes of the same type and orientation should feel no force since this should
be a BPS configuration. In fact more generally we can expect a no-force
condition whenever it is possible to supersymmetrically embed the brane in the background \cite{Tseytlin:1997hi}. We will see that this leads to constraints on the
supergravity background and gives us another way to derive the parallel brane
solutions. Consider for simplicity the case of parallel $p$-branes without
a dilatonic coupling, i.e.\ M2-, M5- or D3-branes. The same procedure works for
other D-branes with the added complication that we need to also solve the
equation of motion for the dilaton.

We take the background generated by $N$ branes to be of the form
\begin{eqnarray}
\ud s^2 & = & e^{2A} \eta_{\mu\nu} \ud x^{\mu} \ud x^{\nu} +
	e^{2B} \delta_{\alpha\beta} \ud x^{\alpha} \ud x^{\beta} \\
C_{(p+1)} & = & -H^{-1} \epsilon_{1,p}
\end{eqnarray}
where $\mu$, $\nu$ run over $0, 1, \ldots p$ and $F_{(p+2)} = \ud C_{(p+1)}$. The
idea of a brane probe is now
to add one more brane, ignoring the backreaction. By supersymmetry we can place
such a brane anywhere in the transverse space. So we should find that there is
no force on such a (static) brane. If we allow some (rigid) motion we should
find a flat metric on moduli space in terms of the brane worldvolume theory,
since the amount of supersymmetry does allow a non-trivial metric. This is
because, as discussed in the next section, the brane configuration is related
to (directly for D3-branes, via dimensional reduction for M2- and M5-branes) a
maximally supersymmetric gauge
theory which (from supersymmetry considerations) cannot have a non-trivial
metric on moduli space. I.e.\
if we consider a probe brane with
worldvolume coordinates $\{ \sigma^{\mu} \}$ embedded so that
\begin{equation}
X^{\mu} = \sigma^{\mu} \;\; , \;\; X^{\alpha} = v^{\alpha}\sigma^0
\end{equation}
we should find that the probe action reduces to
\begin{equation}
S = T_p \int \ud^{p+1}\sigma \left(
	-\frac{1}{2}v^2 + \mathcal{O}(v^4) \right)
\end{equation}
where $v^2 \equiv \delta_{\alpha\beta}v^{\alpha}v^{\beta}$.

We start with the minimal action for a $p$-brane coupled to $C_{(p+1)}$
\begin{equation}
S = T_p \int \ud^{p+1}\sigma \sqrt{-G} + T_p \int \mathcal{P}(C_{(p+1)})
\end{equation}
where $G$ is the determinant of the pull-back metric and $\mathcal{P}(C_{(p+1)})$ is the
pull-back of $C_{(p+1)}$ onto the brane worldvolume. For example
\begin{equation}
G_{00} = \partial_{\sigma^0}X^M \partial_{\sigma^0}X^N g_{MN} =
	-e^{2A} + e^{2B} v^2
\end{equation}
So we find
\begin{eqnarray}
-G & = & e^{2(p+1)A} \left( 1 - e^{2B}e^{-2A} v^2 \right) \\
\mathcal{P}(C_{(p+1)}) & = & -H^{-1} \epsilon_{1,p}
\end{eqnarray}
Expanding for small velocity $v$ we find
\begin{equation}
S = T_p \int \ud^{p+1}\sigma \left( (e^{(p+1)A} -H^{-1}
        -e^{(p-1)A}e^{2B}\frac{1}{2}v^2 + \mathcal{O}(v^4) \right)
\end{equation}
So the absence of a static potential requires
\begin{equation}
e^{2A} = H^{-\frac{2}{p+1}}
\end{equation}
and the requirement of a flat moduli space metric requires
\begin{equation}
e^{2B} = H e^{2A}
\end{equation}
Again, as for the method of requiring Killing spinors, we must now use the
supergravity equations of motion for $C_{(p+1)}$ to show that $H$ is a harmonic
function.

It can easily be seen that the above solution agrees with equations
(\ref{branemet}), (\ref{braneF}) and (\ref{braneharm}), and so is correct for
M2-, M5- and
D3-branes which have no dilatonic coupling. For other branes it is more
complicated but it is always fairly simple to check a given solution by this
method. We will come back to this point later for intersecting branes.

\subsection{Branes and gauge theories}
One of the most important and useful applications of branes has been their
connection with gauge theories. This connection comes through the fact that
branes are dynamical objects and their dynamics can be described through a
worldvolume action. In the case of D-branes the worldvolume action is (the
supersymmetric extension of) the $(p+1)$-dimensional Dirac-Born-Infeld (DBI)
action \cite{Leigh:1989jq, Dai:1989ua}
\begin{equation}
\label{DBI}
S_{\mathrm{DBI}}^{(p+1)} = T_{Dp} \int \ud^{p+1}\sigma e^{-\phi} \sqrt{-\det{(G_{\mu\nu}+\mathcal{F}_{\mu\nu})}}
\end{equation}
together with Wess-Zumino couplings \cite{Douglas:1995bn} (see also
\cite{Green:1997dd} for additional gravitational couplings which we don't
consider here)
\begin{equation}
\label{WZ}
S_{\mathrm{WZ}}^{(p+1)} = T_{Dp} \int \sum_n \mathcal{P}(C_{(n)}) \wedge e^{\mathcal{F}}
\end{equation}
where $\mathcal{F} = 2\pi l_s^2F-\mathcal{P}(B)$ is a linear combination of the pullback of
the spacetime NS-NS 2-form potential $B$ and a worldvolume 2-form U(1) field
strength $F$. In the WZ terms
the sum is over all the RR potentials present in the given supergravity and the integral
is understood to include only $(p+1)$-forms. In the absence of a $B$-field the low energy limit of this action is
simply the $(p+1)$-dimensional U(1) maximally supersymmetric Yang-Mills action.
The case of a non-zero constant $B$-field has also been a topic of recent
interest \cite{Connes:1998cr, Douglas:1998fm, Hofman:1998iy, Chu:1998qz, Chu:1999gi, Seiberg:1999vs}. The low energy limit is a noncommutative gauge theory which
is a generalisation of usual field theory where products of fields are taken
using a noncommutative Moyal product \cite{Moyal:1949sk}.
Scalar fields in the worldvolume action arise from  the coordinates transverse to the brane
\begin{equation}
\Phi^{\alpha}(\sigma^{\mu}) \equiv \frac{1}{2\pi l_s^2} X^{\alpha}(\sigma^{\mu})
\label{transvscalar}
\end{equation}
Expanding the DBI action in flat space for small $l_s$ and removing the
constant term we get
\begin{equation}
S \approx 4\pi^2 l_s^4 T_{Dp} \int \ud^{p+1}\sigma \left(
	\frac{1}{4}F_{\mu\nu}F^{\mu\nu} + \frac{1}{2} (\partial_{\mu}\Phi^{\alpha})(\partial^{\mu}\Phi^{\alpha}) \right)
\end{equation}
We see that the Yang-mills coupling is
given by
\begin{equation}
\frac{1}{g_{YM}^2} = 4\pi^2 l_s^4 T_{Dp}
\end{equation}
We can consider also a collection of $N$ coincident D$p$-branes. The low energy
dynamics will again be described by maximally supersymmetric Yang-Mills but now
with gauge group U($N$). This can be shown by expanding a non-Abelian
generalisation of the DBI action. However, unfortunately it is not so easy to
calculate such an action from string theory although some progress has been
made \cite{Tseytlin:1997cs1, Tseytlin:1999dj, Myers:1999ps}. Note in particular
that simply replacing the Abelian field strength $F$ with a non-Abelian one in
the DBI action does not give a gauge invariant action.

The simplest way to understand the origin of a non-Abelian gauge group is to
start with $N$ separated parallel branes. The low energy effective action is
that of the massless open string states. Each brane has a U(1) factor arising
from open strings with both endpoints on the brane. Now there are also
open strings with endpoints on each pair of branes. These are massive since
they have a minimal length given by the separation of the branes. However,
when we move the branes together we will gain extra massless states from
these strings. Taking account of the fact that the open strings are oriented
we have in total $N^2$ different types of strings which fill out the adjoint
representation of U($N$). Note that labelling the endpoints according to which
brane they are attached is equivalent to introducing Chan-Paton factors. Also
we are lead to a noncommutative target space since the scalars become
$N \times N$ matrices, as do the coordinates through
equation~(\ref{transvscalar}). The $N$ eigenvalues are interpreted as giving
the positions of the $N$ branes. The genuine noncommutative effects arise
when the matrices cannot be simultaneously diagonalised.

One advantage of thinking about gauge theories in terms of branes is that
many properties can be understood geometrically. We have already mentioned
the simplest of these - separating D-branes gives mass to some states,
proportional to the separation. This is just the Higgs effect. Consider two
coincident D$p$-branes so that the gauge group is U(2). When we separate them
the gauge group is broken to U(1)$^2$ and two states (both orientations of open
strings with one end on each brane) get a mass, $m$, proportional to the
separation, $L$,
$ m \sim L / l_s^2$ -- these are the W-bosons\footnote{When we are talking
about states and W-bosons, we are implicitly referring to complete
supermultiplets since supersymmetry is not being broken.}. At the same time
we give an expectation value to a scalar field because we are now expanding
around a configuration with $X = \mathrm{diag}(L / 2, -L / 2)$.

We will consider more complicated and more interesting brane configurations
related to gauge theories in later sections. The main point to make here is
that field theory states can be identified with open strings and that $N$
parallel D-branes means gauge group U($N$) which can be broken by separating
the branes. It should also be noted that the U(1) associated to the centre of
mass of the branes decouples and so typically we refer to the gauge group for
this system as SU($N$).

\subsection{AdS/CFT correspondence}
\label{AdSCFT}

The AdS/CFT correspondence \cite{Maldacena:1998re, Witten:1998qj, Gubser:1998bc}
describes a duality between string theory or M-theory and gauge theories. The
most useful relations are in the limits where supergravity is a good
approximation of string theory or M-theory. Then specific calculations can be
performed in supergravity backgrounds which are related to properties of
strongly coupled gauge theories. There are many reviews of this topic (e.g.\
\cite{Aharony:1999ti}) and we will not go into much detail here. However, this
is one of the motivations for trying to find supergravity solutions for branes
ending on branes. So we will briefly describe the case of $\mathcal{N} = 4$
four-dimensional SU($N$) Yang-Mills and how the brane solution is used to show
that the supergravity dual is $AdS_5 \times S^5$. The procedure will
be essentially the same when we later consider intersecting brane
configurations.

We start by choosing a gauge theory, here $\mathcal{N} = 4$ four-dimensional
SU($N$) Yang-Mills. We then consider a brane configuration which describes
this theory -- $N$ coincident D3-branes. We now ask what the supergravity
description of this brane configuration is, in the limit appropriate to
describing the field theory. The field theory limit is $l_s \rightarrow 0$
in order to decouple gravity and massive string states. We must however keep
certain quantities fixed when taking this limit, in particular the gauge
coupling, or more conveniently the 't Hooft coupling
\begin{equation}
g_{YM}^2 N = 2\pi g_s N
\end{equation}
and gauge theory masses. Gauge theory masses
and VEVs are fixed by keeping (see equation~(\ref{transvscalar}))
\begin{equation}
U \equiv \frac{r}{l_s^2}
\end{equation}
fixed where $r$ is the radial coordinate transverse to the branes. In this limit
the harmonic function in the D3-brane solution is
\begin{equation}
H = 1 + \frac{4\pi N g_s l_s^4}{r^4} \rightarrow \frac{2g_{YM}^2 N}{U^4l_s^4}
\end{equation}
and hence the metric becomes
\begin{equation}
\frac{1}{l_s^2} \ud s^2 =  \frac{U^2}{L^2}\ud x_{(1,3)}^2 + \frac{\ud U^2}{U^2} +
	L^2 \ud\Omega_5^2
\end{equation}
which is the metric for $AdS_5 \times S^5$ where the $AdS_5$ and $S^5$ both have
radius
\begin{equation}
L = (2g_{YM}^2 N)^{1/4}
\end{equation}
The supergravity description is valid when there are no large curvatures, i.e.\
when $L$ is large, whereas the gauge theory calculations can be performed for
small 't Hooft coupling, i.e.\ when $L$ is small. Hence we have two dual
descriptions.

\section{Intersecting branes and branes ending on branes}
\label{IntandEnd}

In this section we will discuss some general properties of configurations
of intersecting branes, in particular the amount of supersymmetry preserved.
We will see how intersecting branes of the same type are related to a brane wrapping a
smooth cycle and how the conditions for preservation of supersymmetry can be
expressed in terms of calibrations.
We will also consider the closely related generalisation to branes ending on
branes. The existence of such configurations can be understood in several
ways including via dualities of a fundamental string ending on a D-brane or
by deforming an intersecting brane configuration by splitting the `smaller'
brane. The conditions for which branes can end on which other branes can also
be directly derived from charge conservation conditions
\cite{Strominger:1996ac, Townsend:1997em, Argurio:1997qv}. We will see
how branes ending on branes can be described from the brane worldvolume point
of view, using the Born-Infeld action. We will postpone
the discussion of the corresponding supergravity solutions and various
applications to later sections.

\subsection{Orthogonally intersecting branes}
\label{OrthogInt}

The concept of orthogonally intersecting branes is very simple. All the
essential features can be described using the simplest example of two
branes, say a $(p+q_1)$-brane and a $(p+q_2)$-brane embedded as
\begin{displaymath}
\begin{array}{lcccc}
 & p+1 & q_1 & q_2 & \tilde{D} \\
(p+q_1)\mathrm{-brane:} & \overbrace{\mathrm{X} \cdots \mathrm{X}} &
	\overbrace{\mathrm{X} \cdots \mathrm{X}} &
	\overbrace{\mathbf{-} \cdots \mathbf{-}} &
	\overbrace{\mathbf{-} \cdots \mathbf{-}} \\
(p+q_2)\mathrm{-brane:} & \mathrm{X} \cdots \mathrm{X} & 
	\mathbf{-} \cdots \mathbf{-} & 
	\mathrm{X} \cdots \mathrm{X} &
	\mathbf{-} \cdots \mathbf{-}
\end{array}
\end{displaymath}
The two branes have a $(p+1)$-dimensional common worldvolume when they
intersect, i.e.\ when they are at the same position in the totally transverse
$\tilde{D}$-dimensional space. The most important features are related to the 
relative transverse space, i.e.\ the $(q_1+q_2)$--dimensional space with some
directions spanned by one brane with the other brane located at a point. Obviously we can
consider intersections of more types of branes and the relative transverse
space will be described by the numbers of dimensions parallel to some branes
and perpendicular to the others. It is the structure of the relatively
transverse space (the numbers $q_1$ and $q_2$ in the case of two branes) which
determines the amount of supersymmetry preserved. We will now consider this
case of two branes (or more generally two different sets of parallel branes) in
detail. In all cases it is the orientation, not the position, of the branes
which determines the amount of supersymmetry preserved. Indeed, the branes will
not actually intersect unless they are at the same location in the overall
transverse space, although we will generically refer to all configurations
containing non-parallel branes as intersecting brane configurations.

We know which supersymmetries are preserved by a single brane. For two branes
(with an obvious generalisation to more branes) we need to consider which
supersymmetries survive both projection conditions. So for brane 1 we have
$\Gamma^{(1)}\epsilon = \epsilon$ while for brane 2 we have
$\Gamma^{(2)}\epsilon = \epsilon$. In a trivial background, for the cases we
will consider with no non-trivial worldvolume fields, $\Gamma^{(1)}$ and
$\Gamma^{(2)}$ are essentially just a product of $\hat{\Gamma}$-matrices, with
specific expressions given in section~\ref{KappaSym}. So either
$\Gamma^{(1)}\Gamma^{(2)} = \Gamma^{(2)}\Gamma^{(1)}$ or
$\Gamma^{(1)}\Gamma^{(2)} = -\Gamma^{(2)}\Gamma^{(1)}$ and (other than the
trivial case where $\Gamma^{(1)} = \Gamma^{(2)}$ which preserves half the
supersymmetry) we have $tr(\Gamma^{(1)}\Gamma^{(2)}) = 0$. In the case where
$\Gamma^{(1)}$ and $\Gamma^{(2)}$ anti-commute
$$ \epsilon = \Gamma^{(1)}\epsilon = \Gamma^{(1)}\Gamma^{(2)}\epsilon =
	-\Gamma^{(2)}\Gamma^{(1)}\epsilon = -\Gamma^{(2)}\epsilon =
	-\epsilon $$
so clearly no supersymmetry is preserved. In the cases where the projections
commute, we will show that one quarter supersymmetry is preserved. Because $\Gamma^{(1)}$ and
$\Gamma^{(2)}$ commute they can be simultaneously diagonalised and because they
are traceless and square to 1, they have equal numbers of (i.e.\ 16) $+1$ and $-1$
eigenvalues. So we have say $n_{+-}$ simultaneous eigenstates of $\Gamma^{(1)}$
and $\Gamma^{(2)}$ with eigenvalues
$+1$ and $-1$ respectively, $n_{++}$ with eigenvalues $+1$ and $+1$ etc.\ where
$$ n_{++} + n_{+-}  = n_{-+} + n_{--} = n_{++} + n_{-+}  = n_{+-} + n_{--} = 16$$
So in particular
$$ n_{++} = n_{--} \; \mathrm{and} \; n_{+-} = n_{-+} $$
Then since $\Gamma^{(1)}\Gamma^{(2)}$ is traceless it is easy to see that
$$ n_{++} + n_{--} = 16 = n_{+-} + n_{-+} $$
and so
$$ n_{--} = n_{+-} = n_{++} = n_{--} = 8 $$
So the number of supersymmetries preserved is $n_{++} = 8$, i.e.\ one quarter of the 32 supersymmetries.

\subsubsection{Simple examples}
In the cases considered here with no worldvolume fieldstrengths it is easy to
see that, for two D-branes in type IIA or type IIB, or for any two branes of
the same type,
\begin{eqnarray}
\left[ \Gamma^{(1)}, \Gamma^{(2)} \right] = 0  & \mathrm{if} & q_1 + q_2 = 0 \pmod 4 \\
\left\{ \Gamma^{(1)}, \Gamma^{(2)} \right\} = 0  & \mathrm{if} & q_1 + q_2 = 2 \pmod 4
\end{eqnarray}
So we see that a common condition for preserving one quarter supersymmetry is
that the branes have 4 relative transverse dimensions. We will discuss the
case where the branes are of the same type in section~\ref{HolomInt}.

\subsubsection{General orthogonal intersections}

If we have more than two types of orthogonally intersecting branes then we can
similarly analyse the amount of supersymmetry preserved. Clearly the result
depends on the types and the orientations of the branes but we can comment on
some general features.

Obviously if the whole configuration of, say $m$ types of, orthogonally
intersecting branes is to preserve any supersymmetry then a necessary condition
is that each pair of branes must preserve (one quarter) supersymmetry. This
is in fact almost a sufficient condition as can be seen by performing a similar
analysis of the simultaneous eigenstates of the $m$ operators $\Gamma^{(i)}$, as
above for $m=2$. It can be seen that provided the product of any number of
these (distinct) operators is traceless, then precisely $1/2^m$
supersymmetry is preserved. This will typically be true for the cases we are
discussing where each such operator is simply a product of $\hat{\Gamma}$-matrices.
However, it is possible that some product of these operators is plus or minus
the identity, rather than a (traceless) product of $\hat{\Gamma}$-matrices. In this case
with the plus sign, one of the projection conditions is already imposed by the
others and so does not further break supersymmetry. I.e.\ if
$\Gamma^{(1)}\Gamma^{(2)}\Gamma^{(3)} = 1$ then
$$ \Gamma^{(2)}\epsilon = \epsilon \;\; \mathrm{and} \;\;
\Gamma^{(3)}\epsilon = \epsilon \; \Rightarrow \; \Gamma^{(1)}\epsilon = \epsilon $$
The case with the minus
sign breaks all supersymmetry. However, this sign can be changed
by reversing the orientation of one of the branes.

So to summarise, in the case of $m$ orthogonally intersecting branes in
Minkowski spacetime, with no non-trivial worldvolume fields, the condition for
preserving supersymmetry is that all pairs of the projection operators commute.
The amount of supersymmetry preserved is $1/2^l$ where $2 \le l \le m$, with the
proviso that if $l<m$ then the worldvolume orientations of $m-l$ branes are
fixed in terms of the others.

\subsubsection{More Examples}

If we have a configuration of three sets of orthogonally intersecting M5-branes 
with worldvolume directions 012345, 012367 and 012389 then the projection
conditions are
\begin{eqnarray}
\hat{\Gamma}_{012345} \epsilon & = & \epsilon \\
\hat{\Gamma}_{012367} \epsilon & = & \epsilon \\
\hat{\Gamma}_{012389} \epsilon & = & \epsilon
\end{eqnarray}
It can easily be seen that these three conditions are compatible and independent
and so such a configuration will preserve one eighth supersymmetry.

If instead we consider the two types of M5-branes with worldvolume directions
012345 and 012367 then we would preserve one quarter supersymmetry. However, we
can still add another brane without breaking any more supersymmetry. This is
because
\begin{equation}
\epsilon = \hat{\Gamma}_{012345} \epsilon = \hat{\Gamma}_{012345} \hat{\Gamma}_{012367} \epsilon = -\hat{\Gamma}_{4567} \epsilon
\end{equation}
which is the projection condition for a KK6-brane. Since
$\hat{\Gamma}_0 \cdots \hat{\Gamma}_{(10)} = 1$ we can equivalently write this
projection condition as
\begin{equation}
\hat{\Gamma}_{012389(10)} \epsilon = -\epsilon
\end{equation}
Note that the orientation of one set of branes is fixed in terms
of the others. I.e.\ if $s_1 = \pm 1$ and $s_2 = \pm 1$ correspond to the
choice of orientation of the two types of M5-branes then we have one quarter
supersymmetry preserving projection conditions
\begin{eqnarray}
\hat{\Gamma}_{012345} \epsilon & = & s_1\epsilon \\
\hat{\Gamma}_{012367} \epsilon & = & s_2\epsilon
\end{eqnarray}
We still preserve one quarter supersymmetry if the orientation of the KK6-branes
is chosen so that
\begin{equation}
\hat{\Gamma}_{012389(10)} \epsilon = -s_1s_2\epsilon
\end{equation}
while all supersymmetry is broken if we chose the opposite orientation
\begin{equation}
\hat{\Gamma}_{012389(10)} \epsilon = s_1s_2\epsilon
\end{equation}

Upon reduction to type IIA along the isometry direction, say $x^7$, this gives
a quarter BPS configuration of orthogonally intersecting NS5-branes, D4-branes
and D6-branes with worldvolume directions 012345, 01236 and 012389(10)
respectively.

\subsection{Holomorphic intersections and calibrations}
\label{HolomInt}

Here we discuss the special case of two $p$-branes intersecting with $p-1$
common worldvolume directions. This preserves one quarter supersymmetry since,
in the notation of section~\ref{OrthogInt}, $q_1 = q_2 = 2$. Let us for
definiteness consider M2-branes. To start
with we will consider two M2-branes with worldvolume directions 012 and 034, in
flat spacetime. Then we have the following projection conditions
\begin{equation}
\hat{\Gamma}_{012} \epsilon = \hat{\Gamma}_{034} \epsilon = \epsilon
\end{equation}
Note that these relations mean that
\begin{equation}
\hat{\Gamma}_{1234} \epsilon = -\epsilon
\end{equation}
which leads to several relations of the following form
\begin{equation}
\hat{\Gamma}_{013} \epsilon = -\hat{\Gamma}_{024} \epsilon
\end{equation}
If we now define complex coordinates $z^m$, $m=1,2$ by
\begin{equation}
z^1 = x^1 + i x^2 \;\; , \;\; z^2 = x^3 + i x^4
\label{cplxM2coords}
\end{equation}
then we can concisely express the above relations as
\begin{equation}
\hat{\Gamma}_{0m\overline{n}} \epsilon = i \delta_{m\overline{n}}\epsilon
\label{HolomProj}
\end{equation}
We use conventions where $\ud s^2 = 2g_{m\overline{n}}\ud z^m\ud z^{\overline{n}}$ so
that $\delta_{1\overline{1}} = 1/2$ etc.

Now we can easily check that we can add an M2-brane with embedding defined by
an arbitrary holomorphic curve without breaking any more supersymmetry. I.e.\ we
embed the M2-brane in the 1234 directions as the zeroes of a holomorphic
function $f(z^1, z^2)$. This is equivalent to defining a complex coordinate
$z = \sigma^1 + i \sigma^2$ on the brane worldvolume and embedding the brane
so that $z^m = Z^m(z)$. The induced metric on the worldvolume of such a brane
has non-zero components
\begin{equation}
G_{00} = -1 \;\; , \;\; G_{z\overline{z}} = (\partial_z Z^m) (\overline{\partial_z Z^n}) \delta_{m\overline{n}}
\end{equation}
So the projection operator for such a brane is given by
\begin{equation}
\Gamma = \gamma_{(3)} = -i G_{z\overline{z}}^{-1} (\partial_z Z^m) (\overline{\partial_z Z^n}) \hat{\Gamma}_{0m\overline{n}}
\end{equation}
Now using the projections conditions~(\ref{HolomProj}) which preserve one
quarter supersymmetry we see that we don't introduce any extra constraints
and so will still preserve one quarter supersymmetry with this arbitrary
holomorphic embedding
\begin{equation}
\gamma_{(3)} \epsilon = -i  G_{z\overline{z}}^{-1} (\partial_z Z^m) (\overline{\partial_z Z^n}) \hat{\Gamma}_{0m\overline{n}} \epsilon =
	G_{z\overline{z}}^{-1} (\partial_z Z^m) (\overline{\partial_z Z^n}) \delta_{m\overline{n}} \epsilon =
	\epsilon
\end{equation}
It is easy to check that the same results hold in the case where we have an
arbitrary background complex Hermitian metric $g_{m\overline{n}}$, i.e.\
\begin{equation}
\ud s^2 = g_{00} \ud x_0^2 + 2g_{m\overline{n}}\ud z^m \ud z^{\overline{n}} + \ud s_{\perp}^2
\end{equation}
In this case we must also check that the background preserves supersymmetry
even without the branes. For example, without any background field strengths
this will require $g_{m\overline{n}}$ to be a Calabi-Yau metric in order that
there will be covariantly constant spinors. The inclusion of background
fields leads to more complicated restrictions which are still not fully
classified (though see e.g.\ \cite{Gauntlett:2001ur, Gauntlett:2002sc, Gauntlett:2002nw}.) However, in the cases where the background geometry is generated by
the branes themselves, the background will preserve the same supersymmetries as
the brane in flat space. In other words, considerations of supersymmetry
preservation give the same results whether or not we include the backreaction
of the branes. This method was used for this case of branes wrapping
supersymmetric 3-cycles as well as branes wrapping supersymmetric 2-cycles in
Calabi-Yau 3-folds in \cite{Becker:1995kb}. See also
\cite{Townsend:1998mk, SheikhJabbari:1998cv} for an analysis, in terms of
$\kappa$-symmetry projection conditions, of the allowed
supersymmetry-preserving angles between intersecting branes, and
\cite{Cvetic:1998yf} for some further analysis for intersecting and wrapped
branes.

There are other useful way of understanding the geometry of
supersymmetry-preserving embeddings of branes. In the case of D-branes we can
consider the consistent boundary conditions which can be imposed in the string
worldsheet SCFT \cite{Ooguri:1996ck, Berkooz:1996km}. A powerful method we will review
is that of calibrations \cite{HarveyLawson,Harvey:Book} which has been used to
classify the supersymmetric cycles which branes can wrap in various special
holonomy manifolds \cite{Becker:1996ay, Becker:1995kb, Bershadsky:1996qy}. This
construction involves a calibrating form $\Omega$. In a background with
vanishing fieldstrengths, this is a $p$-form which is closed
\begin{equation}
\ud\Omega = 0
\end{equation}
and such that at every point in the manifold the pullback of $\Omega$ to any
tangent $p$-plane is less than or equal to the volume form. We further require
that at any point there exists some $p$-plane for which this bound is
saturated. If we have a calibrating $p$-form then we can use it to find
minimal volume $p$-dimensional submanifolds. To see this consider two
$p$-dimensional submanifolds $\mathcal{M}_1$ and $\mathcal{M}_2$ of volume
$V(\mathcal{M}_1)$ and $V(\mathcal{M}_2)$ respectively which share
the same boundary. Then because $\Omega$ is closed we have
\begin{equation}
\int_{\mathcal{M}_1} \Omega = \int_{\mathcal{M}_2} \Omega
\end{equation}
and because the pullback of $\Omega$ is bounded by the volume form at each point
on $\mathcal{M}_1$ and $\mathcal{M}_2$ we have
\begin{equation}
\int_{\mathcal{M}_1} \Omega \le V(\mathcal{M}_1) \;\; , \;\;
\int_{\mathcal{M}_2} \Omega \le V(\mathcal{M}_2)
\end{equation}
We say that $\mathcal{M}_1$ is a calibrated submanifold if
\begin{equation}
\int_{\mathcal{M}_1} \Omega = V(\mathcal{M}_1)
\end{equation}
The claim is that a calibrated submanifold is a minimal volume submanifold
(with given boundary) which can be easily checked
\begin{equation}
V(\mathcal{M}_1) = \int_{\mathcal{M}_1} \Omega = \int_{\mathcal{M}_2} \Omega
	\le V(\mathcal{M}_2)
\end{equation}
Note that the condition for a calibrated submanifold is a local one - that the
pullback of $\Omega$ is equal to the volume form at each point. Also, for static
branes with no coupling to background fields, minimising the volume is
equivalent to minimising the energy (or Hamiltonian.)

For K\"ahler manifolds we have calibrating $2p$-forms defined in terms of the
K\"ahler form $\omega = ig_{m\overline{n}}\ud z^m \wedge \ud z^{\overline{n}}$ by
\begin{equation}
\Omega = \frac{1}{p!}\omega^p
\end{equation}
The Wirtinger theorem is then the statement that all complex submanifolds are
calibrated submanifolds.
In the above M2-brane example we would have $\Omega = \omega$ (or
$\Omega = \sqrt{-g_{00}} \ud x^0 \wedge \omega$ if we include the part of the
brane world volume which is trivially embedded.) Then clearly for a
holomorphic embedding the pullback of $\omega$ is the same as the volume
form
\begin{equation}
\mathcal{P}(\omega) = i(\partial_z Z^m) (\overline{\partial_z Z^n}) g_{m\overline{n}} \ud z \wedge \ud\overline{z} = iG_{z\overline{z}} \ud z \wedge \ud\overline{z}
\end{equation}

Now we can also easily derive the relation between the methods of using the
$\kappa$-symmetry projection conditions on covariantly constant spinors (we
restrict to the case of no background field strengths) and calibrations. We
will show how the calibrating form can be constructed from the spinor
\cite{Harvey:Book, Becker:1995kb, Becker:1996ay, Gauntlett:1998vk, Acharya:1998en, Figueroa-O'Farrill:1998su}.
We start from the projection condition
\begin{equation}
\frac{1}{2}(1 - \Gamma)\epsilon = 0
\label{calproj}
\end{equation}
where in the absence of any background field strengths $\epsilon$ is a
covariantly constant spinor which we can therefore normalise
\begin{equation}
\epsilon^{\dagger}\epsilon = 1
\end{equation}
Now note that for a static brane, $\Gamma$ is
Hermitian -- for example for M2-branes $\Gamma$ is a (real) linear combination of
$\hat{\Gamma}_0 \hat{\Gamma}_i \hat{\Gamma}_j$ and
\begin{equation}
\left(\hat{\Gamma}_0 \hat{\Gamma}_i \hat{\Gamma}_j \right)^{\dagger} =
  \hat{\Gamma}_j^{\dagger} \hat{\Gamma}_i^{\dagger} \hat{\Gamma}_0^{\dagger} =
  \hat{\Gamma}_j \hat{\Gamma}_i (-\hat{\Gamma}_0) =
  \hat{\Gamma}_0 \hat{\Gamma}_i \hat{\Gamma}_j
\end{equation}
So we now have the following inequality from the projection
condition~(\ref{calproj})
\begin{equation}
0 \le \left( \frac{1}{2}(1 - \Gamma)\epsilon \right)^{\dagger} \left( \frac{1}{2}(1 - \Gamma)\epsilon \right) =
  \epsilon^{\dagger} \frac{1}{2}(1 - \Gamma) \frac{1}{2}(1 - \Gamma) \epsilon =
  \epsilon^{\dagger}\frac{1}{2}(1 - \Gamma)\epsilon
\end{equation}
which can obviously be rearranged to give
\begin{equation}
\epsilon^{\dagger} \Gamma\epsilon \le \epsilon^{\dagger}\epsilon = 1
\end{equation}
This is essentially the same as the inequality for the pullback of a calibrating form. More
precisely we have, for any submanifold $\mathcal{M}$ corresponding to the
embedding of a static brane,
\begin{equation}
V(\mathcal{M}) = \int_{\mathcal{M}} \ud^{p+1}\sigma \sqrt{|G|} \ge
  \int_{\mathcal{M}} \ud^{p+1}\sigma \sqrt{|G|}\epsilon^{\dagger}\Gamma\epsilon
	\equiv \int_{\mathcal{M}} \mathcal{P}(\Omega)
\end{equation}
Note that from the definition of $\Gamma$ it is clear that
$\epsilon^{\dagger} \Gamma\epsilon \ud\sigma^0 \wedge \cdots \wedge \ud\sigma^p$
will be the pullback of a $p+1$-form
\begin{equation}
\Omega \sim \left( \epsilon^{\dagger} \Gamma_{\mu_0 \cdots \mu_p}\epsilon \right) \ud x^{\mu_0} \wedge \cdots \wedge \ud x^{\mu_p}
\end{equation}
to $\mathcal{M}$. Hence, allowing for arbitrary $\mathcal{M}$ this construction
does indeed define a spacetime $p+1$-form, $\Omega$. Furthermore $\Omega$ is
closed because $\epsilon$ is covariantly constant.

Consider the above M2-brane example. We have
\begin{equation}
\epsilon^{\dagger} \hat{\Gamma}_{012} \epsilon =
	\epsilon^{\dagger} \hat{\Gamma}_{034} \epsilon =
	\epsilon^{\dagger}\epsilon = 1
\end{equation}
and otherwise 
\begin{equation}
\epsilon^{\dagger} \hat{\Gamma}_{0ij} \epsilon = 0
\end{equation}
since for example
\begin{equation}
\epsilon^{\dagger} \hat{\Gamma}_{013} \epsilon = 
	\epsilon^{\dagger} \hat{\Gamma}_{013} (\hat{\Gamma}_{012} \epsilon) = 
	-\epsilon^{\dagger} \hat{\Gamma}_{012} \hat{\Gamma}_{013} \epsilon =
	-\left( \hat{\Gamma}_{012} \epsilon \right)^{\dagger} \hat{\Gamma}_{013} \epsilon =
	-\epsilon^{\dagger} \hat{\Gamma}_{013} \epsilon
\end{equation}
So we can now see that, using the vielbein $e_I^i$ for the K\"ahler metric
$g_{IJ}$ on the 1234 space,
\begin{eqnarray}
\sqrt{|G|}\epsilon^{\dagger}\Gamma\epsilon & = &
\epsilon^{\dagger}\sqrt{-g_{00}}e_I^ie_J^j\hat{\Gamma}_{0ij}\epsilon \partial_{\sigma^0}X^0 \partial_{\sigma^1}X^I \partial_{\sigma^2}X^J \\
 & = & \sqrt{-g_{00}}e_I^ie_J^j \delta_{ij} \partial_{\sigma^0}X^0 \partial_{\sigma^1}X^I \partial_{\sigma^2}X^J \\
 & = & \sqrt{-g_{00}}g_{IJ} \partial_{\sigma^0}X^0 \partial_{\sigma^1}X^I \partial_{\sigma^2}X^J
\end{eqnarray}
and so
\begin{equation}
\sqrt{|G|}\epsilon^{\dagger}\Gamma\epsilon \ud\sigma^0 \wedge \ud\sigma^1 \wedge \ud\sigma^2 = \mathcal{P}(\Omega)
\end{equation}
where $\Omega$ can be expressed in terms of the K\"ahler form, $\omega$,
giving the expected result
\begin{equation}
\Omega = \sqrt{-g_{00}} \ud x^0 \wedge \omega
\end{equation}

Note that there are obvious generalisations to the M2-brane wrapping a
2-cycle in an $n$-(complex) dimensional K\"ahler manifold. Similarly for any
$p$-brane we would get the result
\begin{equation}
\Omega = \ud V_{p-1} \wedge \omega
\end{equation}
where $\ud V_{p-1}$ is the $(p-1)$-dimensional volume form for the part of the
$p$-brane trivially embedded.

The above construction crucially depends on the existence of a covariantly
constant spinor which is the same condition required for the background to
preserve supersymmetry. So this construction can be generalised to other
manifolds which
admit covariantly constant spinors. The basic types are listed in
table~\ref{Holonomy} along with the fraction of supersymmetry preserved.
Introducing a brane will break a further one half supersymmetry. Note however
that this is the minimum amount of supersymmetry and more can be preserved in
special cases.
\begin{table}
\begin{center}
\begin{tabular}{c|c|c|c}
Type & Real dimension & Special holonomy & Preserved supersymmetry \\ \hline
Calabi-Yau & $2n$ & SU$(n) \subset$ SO(2n) & $1/2^{n-1}$ \\
Hyper-K\"ahler & $4n$ & Sp$(n) \subset$ SO(4n) & $(n+1)/4^n$ \\
$G_2$ & 7 & $G_2 \subset$ SO(7) & 1/8 \\
Spin(7) & 8 & Spin(7) $\subset$ SO(8) & 1/16
\end{tabular}
\caption{Supersymmetric special holonomy manifolds.}
\label{Holonomy}
\end{center}
\end{table}

For Calabi-Yau manifolds we have $2m$-cycles which are complex submanifolds
calibrated by $\frac{1}{m!}\omega^m$ as already mentioned. These are
collectively referred to as K\"ahler calibrations. There are also
special-Lagrangian submanifolds which are $n$-cycles in the manifolds of complex
dimension $n$, calibrated by
\begin{equation}
\Omega = \mathcal{R}e\left( e^{i\theta} \ud z^1 \wedge \cdots \wedge \ud z^n \right)
\end{equation}
for some constant $\theta$. See \cite{Joyce:2001xt} for a review of Calabi-Yau
manifolds and special Lagrangian submanifolds.

Hyper-K\"ahler manifolds are similar to Calabi-Yau manifolds. The additional
feature is that there are additional calibrating forms corresponding to the
different choices of complex structure. Note that in the case of two complex
dimensional manifolds where a Calabi-Yau manifold is automatically a
hyper-K\"ahler manifold (since SU(2) is the same as Sp(1)) the
special-Lagrangian submanifolds are simply holomorphic curves
with respect to a different choice of complex structure.

There are also calibrations in the cases of exceptional holonomy. In G$_2$
holonomy (seven-dimensional) manifolds we have 3- and 4-cycles calibrated by a
3-form and its Hodge dual 4-form, called associative and coassociative
calibrations respectively. In Spin(7) holonomy (eight-dimensional) manifolds
we have 4-cycles calibrated by a self-dual 4-form, known as Cayley calibrations.

There are various ways to build more general calibrations. For example a
submanifold could be a product of calibrated submanifolds in different spaces.
It is also possible to consider calibrating forms which are linear combinations
of the forms mentioned above. For example in a Calabi-Yau four-fold we can
take a linear combination of the K\"ahler and special Lagrangian calibrating
4-forms \cite{Becker:1996ay}.

Note that the K\"ahler calibrations are particularly simple in that the
embedding conditions are specified by an appropriate number of arbitrary
holomorphic functions. In other cases the embedding conditions for a
supersymmetric submanifold are more complicated and there is no simple
expression of the general solution, e.g.\ see \cite{Karch:1998sj}.

There are also various generalised calibrations which allow for non-trivial
worldvolume fields
\cite{Stanciu:1998sk, Gauntlett:1998wb, Gauntlett:1999aw, Lust:1999pq},
background field strengths \cite{Gutowski:1999iu, Gutowski:1999tu} or both
\cite{Barwald:1999ux}. We can interpret background fields as torsion
\cite{Gauntlett:2001ur, Gutowski:2002bc, Gauntlett:2002sc} due to the way they
enter into the Killing spinor equations $\tilde{D}_{\mu} \epsilon = 0$. In the
cases with background fields (or torsion) the calibrating form is no longer
closed and so calibrated submanifolds are not minimal volume submanifolds.
However, they are again energy-minimising embeddings of branes, including the
appropriate interaction with the background potentials \cite{Townsend:1999nf}.
It is also possible to lift the restriction on the branes being static
\cite{Acharya:1998st}.

Generalised calibrating forms can be constructed from the Killing spinors.
Although these forms are not closed, they are covariantly constant with respect
to the connection with torsion. The existence of covariantly constant spinors
and vectors with respect to such a connection can be understood in terms of
a reduced holonomy group, again with respect to the connection with torsion
\cite{Strominger:1986uh, Papadopoulos:2000iv, Friedrich:2001nh, Ivanov:2001ma, Friedrich:2001yp}.
Actually, in type II theories there are two different relevant connections
\cite{Papadopoulos:2000iv}.
Supersymmetric solutions can be classified by covariantly constant generalised
calibrating forms with respect to each of these connections, referred to as
G-structures \cite{Gauntlett:2001ur, Gauntlett:2002sc}. Studying the possible
G-structures may be useful in classifying all supersymmetric solutions
\cite{Gauntlett:2002nw} of supergravity theories. So far such a classification
only exists for some four-dimensional
\cite{Tod:1983pm, Gibbons:1982fy, Tod:1995jf} and five-dimensional
\cite{Gauntlett:2002nw} supergravity theories. In fact, only recently have all
maximally supersymmetric solutions of ten- and eleven-dimensional
supergravities been classified
\cite{Figueroa-O'Farrill:2001tw, Figueroa-O'Farrill:2001nz, Blau:2001ne}.

\subsection{Branes intersecting at angles}

A natural question to ask is what are the most general supersymmetry-preserving
configurations of intersecting branes. For simplicity we restrict to the case
of branes with no non-trivial worldvolume fields embedded (statically) in
Minkowski spacetime. The problem essentially reduces to a technically more
complicated analysis using the same methods presented for orthogonal
intersections in section~\ref{OrthogInt}\footnote{It is also possible to
consider the conditions for a stable (BPS or non-BPS) configuration by
requiring the forces on each brane to cancel. E.g.\ this was considered for the
case of four D-branes in \cite{Vancea:2000zu} by summing the interaction forces
between each pair of branes. These forces can be calculated from one-loop open
string amplitudes -- see e.g. \cite{Matusis:1997qp} or analytically continue the
scattering amplitudes for two parallel branes \cite{Bachas:1996kx, Douglas:1997yp}.}.
Several cases were considered in
\cite{Berkooz:1996km} where the conditions for supersymmetric intersections
were derived using the $\kappa$-symmetry projections and also string
worldsheet boundary conditions for the cases involving D-branes. The results
for branes of the same type
were expressed in terms of a generalised holonomy which is equivalent to the
results of section~\ref{HolomInt} expressed in terms of calibrated submanifolds.
The field theory interpretation of branes intersecting at angles and the
appearance of chiral fermions was discussed in \cite{Berkooz:1996km, Hashimoto:1997gm}.

Take the case of two planar M5-branes as an example, with one of them having
worldvolume directions 012345. The embedding of the second M5-brane is related
by a rotation of the spatial 5-plane (for static configurations) in the
ten-dimensional space. This can be parameterised in terms of five angles
describing the rotations in each of the 2-planes spanning e.g.\ directions 16, 27, etc.
The supersymmetry projection conditions can be analysed with the result
\cite{Ohta:1998fr} that for various constraints on the angles the possible
fractions of supersymmetry which can be preserved are
$$ \frac{1}{32}, \frac{1}{16},\frac{3}{32},\frac{1}{8},\frac{5}{32},\frac{3}{16},\frac{1}{4}, \frac{1}{2}$$

The same example as well as other configurations of intersecting branes of the
same type can also be interpreted in terms of calibrations
\cite{Gibbons:1998hm, Gauntlett:1998vk, Acharya:1998en, Figueroa-O'Farrill:1998su}.
A related group-theoretic formulation of the above problem for two M5-branes
was presented \cite{Acharya:1998yv}
in terms of finding subgroups of $Spin(10)$ which leave spinors invariant.
Branes related by a rotation in such a subgroup impose the same projection
on the singlet spinor(s). In terms of calibrations this is because the Killing
spinors, and so the calibrating forms (which are built out of them as in
section~\ref{HolomInt},) are singlets of such a subgroup. Hence, given any
calibrated submanifold, there is a whole family of calibrated submanifolds,
related by arbitrary rotations under this subgroup of $Spin(10)$. This
approach can be generalised to the case of more than two sets of intersecting
branes \cite{Acharya:1998yv} and non-static intersecting branes
\cite{Acharya:1998st}. For example, M5-branes with a common
$(1+1)$-dimensional worldvolume and the other four worldvolume directions
related by SU(4) rotations preserve 1/16 supersymmetry
\cite{Ohta:1998fr, Acharya:1998yv} as expected from the
discussion of table~\ref{Holonomy} in section~\ref{HolomInt}. Note however, that
in many cases there is more supersymmetry preserved than the minimum amount.
The cases of branes intersecting at angles in pp-wave backgrounds has recently
been discussed in \cite{Biswas:2002yz, Nayak:2002ty}.

We can think of all static supersymmetric configurations of intersecting
branes of the same type (without worldvolume or background fields) as special (singular)
examples of calibrated submanifolds. For example in the case of two branes we
parameterise the relative orientation of the branes by some angles. We choose
an appropriate calibrating form so that one brane (which we consider a fixed
plane) is calibrated and then demanding that the second brane is also a
calibrated plane will lead to certain conditions on the angles. The cases of
K\"ahler calibrations are particularly simple. We have already seen that the
condition for the static embedding of the spatial part of an M2-brane in a
two (complex) dimensional space is that the embedding is given by the zeroes of
a holomorphic function $f(z^1, z^2) = 0$. The cases where this function
factorises are singular limits of manifolds which describe intersecting branes. For
example $f = z^1z^2$ describes two orthogonally intersecting branes embedded
at $z^1 = 0$ and $z^2 = 0$ whereas $f = z^1z^2 + c$ would describe a smoothly
wrapped M2-brane (for $c \ne 0$) with the same asymptotic form as the two
orthogonally intersecting branes. Similarly we can describe two branes with an
intersection parameterised by an angle $\theta$ by choosing
$f = z^1(\cos\theta z^2 + \sin\theta z^1)$. Again this can be viewed as the
singular limit of a smooth manifold by adding a constant term to $f$.

We can clearly generalise to an arbitrary number, say $n$, of M2-branes by
taking $f$ to be a product of $n$ linear factors. There are various subleading
terms we can add to $f$ to describe a smooth configuration with the same
asymptotic behaviour. Similar results immediately apply to any co-dimension
two K\"ahler calibration which is again specified by a single holomorphic
function (of the appropriate number of complex coordinates.) We can also
generalise to other calibrations although it may not be possible to describe the
related smooth cycles exactly.

The cases where we have different types of branes intersecting cannot be
directly related (though they may be indirectly related via dualities) to a
single brane wrapping a smooth cycle. The conditions for supersymmetry are most
easily analysed using the $\kappa$-symmetry projection conditions although
such cases can presumably by analysed using generalised calibrations if a
suitable supergravity background is known. I.e.\ in the case of two types of
branes we could consider generalised calibrations to determine how the second
brane could be embedded into the background of the first brane.

An interesting example of different types of branes intersecting at angles is
that of a $(p,q)$5-brane web \cite{Aharony:1997ju, Aharony:1998bh} or
$(p,q)$-string web \cite{Aharony:1996xr, Schwarz:1997bh, Gaberdiel:1998ud, Dasgupta:1998pu, Sen:1998xi, Krogh:1998dx, Matsuo:1998jw}. The simplest case is a static
configuration of strings, all lying within a 2-plane, although there are also
non-planar supersymmetric configurations \cite{Bhattacharyya:1998vr}. A 5-brane web is
essentially the same, with the 5-branes having a common $(4+1)$-dimensional
worldvolume. Three $(p_i, q_i)$-strings can meet at a
`string junction' provided the charges are conserved
\cite{Schwarz:1997bh, Gaberdiel:1998ud}, i.e.\ provided (with some appropriate
definition of the orientation of strings at each junction to distinguish
between $(p,q)$-strings and $(-p,-q)$-strings)
\begin{equation}
\sum_{i=1}^3 p_i = 0 = \sum_{i=1}^3 q_i
\end{equation}
In this way complicated webs of strings can be constructed. Such a
configuration will preserve one quarter supersymmetry provided each
$(p,q)$-string lies at an angle in the plane given by $\theta$ in the
$\kappa$-symmetry projector equation~(\ref{SUSYpqstring}) \cite{Sen:1998xi}.
This is precisely the same condition for the forces due to the string tensions
to balance at each junction \cite{Dasgupta:1998pu}. The condition can also be derived
from the dual description of a string junction as an M2-brane with a
holomorphic embedding \cite{Aharony:1996xr, Krogh:1998dx, Matsuo:1998jw}. String junctions
ending on 5-brane webs \cite{Kol:1998cf} or other type IIB branes
\cite{Bergman:1998yw, Bergman:1998br, Hashimoto:1998zs, Bergman:1998gs, Bergman:1998ej, Ohtake:1998sd}
correspond to BPS states in the corresponding field theory. The presence of
7-branes is particularly interesting since a $(p,q)$-string winding around a
7-brane undergoes a monodromy transformation which maps it to a different
$(p,q)$-string. This gives a complicated BPS spectrum which can lead to the
appearance of exceptional symmetry groups
\cite{Gaberdiel:1998ud, DeWolfe:1998zf, Iqbal:1998xb, DeWolfe:1998yf, DeWolfe:1998pr, Hauer:1999pt, Mohri:2000wu, Ohtake:2001fv}.

\subsection{Worldvolume description}
\label{BIons}

Before we discuss supergravity solutions for intersecting branes, it is
interesting to consider the worldvolume description. We will see that if we
have a brane ending on another brane then we can use the worldvolume theory
of either brane. The `smaller' brane then appears as a spike on the worldvolume
of the `larger' brane
\cite{Callan:1998kz, Howe:1998ue, Gibbons:1998xz, Lee:1998xh} or equivalently
the `larger' brane appears as a funnel expanding from the worldvolume of a large
number of `smaller' branes \cite{Constable:1999ac, Diaconescu:1997rk}, rather
similar to the Myers dielectric effect \cite{Myers:1999ps}. Such solitonic
solutions of the DBI action are often called BIons \cite{Gibbons:1998xz}.
To be concrete we will consider the case of $N$ coincident D1-branes with
worldvolume directions 04 ending on a D3-brane with worldvolume directions
0123. Note that we expect this configuration to preserve one quarter
supersymmetry since the D-branes have a four-dimensional relative transverse
space. We closely follow the discussion of \cite{Constable:2001kv}.

\subsubsection{D3-brane worldvolume description}
\label{D3WV}

Let's start with the D3-brane worldvolume theory. The picture is that the
D1-branes can be described as a spike extending from the D3-brane. To see this,
consider the DBI action, equation~(\ref{DBI}), for a D3-brane in Minkowski spacetime with
a static embedding $x^\mu = X^\mu(\sigma^0, \ldots , \sigma^3)$ where
\begin{equation}
X^{0, 1, 2, 3} = \sigma^{0, 1, 2, 3} \;\; , \;\; X^4 = X^4(\sigma^1, \sigma^2, \sigma^3) \;\; , \;\; X^{5, 6, 7, 8, 9} = 0
\end{equation}
From the D3-brane worldvolume point of view a D1-brane ending on it is a magnetically
charged particle (monopole) so we expect a solution where the worldvolume
electric field vanishes
\begin{equation}
E_i = F_{0i} = 0
\end{equation}
but we have a non-trivial magnetic field
\begin{equation}
B_i \equiv \frac{1}{2} \epsilon_{ijk} F^{jk}
\end{equation}
In this case, with vanishing NS-NS B-field and defining $\Phi$ as in
equation~(\ref{transvscalar})
\begin{equation}
\Phi = \frac{1}{\lambda}X^4 \;\; , \;\; \lambda = 2\pi l_s^2
\end{equation}
we have
\begin{eqnarray}
-\det(G_{\mu\nu} + \mathcal{F}_{\mu\nu}) & = &
	1 + \lambda^2(|\nabla\Phi|^2 + |\mathbf{B}|^2) + \lambda^4(\mathbf{B}.\nabla\Phi)^2 \nonumber \\
 & = & (1 \pm \lambda^2(\mathbf{B}.\nabla\Phi))^2 + \lambda^2|\nabla\Phi \mp \mathbf{B}|^2
\end{eqnarray}
So we see the appearance of a BPS bound which is saturated when
\begin{equation}
\mathbf{B} = \pm \nabla\Phi
\end{equation}
Using the Bianchi identity for $F$ with a magnetic source of charge $N$ at
$r\equiv \sqrt{\sigma_1^2 + \sigma_2^2 + \sigma_3^2} = 0$ we have, taking the
`$+$' sign
\begin{equation}
\nabla^2 \Phi = -2\pi N \delta^3(r)
\end{equation}
with solution (for asymptotically vanishing $\Phi$)
\begin{equation}
\Phi = \frac{N}{2r} \;\; , \;\; \mathbf{B} = -\frac{N}{2r^2}\hat{\mathbf{r}}
\end{equation}
So as $r \rightarrow 0$, $\Phi$, and so $X^4$, diverges. This spike is
interpreted as $N$ coincident D1-branes ending on the D3-brane at $r=0$. Indeed
we can easily find the total energy of this static BPS configuration from the
DBI action
\begin{equation}
E = T_{D3}\int \ud^3\sigma \left( 1 + \lambda^2(\mathbf{B}.\nabla\Phi) \right) =
	T_{D3}\int \ud^3\sigma + NT_{D1} \int_0^{\infty} \ud X^4
\end{equation}
which gives precisely the energy of a D3-brane with $N$ D1-branes ending on it
and extending to $X^4 = \infty$.

There are various simple generalisations of this configuration which are
related by dualities. Configurations with D$q$-branes ending on a D$p$-brane
(with $q \le p$) are related to this system by T-duality and similarly have a
D$p$-brane worldvolume description
\cite{Callan:1998kz, Howe:1998ue, Gibbons:1998xz, Howe:1998et, Gauntlett:1998ss, Lambert:2002ms}. The case where $q=p$ can be generalised to describing a
D$p$-brane wrapping a smooth supersymmetric cycle.
By SL(2,$\mathbf{Z}$) duality of the D1-branes ending on a D3-brane we can
describe any $(p,q)$-strings ending on a D3-brane. These solutions were
found in the same way by introducing dyonic sources on the D3-brane
\cite{Howe:1998ue, Gauntlett:1998ss}. The scattering of parallel $(p,q)$-strings
was analysed and the moduli space metric was found in \cite{Gutowski:1998xt}.
In the case of fundamental
strings ending on a D3-brane we can use T-duality to relate this to fundamental
strings ending on any D$p$-brane. Such solutions can be found using the
$(p+1)$-dimensional DBI action \cite{Callan:1998kz, Gibbons:1998xz, Gauntlett:1998ss}. It can also be seen that the case of
fundamental strings ending on a D4-brane can be derived from M2-branes ending on
an M5-brane \cite{Howe:1998ue, Gauntlett:1998ss, Callan:1998kz}.

By analysing fluctuations around these configurations it is possible to perform
further checks on the identification of these worldvolume solitons as other
branes. For example by analysing the reflection of waves along the supposed
fundamental strings ending on a D$p$-brane it can be seen that the appropriate
Dirichlet and Neumann boundary conditions are enforced
\cite{Callan:1998kz, Lee:1998xh, Savvidy:1999wx, Kastor:1999ag}. The behaviour
of such fluctuations along a spike can also be seen to be the same as modes
propagating along a probe string in the D$p$-brane supergravity background
\cite{Lee:1998xh, Kastor:1999ag}. It can also be
checked that these BPS configurations are supersymmetric with the expected
spinor projection conditions and central charges \cite{Lee:1998xh, Gauntlett:1998ss}.
In fact it is possible to see that
BPS solutions of the worldvolume action (or low energy $\sigma$-model) are
related to calibrated embeddings \cite{Gibbons:1998xz, Townsend:1999hi, Gauntlett:2000de, Portugues:2002ih}.

A simple generalisation of the case of $N$ coincident branes ending on another
brane is to separate the $N$ branes \cite{Callan:1998kz}. For example in the case of D1-branes
ending on a D3-brane we can do this by choosing $\Phi$ to be a multi-centred
harmonic function. Note that in terms of each D1-brane the sign of $\Phi$
(near its core) corresponds to whether it extends to $X^4 = \pm \infty$. The
sign of $\mathbf{B}.\hat{\mathbf{r}}$ corresponds to the sign of the charge
of the monopole which is equivalent to whether the orientation of the D1-brane
is towards or away from the D3-brane. Hence we see that the relative sign between
$\mathbf{B}$ and $\nabla \Phi$, which is the same for all D1-branes in a BPS
configuration corresponds to a choice of orientation of the D1-branes. So, as
expected in a BPS configuration all the D1-branes are parallel and there are no
anti-D1-branes.

It is also possible to describe more complicated configurations of
$(p,q)$-string junctions ending on a D3-brane \cite{Gauntlett:1999xz}. It was
also shown that these solutions for strings ending on D3-branes (and M2-branes
ending on M5-branes) could be described in the appropriate parallel D3-brane
(or M5-brane) supergravity background \cite{Gauntlett:1999xz}. Some cases where
worldvolume fields are non-zero (even in the absence of the BIon) or there is
a background NS-NS B-field (or 3-form in the case of M5-branes) have been
considered in
\cite{Gibbons:1998xz, Lust:1999pq, Michishita:2000hu, Youm:2000kr, Karczmarek:2001pn}.
For example, solutions describing non-orthogonal intersections can also be
found, e.g.\ a D1-brane at an angle $\pi/2 - \alpha$ arises when a constant
NS-NS B-field with non-zero component $\lambda B_{12} = \tan\alpha$ is
introduced \cite{Karczmarek:2001pn}.

One process which can be described using the D3-brane worldvolume theory is that
of D1-branes intersecting the D3-brane and then splitting into separated
D1-branes ending on the D3-brane (with equal numbers extending to positive and
negative $X^4$.) This similarly applies to the other configurations mentioned
above. This supports the idea that the conditions for preservation of
supersymmetry are the same for branes ending on branes as for intersecting
branes since the former
configuration, at least at the level of the DBI action, is a smooth BPS
deformation of the latter one. The constraint of equal numbers of branes ending
from each side is not important since we can simply move some branes to infinity
(provided the scalars have at least a $1/r$ fall-off.) It is also possible
to analyse the interaction between different strings ending on a D3-brane
\cite{Bak:1998xp}.

There are also similar solutions when there are several
parallel `larger' branes \cite{Lee:1998xh, Brecher:1998tv}. The BPS properties
essentially guarantee that the BIon solutions of the DBI action are also
solutions of a non-Abelian generalisation. There are also solutions describing
fundamental strings stretching between two parallel D-branes \cite{Gibbons:1998xz}.

There are similar non-BPS configurations describing, for example, D1-branes
ending on D5-branes \cite{Constable:2001ag, Constable:2001kv} or fundamental strings
stretching between a D-brane and an anti-D-brane \cite{Callan:1998kz}. The latter system
is unstable and the annihilation of the branes can be described in this way since
the tube connecting the branes (the string BIon) will expand \cite{Savvidy:1998xx}
(see also \cite{Gorsky:1999gk}.)

\subsubsection{D1-brane worldvolume description}
\label{D1WV}

We now turn to the description of the same system using the worldvolume action
of the $N$ coincident D1-branes \cite{Constable:1999ac, Diaconescu:1997rk, Hashimoto:1998qh}. This requires a non-Abelian DBI action which
is not known in full. However, we can use the symmetrised trace prescription, which in
the case of no worldvolume field strengths is
\begin{eqnarray}
S & = & T_{D1} \int \ud^2\sigma \mathrm{STr} \sqrt{-\det(\eta_{\mu\nu} + \lambda^2\partial_{\mu}\Phi^iQ_{ij}^{-1} \partial_{\nu}\Phi^j) \det(Q^{ij})} \\
Q^{ij} & = & \delta^{ij} + i\lambda[\Phi^i, \Phi^j] \;\; , \;\; i,j = 1,\ldots,8
\end{eqnarray}
This action was proposed for branes in trivial backgrounds (as considered here)
\cite{Tseytlin:1997cs1, Tseytlin:1999dj} and general backgrounds
\cite{Myers:1999ps}. The action is known to be incomplete
\cite{Hashimoto:1997gm, Bain:1999hu} but it is sufficient for BPS configurations
\cite{Thorlacius:1998zd, Hashimoto:1998px, Bak:1998xp}. The symmetrised trace
prescription, $\mathrm{STr}$, means that we symmetrise over all permutations of the
$N \times N$ matrices $\partial_{\mu}\Phi^i$ and $[\Phi^i, \Phi^j]$ after
expanding the square root of the determinants. Since the $\Phi^i$ are related
to coordinates ($X^i = \lambda \Phi^i$) transverse to the D1-branes we see that the non-Abelian gauge
group leads to a non-Abelian space. The $N$ eigenvalues can be interpreted as
the positions of the $N$ individual D1-branes.

Evaluating the determinants for a static configuration with three non-trivial
scalars $\Phi^i$ ($i = 1, 2, 3$) produces a sum of squares which gives a
BPS condition
\begin{equation}
\partial_{\sigma}\Phi^i = \pm \frac{i}{2}\lambda^2 \epsilon^{ijk}[\Phi^j, \Phi^k]
\label{NahmEqns}
\end{equation}
where $\sigma$ is the spatial coordinate on the D1-branes. These three scalars
will correspond to the three worldvolume spatial directions on a D3-brane.
Note \cite{Diaconescu:1997rk} that equations~(\ref{NahmEqns}) are the Nahm
equations \cite{Nahm:1980yw} for BPS monopoles in SU(2) SYM. Indeed if we
introduce another parallel D3-brane, the D1-branes stretching between the two
D3-branes would have precisely that interpretation.
When these conditions are satisfied the total energy is
\begin{equation}
E = NT_{D1} \int \ud\sigma \pm \frac{i}{3}\lambda^2T_{D1} \int \ud\sigma \partial_{\sigma} Tr(\epsilon^{ijk}\Phi^i\Phi^j\Phi^k)
\end{equation}

It is easy to see that we have a solution to equations~(\ref{NahmEqns}) where
$\Phi^i$ are proportional to SU(2) generators $T^i$
\begin{eqnarray}
\Phi^i & = & \pm \frac{1}{2\sigma}T^i \\
{[}T^i, T^j] & = & 2i\epsilon^{ijk}T^k
\end{eqnarray}
Taking the generators $T^i$ to be the irreducible $N \times N$ representation
we have
\begin{equation}
(T^1)^2 + (T^2)^2 +(T^3)^2 = (N^2 - 1) I_{N \times N}
\end{equation}
and so for fixed $\sigma$ the three transverse scalars $X^i = \lambda \Phi^i$
parametrise a non-commutative or fuzzy two-sphere of radius
\begin{equation}
R(\sigma) = \sqrt{\frac{1}{N} Tr((X^i)^2)} = \frac{\pi l_s^2}{\sigma} \sqrt{N^2 - 1}
\end{equation}
So the whole configuration is that of a fuzzy cone or funnel which looks like
$N$ coincident D1-branes for large $\sigma$ (small $R$) and blows up into a
flat D3-brane as $\sigma \rightarrow 0$ (and $R$ diverges.) Identifying $r$
from section~\ref{D3WV} with $R$ we see that the system can be described in terms of either the D3-brane
or the N D1-branes worldvolume theories. Also, the total energy
agrees, for large $N$, with the interpretation that this is a BPS system of $N$
D1-branes ending on a D3-brane
\begin{equation}
E = NT_{D1} \int_0^{\infty} \ud\sigma + \frac{N}{\sqrt{N^2 - 1}}T_{D3} \int_0^{\infty} 4\pi r^2 \ud R
\end{equation}
It can also be checked \cite{Constable:1999ac} that this configuration carries the expected
(up to a factor $\frac{N}{\sqrt{N^2 - 1}}$) D3-brane charge, due to
non-Abelian Wess-Zumino couplings \cite{Myers:1999ps}.

There are once again many other similar examples. D1-branes between two
D3-branes, $(p,q)$-strings ending on D3-branes and embedding the non-Abelian
worldvolume action into a D3-brane supergravity background were all
considered in \cite{Constable:1999ac}. Solutions where $N$ D$p$-branes expand
into a D$(p+r)$-brane can be found and it is possible to identify the U(1)
fieldstrength on the D$(p+r)$-brane \cite{Karczmarek:2001pn}. Fluctuations of
the solutions were also analysed and the results were in agreement with the
analysis from the D3-brane worldvolume theory. $(p,q)$-string junctions which
can end on D3-branes were described using the D1-branes worldvolume theory
\cite{Hashimoto:1998qh}. Including a constant NS-NS B-field produces a solution
describing a non-orthogonal configuration of D1-branes and D3-brane, as was
seen from the D3-brane worldvolume theory \cite{Karczmarek:2001pn}. In
\cite{Constable:2002yn} it was shown that there are more
general funnel solutions of the D1-string non-Abelian worldvolume theory which
describe D3-branes wrapping calibrated 3-cycles, preserving $1/4, 1/8, 1/16$ or
$1/32$ supersymmetry. Non-BPS configurations can also be analysed -- see
\cite{Constable:2001ag, Constable:2001kv} for D1-branes expanding into D5-branes.

\subsubsection{Comments on dual worldvolume descriptions}

While we can expect the system of $N$ D1-branes ending on a D3-brane to have
a good D3-brane worldvolume description for large $r$ and be well-described by
the D1-brane non-Abelian worldvolume theory for small $r$, it is perhaps rather
surprising that the two descriptions match so well, e.g.\ both giving the
correct total energy (for large $N$) and describing fluctuations along the D1-branes. However,
by considering the validity of the (non-Abelian) DBI action we can easily see
that there should be an overlap between the two descriptions for large $N$.

The DBI action is valid for constant field strengths and (related by
supersymmetry) constant first derivatives of the worldvolume scalars. So we can
expect to neglect higher derivative corrections to these solutions provided
the second derivatives of the scalars are small, i.e.\ schematically
\begin{equation}
l_s \partial^2 \Phi \ll \partial \Phi
\end{equation}
For the D3-brane solution this leads to the constraint $r \gg l_s $ while for
the D1-brane description we get $\sigma \gg l_s $ which is equivalent to
$r \ll Nl_s $. Hence for large $N$ we have a large range of $r$ where both
descriptions are useful.

For the non-Abelian action required for the D1-branes worldvolume theory, we
also have higher commutator corrections. Requiring these to be small, i.e.\
\begin{equation}
l_s [\Phi, \Phi] \ll \Phi
\end{equation}
leads to the same constraint, $r \ll Nl_s $, found for the higher derivative
terms. There is a stronger constraint which arises if we also demand that the
expansion in powers of $l_s$ of square root in the action should be convergent.
This requires in addition that
\begin{equation}
l_s^4|\partial \Phi|^2 \ll 1
\end{equation}
which leads to $r \ll \sqrt{N}l_s $, although this still gives a large range
of overlap of the two descriptions for large $N$. However, it is possible that
many higher order terms vanish for BPS configurations such as we have been
considering \cite{Hashimoto:1997gm, Bain:1999hu} and so the range of overlap
may extend to $l_s \ll r \ll Nl_s$.

Finally we note that since there is no explicit dependence on $g_s$ (except
through $T_{Dp}$) in the solutions, we can always take weak enough coupling so
that $g_sN \ll 1$ and the brane actions will decouple from gravity.

\section{Smeared intersections and black holes}
\label{Smeared}

We have seen in section~\ref{HalfBPS} that solutions for parallel branes are
described by a harmonic function with singularities at the locations of the
branes. It turns out that a large class of intersecting brane solutions can be
described in a similar way by following a set of simple rules for combining
the harmonic functions associated to each type of brane
\cite{Tseytlin:1996bh, Gauntlett:1996pb, Tseytlin:1997hi, Tseytlin:1997cs, Bergshoeff:1997tt, Gauntlett:1997cv, Lu:1998mi}.
Specifically, this method applies to supersymmetry-preserving orthogonal
intersections of branes. However, it is possible to relate orthogonal
intersections to non-orthogonal intersections via boosts and duality
transformations \cite{Balasubramanian:1997uc}. This was used to construct
supergravity solutions for non-orthogonal intersections from solutions for
orthogonal intersections \cite{Behrndt:1997ph, Costa:1997dt}. Such solutions
were also found in \cite{Breckenridge:1997ar, Hambli:1997uq} and expressed as a
generalisation of the harmonic functions rules for orthogonal intersections
\cite{Balasubramanian:1998az, Michaud:1997tk}. See also \cite{Biswas:2002sa} for
examples of 1/4-BPS intersecting D2-branes with additional NS-NS two-form
flux, and T-dual configurations of intersecting D1- and D3-branes.
There are also solutions describing bound states of branes within the
worldvolume of other branes
\cite{Duff:1995yh, Bergshoeff:1996sq, Izquierdo:1996ms, Russo:1997if, Breckenridge:1997tt, Costa:1998zd, Tseytlin:1997cs, Sorokin:1997ps}
such as $(p,q)$-5-branes which are bound states of D5- and NS5-branes or
D$p$-branes within a D$(p+2)$-brane preserving one half supersymmetry. These
non-marginal solutions have non-zero binding energy when interpreted in terms
of constituent branes and are more closely related to parallel brane solutions,
e.g.\ having the interpretation of a D$(p+2)$-brane with non-trivial
worldvolume fields. See also \cite{Kumar:2001ag} for the 1/4-BPS case of a $(p,q)$-string web within a D3-brane worldvolume. The BPS solutions we describe here are marginal, i.e.\
there is no binding energy between the constituent branes. The non-marginal
solutions can be derived from marginal ones by various duality transformations,
see e.g.\ \cite{Tseytlin:1997cs}.

In section~\ref{HarmonicFnRules} we give the general method for constructing a
supergravity solution describing any BPS orthogonal intersection of branes.
However, typically some branes must be smeared or delocalised over some of
their transverse directions. In section~\ref{BH} we will present an example of
an intersecting brane solution which becomes a black hole after toroidal
compactification and briefly review how this was used to calculate the entropy
by counting the microscopic degrees of freedom.

\subsection{Harmonic function rules}
\label{HarmonicFnRules}

The harmonic function rules give a method of constructing intersecting brane
solutions by simply combining the one half BPS solutions for the constituent
branes -- adding the field strengths and multiplying the components of the
diagonal metrics. We will consider in detail the case of M2-branes intersecting
in a way which preserves one quarter supersymmetry in section~\ref{OrthogM2}
before stating the method of constructing more general supersymmetric solutions
describing orthogonal intersections of branes in
section~\ref{GenHarmonicFnRules}. Many other examples of such solutions are
presented in a very useful review by Gauntlett \cite{Gauntlett:1997cv}. We show
that these solutions are consistent with the no-force condition for appropriate
probe branes in section~\ref{HFRProbes}.

\subsubsection{Orthogonal intersecting M2-branes}
\label{OrthogM2}

We will consider the case of a set of parallel M2-branes with worldvolume
directions 012 intersecting with another set of parallel M2-branes with
worldvolume directions 034. We already know that the constituent parallel
branes (those with worldvolume directions either 012 or 034) would be described
in terms of harmonic functions $H_{(1)}$ and $H_{(2)}$ respectively as
\begin{eqnarray}
\ud s^2 & = & -H_{(1)}^{-\frac{2}{3}} \ud t^2 + H_{(1)}^{-\frac{2}{3}} \left( \ud x_1^2 + \ud x_2^2 \right) + \nonumber \\
 & & H_{(1)}^{\frac{1}{3}} \left( \ud x_3^2 + \ud x_4^2 \right) + H_{(1)}^{\frac{1}{3}} \ud x_{\perp}^2 \\
F & = & - \ud(H_{(1)}^{-1}) \wedge \ud t \wedge \ud x^1 \wedge \ud x^2
\end{eqnarray}
and
\begin{eqnarray}
\ud s^2 & = & -H_{(2)}^{-\frac{2}{3}} \ud t^2 + H_{(2)}^{\frac{1}{3}} \left( \ud x_1^2 + \ud x_2^2 \right) + \nonumber \\
 & & H_{(2)}^{-\frac{2}{3}} \left( \ud x_3^2 + \ud x_4^2 \right) + H_{(2)}^{\frac{1}{3}} \ud x_{\perp}^2 \\
F & = & - \ud(H_{(2)}^{-1}) \wedge \ud t \wedge \ud x^3 \wedge \ud x^4
\end{eqnarray}
So the solution for intersecting M2-branes is given by adding the fieldstrengths
and multiplying the metric components (or more precisely the vielbeins in order
to preserve the signature)
\begin{eqnarray}
\ud s^2 & = & -H_{(1)}^{-\frac{2}{3}}H_{(2)}^{-\frac{2}{3}} \ud t^2 + H_{(1)}^{-\frac{2}{3}}H_{(2)}^{\frac{1}{3}} \left( \ud x_1^2 + \ud x_2^2 \right) + \nonumber \\
 & & H_{(1)}^{\frac{1}{3}}H_{(2)}^{-\frac{2}{3}} \left( \ud x_3^2 + \ud x_4^2 \right) + H_{(1)}^{\frac{1}{3}}H_{(2)}^{\frac{1}{3}} \ud x_{\perp}^2 \label{IntM2metric} \\
F & = & - \ud(H_{(2)}^{-1}) \wedge \ud t \wedge \ud x^3 \wedge \ud x^4 - \ud(H_{(2)}^{-1}) \wedge \ud t \wedge \ud x^3 \wedge \ud x^4 \label{IntM24form}
\end{eqnarray}
Note that we expect this solution to preserve one quarter supersymmetry since
the two types of M2-branes have precisely four relative transverse dimensions.
Indeed, it is straightforward to check that this is true. However, first we will
comment on the conditions that this combination of harmonic functions does give
a solution to the supergravity equation of motion.

Consider first the equations for the four-form $F$. Clearly the Bianchi
identity $\ud F = 0$ is a linear equation so we can simply add together solutions
to get a new solution. The equation of motion for $F$ is less trivial since
taking the Hodge dual involves the metric. In components we have the
condition (ignoring source terms)
\begin{equation}
\partial_{\mu} \left( |g|^{\frac{1}{2}} F^{\mu \nu \rho \lambda} \right) = 0
\end{equation}
In terms of the separate solutions, this reduced to the condition that $H_{(1)}$
and $H_{(2)}$ are harmonic functions (with respect to the flat-space Laplacian
$\nabla^2 = \sum_i \frac{\partial^2}{\partial (x^i)^2}$) in the spaces
transverse to each type of M2-branes. Now in the intersecting solution we
want the same conditions even though the metric (appearing through the
determinant and used to raise the indices on $F$) has changed. Considering say
$|g|^{\frac{1}{2}} F^{\mu 012}$ we gain an extra factor of
$H_{(2)}^{\frac{1}{3}}$ from $|g|^{\frac{1}{2}}$ and
$H_{(2)}^{\frac{2}{3}-\frac{1}{3}-\frac{1}{3}} = 1$ from raising the three
indices $012$. So we will have the condition that $H_{(1)}$ is a harmonic
function as before, provided we get a factor $H_{(2)}^{-\frac{1}{3}}$ from
raising the $\mu$ index. This will happen precisely if $\mu$ is an index for
one of the totally transverse directions.

By symmetry we get the same result for $H_{(2)}$. So we see that the equation
of motion for $F$ is satisfied provided $H_{(1)}$ and $H_{(2)}$ are harmonic
functions of the coordinates transverse to both types of M2-branes. In terms of the
constituent M2-branes, say those oriented in the $012$ directions, we can
interpret the form of $H_{(1)}$ as describing a continuous distribution of
such M2-branes in the $34$ directions. We
say that these branes are smeared in the $34$ directions. So the
solution corresponds to the intersection of M2-branes oriented in the $012$
and $034$ directions, smeared over their relative transverse directions.

Of course, we must still check that all the supergravity equations of motion
are satisfied (essentially the Einstein equations.) It turns out that they are
and so we do have a supergravity solution for smeared intersecting branes. In
fact this is essentially guaranteed since we have a diagonal metric
\cite{Kaya:1999mm, Kaya:2000zs}.
This solution was originally found by G\"uven \cite{Gueven:1992hh} in the
special case where the two harmonic functions were the same. The interpretation
of the solution as intersecting branes was given in \cite{Papadopoulos:1996uq}
and the generalisation to different harmonic functions for each brane (and to
other types of intersecting brane solutions) soon followed
\cite{Tseytlin:1996bh, Gauntlett:1996pb}.

We should also consider the appropriate normalisation of the coefficients in
the harmonic functions. This can be fixed by considering parallel brane
configurations. We can explicitly smear a solution, say around a circle of
radius $R$, by placing copies of the branes around the circle, say spaced by
$2\pi R/m$. The harmonic function describing the solution is then a
multi-centred harmonic function as in the case of separated parallel branes.
However, since we are smearing the branes rather than introducing other branes,
we should divide the coefficients in the harmonic function by $m$ -- i.e.\ we
imaging splitting each brane into $m$ equal fractions. Taking the limit
$m \rightarrow \infty$ turns the sum of terms in the harmonic function into an
integral which can be performed. The branes are thus smeared over the circle
and the harmonic function becomes a harmonic function in one dimension lower.
This process can be easily generalised to smearing over any torus.

Equivalently, we can just use the fact that the equations of motion are
satisfied for a lower dimensional harmonic function and find the coefficients
by properly normalising the source terms. Essentially this means that we
require the smeared branes to have the same total charge. Since the charge is
proportional to the integral of $\nabla^2 H$ over the directions transverse to
the branes (including the directions over which the brane has been smeared) it
can easily be seen that we get the correct normalisation by replacing
\begin{equation}
\frac{1}{r^a} \rightarrow \frac{a V(S^{a+1})}{(a-b)V(S^{a-b+1})V_b\tilde{r}^{a-b}}
\label{smearednormalisation}
\end{equation}
when smearing the branes over a $b$-dimensional space of volume $V_b$. Here $r$
($\tilde{r}$) is the radial coordinate in the space transverse to the localised
(smeared) branes.

An obvious question to ask is whether we can find localised solutions using the
ansatz of equation (\ref{IntM2metric}) and (\ref{IntM24form}). From the above
discussion we clearly must relax the condition that $H_{(1)}$ and $H_{(2)}$
are harmonic functions. The result of checking the supergravity equations of
motion is that one of these functions, say $H_{(2)}$, must be independent of
the worldvolume directions of the other brane, i.e.\ $x^1$ and $x^2$ in this
example. So we see that at least one of the branes must be smeared. In this case
$H_{(2)}$ is a harmonic function in the totally transverse space, whereas from
the four-form equation of motion $\ud \ast F = 0$, we find that $H_{(1)}$ must satisfy
the curved-space Laplace equation \cite{Tseytlin:1997cs, Lu:1998mi}
\begin{equation}
\partial_M \left( \sqrt{|g|} g^{MN} \partial_N H_{(1)} \right) = H_{(2)} \left(\partial_3^2 + \partial_4^2\right)H_{(1)} + \left(\partial_5^2 + \cdots + \partial_{10}^2\right)H_{(1)} = 0
\end{equation}
In general it is not possible to find explicit solutions for $H_{(1)}$. The
cases where solutions are known are the above case where both branes are
smeared and the case where we solve in the near-core region of the smeared
brane. We will consider this latter case in section~\ref{semilocalised}.

\subsubsection{General construction}
\label{GenHarmonicFnRules}

There is an obvious generalisation of the above example. For a general
 configuration
of orthogonally intersecting branes we simply combine the solutions for each
constituent brane - i.e.\ add together the field strengths and multiply the
components of the diagonal metrics. When we are considering branes in type IIA
or type IIB, there is also a dilaton which is given as the sum of the solutions
for the dilaton. It turns out that this can provide a solution describing
intersecting branes provided the configuration of intersecting branes is
supersymmetric. It is also possible to include a gravitational wave as one of
the constituent `branes'. We will see an example of this in
section~\ref{5dBH}.

The general (with one exception discussed below) result of checking the equations of motion for
such an ansatz is that for each pair of (sets of parallel) branes, at least one of them must be
smeared over the worldvolume directions of the other. In general this gives
various choices for how we wish to smear the branes. Once we have made such a
choice we have determined on which coordinates each `harmonic' function can
depend. Now these `harmonic' functions must satisfy, not the flat-space Laplace
equation but the curved-space Laplace equation \cite{Tseytlin:1997cs, Lu:1998mi}
(see also \cite{Yang:1999ze}) which arises from $\ud \ast F = 0$, with appropriate localised or smeared source
terms. It is not usually possible to find explicit solutions to these coupled
equations but there are some (fully and partially) localised examples which we discuss
in section~\ref{semilocalised}.

We can find explicit solutions in the simplest case where we smear all the
branes over the relative transverse coordinates. In this case, as for two
intersecting branes, the solution is simply given by harmonic functions of
the totally transverse coordinates.
 We shall present an example involving D1- and D5-branes in
section~\ref{5dBH}.

The exception to the above curved-space harmonic function rules is when two
branes intersect with eight relative transverse dimensions (e.g.\ D4-branes
intersecting at a point.) In this case we can allow the harmonic functions to
depend on the relative transverse coordinates provided they are independent of
the overall transverse direction (if there is one.) We will discuss such
solutions in section~\ref{relativelocalised}.

\subsubsection{Brane probes}
\label{HFRProbes}

We can use brane probe techniques to check some of the features of these
intersecting brane solutions. The idea is the same as for parallel branes in
section~\ref{ParBraneProbe}. The difference here is that the reduced
supersymmetry allows for a non-trivial metric on moduli space so the only
requirement we have is that the static potential should be constant. One application
of this method is to take a known solution and then probe with a different type
(or orientation) of brane. If the static potential vanishes then it is possible
to introduce the probe brane into the background. Hence this gives an
alternative derivation of the allowed configurations of intersecting branes
which preserve supersymmetry \cite{Tseytlin:1997hi}\footnote{The assumption is that these are
supersymmetric since we will not have
a moduli space or a no-force condition for non-supersymmetric configurations.}.

Consider the case of $N$ parallel M2-branes with worldvolume directions 012.
The supergravity solution is given by equations (\ref{M2metric}),
(\ref{M24form}) and (\ref{M2harmonic}). We know from section~\ref{ParBraneProbe}
that we can introduce parallel M2-brane probes but here we will consider a probe
M2-brane with worldvolume directions 034. This probe will not couple to the
background 3-form potential so the static potential is (up to a constant factor
involving the brane tension) simply given by the determinant
of the pullback metric
\begin{equation}
\sqrt{-G} = \sqrt{-G_{00}G_{33}G_{44}} = \sqrt{H^{-2/3}H^{1/3}H^{1/3}} = 1
\end{equation}
Hence we see that, as alternatively derived from $\kappa$-symmetry considerations
in section~\ref{HolomInt}, it is consistent to have such an intersection of
M2-branes.

It is equally simple to see that we cannot for example have a supersymmetric
intersection of M2-branes with worldvolume directions 012 and 013. In this
case the probe brane would have a static potential given by
\begin{equation}
\sqrt{-G} = \sqrt{-G_{00}G_{11}G_{33}} = \sqrt{H^{-2/3}H^{-2/3}H^{1/3}} = H^{-1}
\end{equation}
which is clearly not constant.

Finally we can perform a consistency check on a solution for intersecting branes by probing with any
of the constituent branes. For example probing the intersecting M2-brane
solution of equations (\ref{IntM2metric}) and (\ref{IntM24form}) with a
probe M2-brane with worldvolume directions 034 we find a static potential proportional to
\begin{equation}
\sqrt{-G} - H_{(2)}^{-1} = 0
\end{equation}
which is constant, as expected.

\subsection{Application to Black holes}
\label{BH}

While the harmonic function rules provide a method of constructing large
classes of solutions which are related to intersecting branes, the fact that
the branes are smeared over the relative transverse directions means that it is
not obvious that these solutions actually describe what happens at the
intersection. Indeed there are certainly important features which cannot be
described by these solutions such as the relative separations of the branes in
directions over which they are smeared. As we will discuss
in section~\ref{HananyWitten} such parameters are important for certain
intersecting brane configurations which describe gauge theories.

However, there is one obvious situation where the smearing of the branes is not
important and indeed is even a necessary feature of the supergravity solution.
That is when we wish to compactify the directions along which the branes are
smeared. If the branes were not smeared we would anyway have to effectively
construct the smeared solution in order to obtain the necessary isometries for
the reduction. When we perform such a reduction we end up with a $p$-brane
solution of a lower dimensional supergravity, where $p+1$ is the number of
common worldvolume dimensions of the intersecting branes. We can, of course,
further compactify some or all of these $p$ spatial directions. The most important
application of these solutions has been the case where we compactify all $p$
directions to end up with a particle. Such solutions describe black holes
with various charges specified by the constituent intersecting branes.

Although charged black hole solutions can easily be constructed in supergravity
theories, the important point in constructing them from intersecting brane
solutions in ten or eleven dimensions is that we automatically have a string theory (or M-theory)
interpretation. In particular the interpretation of a black hole as a
particular configuration of branes allows us to calculate the entropy of the
black hole by considering the number of massless degrees of freedom in string
theory. This can then be compared to the area of the lower dimensional black
hole horizon to provide a microscopic derivation \cite{Strominger:1996sh} of the Bekenstein-Hawking
\cite{Bekenstein:1972tm, Bekenstein:1973ur, Bardeen:1973gs, Hawking:1974rv, Bekenstein:1974ax, Hawking:1975sw}
black hole entropy. This is a large subject which has been reviewed in detail
in \cite{Maldacena:1996ky, Youm:1997hw, Peet:1998es, Mohaupt:2000mj}.
We will just consider one of the simplest cases \cite{Callan:1996dv, Maldacena:1996ds} in detail in
section~\ref{5dBH} to illustrate the
application of the harmonic function rules. This concerns black holes in
five-dimensional $\mathcal{N} = 8$ supergravity. We will mention some aspects
of other black hole solutions and entropy counting in section~\ref{OtherBH}.

\subsubsection{Five-dimensional extremal black hole entropy}
\label{5dBH}

We can construct a black hole preserving one eighth supersymmetry from a brane
configuration involving the intersection of $N_1$ D1-branes with $N_5$
D5-branes. Such a system would preserve one quarter supersymmetry provided the
D1-branes, say with worldvolume directions 05, are parallel to the D5-branes
which we can therefore choose to have worldvolume directions 056789. Using the
harmonic function rules the metric for this system is
\begin{eqnarray}
\ud s^2 & = & H_1^{-\frac{1}{2}}H_5^{-\frac{1}{2}} (-\ud t^2 + \ud x_5^2) +
	H_1^{\frac{1}{2}}H_5^{\frac{1}{2}} (\ud x_1^2 + \cdots + \ud x_4^2) + \nonumber \\
 & & H_1^{\frac{1}{2}}H_5^{-\frac{1}{2}} (\ud x_6^2 + \cdots + \ud x_9^2) \\
e^{-\phi} & = & H_1^{-\frac{1}{2}}H_5^{\frac{1}{2}}
\end{eqnarray}
where
\begin{eqnarray}
H_1 & = & 1 + \frac{c_1N_1}{r^2} \\
H_5 & = & 1 + \frac{c_5N_5}{r^2} \\
r^2 & = & x_1^2 + \cdots + x_4^2
\end{eqnarray}
where, with the D1-branes smeared over a $T^4$ in the 6789 directions of
volume $V_4$ and the 5 direction compactified on a circle of radius $R_5$, we
have
\begin{eqnarray}
c_1 & = & \frac{4G_5R_5}{\pi g_s l_s^2} \\
c_5 & = & g_s l_s^2
\end{eqnarray}
where the five-dimensional Newton's constant, $G_5$, is related to the
ten-dimensional Newton's constant by the volume of the five compact dimensions,
$V_5 = 2\pi R_5V_4$, by $G_5 = G_{10}/V_5$.
By toroidally compactifying the directions 56789 we can construct a metric for
a point-like mass in five dimensions. However, it turns out that this would
describe a black hole with (classically at least) a horizon of zero area. In
order to get a macroscopic black hole we need to break more supersymmetry and
we can do this by introducing a wave with momentum $P$ along the D1-branes.
The supersymmetry projection condition for a wave carrying
momentum in the 5 direction is
\begin{equation}
\hat{\Gamma}_{05} \epsilon = \epsilon
\end{equation}
The metric for this one eighth BPS system is
\begin{eqnarray}
\ud s^2 & = & H_1^{-\frac{1}{2}}H_5^{-\frac{1}{2}}
	\left( -\ud t^2 + \ud x_5^2 + (H_W - 1)(\ud t-\ud x_5)^2 \right) + \nonumber \\
 & & H_1^{\frac{1}{2}}H_5^{\frac{1}{2}} (\ud x_1^2 + \cdots + \ud x_4^2) +
        H_1^{\frac{1}{2}}H_5^{-\frac{1}{2}} (\ud x_6^2 + \cdots + \ud x_9^2) \\
e^{-\phi} & = & H_1^{-\frac{1}{2}}H_5^{\frac{1}{2}}
\end{eqnarray}
where
\begin{equation}
H_W = 1 + \frac{c_WN_W}{r^2} \;\; , \;\; c_W = \frac{4G_5}{\pi R_5}
\end{equation}
In this metric the wave is also smeared over the D5-brane worldvolume. Since the
momentum is around a circle of radius $R_5$, it quantised as $P = N_W / R_5$
where $N_W$ is a positive integer. We will describe how to calculate the
microscopic entropy but first we will show precisely how the intersecting
brane solution is related to a five-dimensional black hole and what its horizon
area is.

By rewriting the two dimensional `wave part' of the metric as
\begin{equation}
-\ud t^2 + \ud x_5^2 + (H_W - 1)(\ud t-\ud x_5)^2 = -H_W^{-1}\ud t^2 + H_W \left(
	\ud x_5 - (1 - H_W^{-1})\ud t \right)^2
\end{equation}
we can perform the dimensional reduction in the 56789 directions. We see also
from the above form of the wave that there will be a Kaluza-Klein gauge
potential determined by $H_W$. This shows that the five dimensional solution
will have a U(1) charge $N_W$. The solution will also have charges $N_1$ and
$N_5$ (under different U(1) groups) coming directly from the RR-charges in ten
dimensions sourced by the D1- and D5-branes. The five-dimensional Einstein
metric is
\begin{equation}
\ud s_E^2 = \left( H_1 H_5 H_W \right)^{-\frac{2}{3}} \ud t^2 +
	\left( H_1 H_5 H_W \right)^{\frac{1}{3}} (\ud r^2 + r^2 \ud \Omega_3^2)
\label{5dBHmetric}
\end{equation}
The rescaling involved ensures that the ten-dimensional action with the
string-frame metric reduces to the five-dimensional Einstein-Hilbert action.
\begin{equation}
S \sim \frac{1}{2\kappa_{10}^2} \int \ud^{10}x e^{-2\phi} \sqrt{|g|} R(g) =
\frac{V_5}{2\kappa_{10}^2} \int \ud^5x e^{-2\phi} \sqrt{g_I} \sqrt{|g_5|} R(g_5)
 = \frac{1}{2\kappa_5^2} \int \ud^5x \sqrt{|g_E|} R(g_E)
\end{equation}
where $g_I$ is the determinant of the internal (compact) part of the string metric
\begin{equation}
g_I = \left( H_1^{-\frac{1}{2}} H_5^{-\frac{1}{2}} H_W \right)
	\left( H_1^{\frac{1}{2}} H_5^{-\frac{1}{2}} \right)^4 =
	H_1^{\frac{3}{2}} H_5^{-\frac{5}{2}} H_W
\end{equation}

We clearly see that the three charges appear on the same footing in the
five-dimensional Einstein metric, equation~(\ref{5dBHmetric}) even though they
have a different ten-dimensional origin. (Actually, this is not too surprising
since the charges can be permuted by performing various T- and S-duality
transformations.) Since
\begin{equation}
\lim_{r \rightarrow 0} \left( H_1 H_5 H_W \right)^{\frac{1}{3}} r^2 =
	\left( c_1c_5c_WN_1N_5N_W \right)^{\frac{1}{3}}
\end{equation}
we see that the five-dimensional black hole has a (3-sphere) horizon of radius
$\left( 16G_5^2N_1N_5N_W/\pi^2 \right)^{\frac{1}{6}}$ and hence a horizon area
\begin{equation}
A = 8\pi G_5\sqrt{N_1N_5N_W}
\end{equation}
We will now briefly explain the counting of microscopic states in string theory
which leads to a calculation of the black hole entropy, $S_{BH}$, which supports the
Bekenstein-Hawking relation between entropy and horizon area (in five dimensions)
\begin{equation}
S_{BH} = \frac{A}{4G_5} = 2\pi \sqrt{N_1N_5N_W}
\end{equation}

The momentum of the wave in the brane configuration can be viewed as the
momentum carried by fundamental open strings moving around the $x_5$ circle.
The entropy of the system is determined by the number of ways this momentum
can be partitioned among an arbitrary number of fundamental strings. It turns
out that the important strings are those ending on both a D1-brane and a
D5-brane. There are four such bosonic strings for each D1/D5 pair corresponding
to the four independent directions the D1-brane can be moved (within the
D5-brane worldvolume) without reducing the number of massless degrees of freedom
(e.g. moving a D1-brane away from a D5-brane makes even lightest fundamental
string state connecting those branes massive.) In a typical configuration the
D1-branes will be separated, as will the D5-branes, so there will be no massless
open strings with ends on two different D1-branes or two different D5-branes.
So in total there are $4N_1N_5$ massless bosonic degrees of freedom (and by
supersymmetry the same number of massless fermionic degrees of freedom) which
can carry the momentum $P = N_W/R_5$. Each state carries momentum $n/R_5$ around
the circle, for some positive integer $n$ (modes with $n<0$ would break
all supersymmetry.) Alternative methods of counting are to consider the
dimension of the moduli space of $N_1$ instantons (the D1-branes) in a U($N_5$)
gauge theory (on the worldvolume of the D5-branes,) or, to calculate the
central charge of the $(1+1)$-dimensional $\sigma$-model which is the effective
action of the system after compactification on the $T^4$ in the 6789 directions
(i.e.\ in the limit $R_5 \gg V_4^{1/4}$.)

So we are interested in counting the number of ways we
can assign positive integers, totalling $N_W$, to $N_B = N_F = 4N_1N_5$
bosonic and fermionic states. This can be calculated as the coefficient
$d(N_W)$ of $q^{N_W}$ in the partition function
\begin{equation}
Z = \left( \prod_{n=1}^{\infty} (1 + q^n) \right)^{N_F}
	\left( \prod_{n=1}^{\infty} (1 - q^n) \right)^{-N_B} \equiv
	\sum_{m=0}^{\infty} d(m) q^m
\end{equation}
For $N_W \gg N_1N_5 \gg 1$ this gives the entropy of the system
\begin{equation}
S_{BH} = \ln\left( d(N_W) \right) \approx 2\pi \sqrt{N_1N_5N_W}
\end{equation}

It is interesting to see how the entropy can also be calculated when the three
charges are of the same order. In this case the counting above gives a much
lower value for the entropy, $S_{BH} \sim N \ln N$ if $N_1 \sim N_5 \sim N_W \sim N \gg 1$
rather than the expected value of $N^{3/2}$ from the black hole area. The
important idea to get the correct entropy is that we can consider a system of
$N_1$ coincident D1-branes, each wrapped once around a circle or equivalently a
single D1-brane wrapped $N_1$ times around the same circle (or various
intermediate possibilities giving total winding number $N_1$.) Similarly we
can consider a single D5-brane with winding number $N_5$. In this case there
is only one D1-brane and one D5-brane and so by the previous counting we
have only $N_B = N_F = 4$ bosonic and fermionic degrees of freedom. However,
an open string with one end on each brane now sees an effective circle of radius
$N_1N_5R_5$ since each time it moves around the circle of radius $R_5$ its ends
have moved along to the next loop of the D1- and D5-branes. So for $N_1$ and
$N_5$ relatively prime the open string only returns to the same position after
moving round the circle $N_1N_5$ times. If $N_1$ and $N_5$ are not relatively
prime then we simply consider the system to consist of two D1-branes with
winding numbers $\tilde{N}_1$ and $n_1 = N_1 - \tilde{N}_1 \ll N_1$, and two
D5-branes with winding numbers $\tilde{N}_5$ and
$n_5 = N_5 - \tilde{N}_5 \ll N_5$ such that $\tilde{N}_1$ and $\tilde{N}_5$ are
relatively prime. To leading order we only need to consider the branes with
winding numbers $\tilde{N}_1$ and $\tilde{N}_5$. So in all cases to leading
order we have $N_B = N_F = 4$ massless degrees of freedom moving on a circle
of radius $N_1N_5R_5$. The point is that now these modes carry momentum
quantised in units of $1/(N_1N_5R_5)$ and so the total momentum consists of
$N_1N_5N_W$ units which can be partitioned in $d(N_1N_5N_W)$ ways which can be
calculated from the partition function
\begin{equation}
Z = \left( \prod_{n=1}^{\infty} (1 + q^n) \right)^{4}
        \left( \prod_{n=1}^{\infty} (1 - q^n) \right)^{-4} \equiv
        \sum_{m=0}^{\infty} d(m) q^m
\end{equation}
This gives the entropy of the system
\begin{equation}
S_{BH} = \ln\left( d(N_1N_5N_W) \right) \approx 2\pi \sqrt{N_1N_5N_W}
\end{equation}
which agrees with the result from the black hole area. Note that in the
original limit $N_W \gg N_1N_5$ it is not important whether we consider singly
or multiply wound D1- and D5-branes. However, the multiply wound branes lead to
the correct quantisation for the energy of excited states
\cite{Strominger:1996sh, Breckenridge:1997is, Breckenridge:1996sn, Maldacena:1996ds}.

It should be noted that it is not obvious that we should find agreement between
the Bekenstein-Hawking entropy and the counting of microscopic states as
described above. The reason is that the counting was implicitly done as weak
coupling (small values for parameters such as $g_sN$ which correspond to the
`t Hooft coupling in the U($N$) gauge theory on the brane worldvolume,) where
we counted essentially free fundamental string states in a
fixed background of branes. However, the interpretation of the brane
configuration as a macroscopic black hole is valid at large coupling and so
comparison between the two limits requires some sort of non-renormalisation
theorem, for example forbidding large corrections to the mass as we vary the
couplings (at fixed charges.) Such situations may be expected to arise when
supersymmetry is preserved. However it appears that supersymmetry is not
necessarily the important factor since in some supersymmetric cases there is
only agreement up to a numerical factor \cite{Gubser:1996de} whereas there can be
exact agreement in non-extremal (close to the supersymmetric limit) or extremal
(saturating a classical BPS condition) but non-supersymmetric examples which
we will briefly mention in the section~\ref{OtherBH}. See
\cite{Horowitz:1996rn} for an overview of these issues.

\subsubsection{Other black holes}
\label{OtherBH}

It is possible to construct other black hole solutions from intersecting branes.
The method is the same -- simply construct an intersecting brane configuration
using the harmonic function rules and toroidally compactify the relative
transverse space over which the branes are smeared. Again, the most useful
cases are where there is classically a non-zero horizon area. The method of
counting the microscopic degrees of freedom depends on the type of brane
configuration. Typically, open string counting methods similar to those of the five-dimensional
example discussed in section~\ref{5dBH} can be used for ten-dimensional
constructions. However, for eleven-dimensional configurations more general
$\sigma$-model methods are required.

Solutions describing four-dimensional black holes, which can be interpreted
as intersections of NS5-branes, fundamental strings, waves and KK-monopoles,
were constructed in 
\cite{Cvetic:1996uj, Cvetic:1996yq} and used to count the microscopic entropy
in \cite{Cvetic:1996bj}. See also \cite{Cvetic:1996gq} for non-extremal
generalisations. A dual description, similar to the five-dimensional example
of section~\ref{5dBH} is given by
$N_6$ D6-branes with worldvolume
directions 0456789, $N_5$ NS5-branes with worldvolume directions 046789, $N_2$ D2-branes
with worldvolume directions 045 and momentum $P_4 = N_W/R_4$ along the 4 direction. The
$1/8$-BPS supergravity solution can easily be derived and reduced to four dimensions.
The resulting black hole has a non-zero horizon area which corresponds to an
entropy
\begin{equation}
S_{BH} = 2\pi \sqrt{N_2N_5N_6N_W}
\end{equation}
We can again picture the momentum as being carried by fundamental open strings
with ends on a D2-brane and a D6-brane. The role of the
NS5-branes is to split the D2-branes, i.e.\ rather than each D2-brane wrapping
the 5 direction, it is split into $N_5$ pieces which stretch between
consecutive NS5-branes. In this way there are $N_2N_5N_6$ distinct
possibilities for an open string to have ends on a D6-brane and one of the
$N_2N_5$ D2-brane segments. Once again each such string has four bosonic and four
fermionic degrees of freedom and so the entropy precisely matches the
prediction from the horizon area. The interpretation of the
system as consisting of branes wrapping many times around the 4 direction is
important for calculating the correct entropy when $N_W$ is not large, in this case
compared to $N_2N_5N_6$. As for the five-dimensional black hole in section~\ref{5dBH} we could also
count the degrees of freedom by viewing the (segments of) D2-branes as
instantons within the D6-branes worldvolume.

This system can also be studied in eleven dimensions. The direct lift is not so
convenient since it involves KK6-branes in eleven dimensions. However, if we
first T-dualise in the 8 and 9 directions and then lift to eleven dimensions we
get a configuration involving three sets of orthogonally intersecting M5-branes
with worldvolume directions 04567(10), 046789 and 04589(10) with momentum along
the 4 direction. Once we have reduced to four dimensions we get the same black
hole solution. So we have a choice of many possible (duality related) ten- and
eleven-dimensional intersecting brane configurations to describe the same
lower dimensional black holes. We can in principle use any of these
configurations to count the microscopic degrees of freedom. An interesting
question is how do we calculate the microscopic entropy from this
eleven-dimensional configuration of intersecting M5-branes. It was proposed
\cite{Klebanov:1996mh} (see also \cite{Balasubramanian:1996rx, Behrndt:1997mm})
that the momentum
is carried by M2-branes, analogous to the fundamental strings in ten-dimensional
configurations. However, in this case the counting requires the M2-branes to
end on three different M5-branes -- such M2-branes would be massless states at
each point where three M5-branes intersect. (Generically the parallel M5-branes
will be separated so there are no other massless states.) It is not clear why
we should
expect to count the states this way so this is really a prediction about
M-theory rather than a derivation of the black hole entropy. The same result
for the entropy can be derived from type IIA string theory by first compactifying along the 4 direction where now the momentum
becomes a number of D0-branes which can be distributed among the points where
three D4-branes intersect \cite{Balasubramanian:1996rx}. The general
four-dimensional black hole solutions arising from toroidal compactification
from ten or eleven dimensions have been constructed \cite{Bertolini:2000ei} and
their microscopic entropy calculated \cite{Bertolini:2000ya}.

The above constructions can also be generalised to the case where the
56789(10)-space is a Calabi-Yau threefold. The choice of Calabi-Yau threefold
determines the intersection numbers of the M5-branes (or D4-branes) and the
proposed rules for counting states again reproduces the entropy expected from
the horizon area \cite{Maldacena:1997qk}. However, it is possible to count the states
precisely without making any assumption about the properties of M2-branes
ending on M5-branes. This can be done by calculating the central charge of the
two-dimensional $\sigma$-model describing an M5-brane wrapped on a 4-cycle in
the Calabi-Yau threefold \cite{Maldacena:1997de} (see also
\cite{Strominger:1996sh, Tseytlin:1996as, Tseytlin:1996qg}.)
This is possible because generically the M5-branes will
not have singular intersections but will be described by a single M5-brane
wrapping a smooth 4-cycle (given by the zeroes of a holomorphic function.) The
possible deformations of such a 4-cycle (see \cite{Vafa:1998gr} for a similar
type IIA description) together with the M5-brane self-dual
worldvolume fieldstrength contribute to the central charge. This gives the
expected entropy and also quantum corrections \cite{Maldacena:1997de, Vafa:1998gr, Harvey:1998bx, LopesCardoso:1999ur, Mohaupt:2000mj}.

There are also generalisations to non-extremal and rotating black holes. It
is often possible to successfully calculate the microscopic entropy and also
analyse Hawking radiation. We refer the reader to the reviews
\cite{Maldacena:1996ky, Peet:1998es, Mohaupt:2000mj} and references therein.

Finally, it is interesting to note that in some cases, even without
supersymmetry, the counting of microscopic degrees of freedom agrees with the
Bekenstein-Hawking entropy \cite{Horowitz:1996ac, Ortin:1998bz, Dabholkar:1997rk, Sheinblatt:1998nt}. These black
holes can be described using intersecting branes but all supersymmetry is
broken since some branes (or waves) have the opposite orientation to that
required for preserving supersymmetry, or simply because the brane
configurations are not supersymmetric (e.g.\ consisting of D0- and D6-branes.)
For example \cite{Dabholkar:1997rk} the
five-dimensional black holes considered in section~\ref{5dBH} can also be
constructed in SO(32) type I string theory (which is a projection of type IIB)
but reversing the direction of the propagating wave breaks all supersymmetry.
Hence such configurations are not even related to
supersymmetric configurations by a small deformation. However, they can be
BPS in the sense of saturating a classical BPS bound relating mass to charge.
The agreement between counting of states and Bekenstein-Hawking entropy implies
that this classical relation should not be modified by quantum corrections
(at least within the regime under consideration of large charges.) Such
quantum corrections (or their absence) and direct comparisons between the
strong and weak coupling regimes have been considered in
\cite{Itzhaki:1997fm, Ortin:1997yn, Dabholkar:1998fc, Pierre:1997zd, Dhar:1998ip, Barbon:1998nx}.

\section{(Partially) localised solutions}
\label{semilocalised}

While the smeared intersecting brane solutions were all that was required for
constructing lower dimensional black holes, it is natural to ask whether we
can construct localised solutions. As discussed in
section~\ref{GenHarmonicFnRules} we can find implicit solutions for partially
localised intersections, in terms of functions which satisfy curved-space
Laplace equations. We will show how explicit solutions can be found in a
near-core limit in section~\ref{nearhorizoncurvedharm}. The special case of
eight relative transverse dimensions is considered in
section~\ref{relativelocalised}. Then, in sections \ref{D2inD6} and
\ref{D4intD6}, we consider intersecting brane solutions involving the
near-core limit of D6-branes which can be derived from orbifolds in eleven
dimensions. A more general approach to finding intersecting brane solutions
from geometry is discussed in section~\ref{IntfromGeom}.

\subsection{Near-horizon solutions from curved-space harmonic function rules}
\label{nearhorizoncurvedharm}

Consider, for example two sets of coincident M5-branes, say $N$ M5-branes with
worldvolume directions 012345 and $M$ with worldvolume directions 012367, all
at the origin of the transverse space, $r=0$. The
metric for such a solution has the form
\begin{eqnarray}
\ud s^2 & = & H_1^{-1/3}H_2^{-1/3}(-\ud x_0^2 + \cdots + \ud x_3^2) +
        H_1^{-1/3}H_2^{2/3}(\ud x_4^2 + \ud x_5^2) + \nonumber \\
 & & H_1^{2/3}H_2^{-1/3}(\ud x_6^2 + \ud x_7^2) +
        H_1^{2/3}H_2^{2/3}(\ud r^2 + r^2\ud\Omega_2^2)
\end{eqnarray}
And if we choose to smear the $N$ M5-branes over the 67 directions then we
have $H_1 = H_1(r)$ and $H_2 = H_2(x_4, x_5, r)$. The curved space Laplace
equations then become (up to appropriately smeared source terms)
\begin{eqnarray}
\nabla_{(3)}^2[r] H_1 & = & 0 \\
H_1(\partial_4^2 + \partial_5^2) H_2 + \nabla_{(3)}^2[r] H_2 & = & 0
\end{eqnarray}
We use the notation $\nabla_{(d)}^2[r]$ for the (flat-space) Laplace operator in
$d$ dimensions with radial coordinate $r$.
We can clearly solve the first of these equations with a solution of the form
\begin{equation}
H_1 = 1 + \frac{c}{r}
\end{equation}
From equations (\ref{smearednormalisation}), (\ref{c_p}) and (\ref{TM2}) we have
\begin{equation}
c = \frac{2 \pi^2 N l_P^3}{V_2}
\end{equation}
where $V_2$ is the volume of the compactified 67 space over which the $N$
branes are smeared.
The equation for $H_2$ cannot easily be solved. However, if we take
the near-horizon limit $r \rightarrow 0$ so that $H_1 \rightarrow c/r$ then the problem simplifies since
we can write
\begin{equation}
\frac{c}{r}(\partial_4^2 + \partial_5^2) + \nabla_{(3)}^2[r] =
        \frac{c}{r} \nabla_{(6)}^2[R]
\end{equation}
where
\begin{equation}
R^2 = x_4^2 + x_5^2 + 4cr
\end{equation}
and we have assumed that $H_2$ only depends on $x^4$, $x^5$ and $r$ through an
effective transverse radial coordinate $R$. So we find the solution
\cite{Youm:1999zs, Fayyazuddin:1999zu, Loewy:1999mn}
\begin{equation}
H_2 = 1 + \frac{C}{R^4}
\end{equation}
where, after some manipulation of the delta-function source, we find
\begin{equation}
C = 6\pi M l_p^2 c = \frac{12\pi^3MNl_p^6}{V_2}
\end{equation}
Notice the unexpected appearance of a six-dimensional Laplacian even though
there are only five dimensions transverse to the brane. We will see a similar
phenomenon in section~\ref{D2inD6} when we consider branes intersecting
Kaluza-Klein monopoles but there we have a natural explanation for this effect
since the solutions originate in one dimension higher via a Kaluza-Klein
compactification.

Many other semi-localised solutions can be constructed by this method
\cite{Youm:1999zs, Loewy:1999mn, Youm:1999tt}. A
particularly interesting case is when one type of brane is contained within the
worldvolume of another. In this case the harmonic function for the larger
branes is automatically smeared over the worldvolume directions of the smaller
branes. Therefore we can construct a fully localised solution for this
system by solving the curved-space Laplace equation for the smaller branes. As
above we can only find explicit solutions in the near-core region of the larger
brane. Several examples of these types of intersections, including pp-waves
on the worldvolume of branes have been discussed, for example, in
\cite{Yang:1999ze, Youm:1999zs, Brecher:2000pa}. We will
present an alternative method of constructing such solutions in
section~\ref{D2inD6} in the special cases where the larger brane is a
Kaluza-Klein monopole. See also
\cite{Gibbons:1996vg, Chu:1998in, Kogan:1998re, Bianchi:1998nk, Park:1998uv}
for the case of a D(-1)-brane in
$AdS_5 \times S^5$ (the near-horizon limit of D3-branes) with various degrees
of localisation. The geometry of the near-horizon limits of several
semi-localised solutions was studies in \cite{Cvetic:2000cj}. As expected from
the AdS/CFT correspondence these geometries are of the form of warped products
of AdS.

\subsection{Localisation in relative transverse directions}
\label{relativelocalised}

In the case where we have two branes intersecting with an eight-dimensional
relative transverse space it is possible to find a solution using the
curved-space harmonic function rules where the branes are fully localised in
the relative transverse space but smeared over any overall transverse
directions. In some ten-dimensional examples this leads to a fully localised
intersecting brane solution since there are no overall transverse directions.
The first example of such a solution was the case of two NS5-branes, say with
worldvolume directions 012345 and 016789 \cite{Khuri:1993cs, Khuri:1993ii}
which was interpreted as a fully localised intersecting brane solution in
\cite{Gauntlett:1996pb}. This solution was generalised to other ten- and
eleven-dimensional cases such as intersecting M5-branes with the same
$(1+1)$-dimensional intersection but smeared over the eleventh dimension \cite{Gauntlett:1996pb}.
The classification of these intersections as the most general cases (within
the context of the curved-space harmonic function rules) with full
localisation of both branes in the relative transverse space (other than branes within branes)
was given in \cite{Tseytlin:1997cs, Edelstein:1998vs}. Generalisations to
including more branes including e.g.\ M2-branes with worldvolume directions
01(10) which still preserve 1/4 supersymmetry were given in
\cite{Edelstein:1998vs, Gauntlett:1998kc}. See also
\cite{Ohta:1997gw, Aref'eva:1998uh} for the case of non-extremal intersecting
branes.

Consider the case of two sets of M5-branes together with M2-branes as above.
The metric is of the form
\begin{eqnarray}
\ud s^2 & = & H_1^{-1/3}H_2^{-1/3}H_{M2}^{-2/3} (-\ud x_0^2 + \ud x_1^2) +
        H_1^{-1/3}H_2^{2/3}H_{M2}^{1/3} \ud x_{(4)}^2 + \nonumber \\
 & & H_1^{2/3}H_2^{-1/3}H_{M2}^{1/3} d\tilde{x}_{(4)}^2 +
        H_1^{2/3}H_2^{2/3}H_{M2}^{-2/3} \ud x_{10}^2
\end{eqnarray}
The constraints allow the M5-branes to be smeared over $x^{10}$ but fully
localised in the eight-dimensional relative transverse space. The constraints
on each set of M5-branes together with the M2-branes are the standard ones so
that one type of brane must be smeared over the worldvolume directions of the
other. In this case both sets of M5-branes are already smeared over the
worldvolume directions of the M2-branes so the M2-branes can be fully localised
and we don't need to smear the M5-branes over any other directions. It is now
easy to check that the harmonic functions for the M5-branes satisfy a flat-space
Laplace equation in the four appropriate relative transverse directions while
the harmonic function for the M2-branes satisfies a curved-space Laplace
equation
\begin{eqnarray}
\tilde{\nabla}_{(4)}^2 H_1 & = & 0 \\
\nabla_{(4)}^2 H_2 & = & 0 \\
\left( H_1 \nabla_{(4)}^2 + H_2 \tilde{\nabla}_{(4)}^2 \right) H_{M2} & = & 0
\end{eqnarray}
So clearly we can find explicit solutions in the case of intersecting M5-branes
alone
but in general will not be able to solve the equation for $H_{M2}$ when we also
include M2-branes. Note that if we reduce to type IIA along $x^{10}$ we will
get fully localised solutions for intersecting NS5-branes together with
fundamental strings. Again we will have an explicit solution only for the case
without any strings.

\subsection{Branes within D6-branes}
\label{D2inD6}

In this section we describe intersecting brane solutions involving D6-branes in
a near-core limit. However, it will be clear that such solutions apply to
other dimensions where we can have branes intersecting Kaluza-Klein monopoles.
First, in section~\ref{D2inD6}, we consider the cases where we have branes
contained within the worldvolume of the D6-branes. As mentioned in
section~\ref{nearhorizoncurvedharm} these are special cases of solutions which can be
constructed using curved-space harmonic function rules. However, the cases
described here have a particularly simple construction which generalises to
the cases of branes intersecting D6-branes such as D4-branes ending on
D6-branes which will be described in section~\ref{D4intD6}.

These solutions describing fully localised intersecting branes were
found by Itzhaki, Tseytlin and Yankielowicz \cite{Itzhaki:1998uz}. One of the
cases considered was D2-branes within D6-branes, preserving
one quarter supersymmetry.
A similar construction produces solutions for a wave or NS5-branes
within D6-branes. There are various other solutions which can be found using
T- and S-duality transformations.

In the near-core or near-horizon limit of the D6-branes the problem is greatly
simplified since there is then an essentially trivial eleven-dimensional
description of the D6-branes.
In the near-core limit, the eleven-dimensional description of $N$ D6-branes is
flat 6+1 dimensions plus an ALE space with $A_{N-1}$ singularity. This ALE
space is simply the orbifold $\mathbf{C}^2/\mathbf{Z}_N$ where $\mathbf{Z}_N$ acts on the complex coordinates as
$z_1 \rightarrow e^{2\pi i/N} z_1$ and $z_2 \rightarrow
e^{-2\pi i/N} z_2$, with the metric given by the flat Euclidean
metric with these $\mathbf{Z}_N$ identifications. To see this explicitly we define
real coordinates $\rho \ge 0$, $\tilde{\theta} \in [0, \pi/2]$ and
$\tilde{\phi},
\tilde{\varphi} \in [0, 2\pi)$ through $z_1 = \rho \cos\tilde{\theta}
e^{i\tilde{\phi}}$ and $z_2 = \rho \sin\tilde{\theta}
e^{i\tilde{\varphi}}$. The eleven-dimensional metric is then
\begin{equation}
\ud s_{(1,10)}^2 = \ud x_{(1,6)}^2 + d \rho^2 + \rho^2 \left( d\tilde{\theta}^2 +
\cos^2\tilde{\theta} d\tilde{\phi}^2 + \sin^2\tilde{\theta}
d\tilde{\varphi}^2 \right)
\end{equation}
with the identification of the coordinates under
\begin{equation}
\left( \tilde{\phi}, \tilde{\varphi} \right) \rightarrow \left(
\tilde{\phi} + \frac{2\pi}{N}, \tilde{\varphi} - \frac{2\pi}{N}
\right)
\end{equation}
Note that this background will preserve supersymmetry since the identifications
are such that the holonomy is (a discrete subgroup of) SU(2). So we have
(a singular limit of) a supersymmetric special holonomy manifold from
table~\ref{Holonomy}.
We can reduce to ten dimensions along the isometry direction given by $\psi
\equiv N\tilde{\phi}$ which has the standard period of $2\pi$. We also
define $\varphi \equiv \tilde{\phi} + \tilde{\varphi}$ which
is invariant under the $\mathbf{Z}_N$ transformations and, for later convenience,
$\theta \equiv 2\tilde{\theta}$ and $U \equiv
\frac{\rho^2}{2Nl_P^3}$. The Kaluza-Klein reduction along $x^{10} = R_{11}\psi$
using the relations equations (\ref{IIA11dconst}) and (\ref{IIA11dmetric})
gives the string frame metric, dilaton and Kaluza-Klein gauge potential
\begin{eqnarray}
\frac{1}{l_s^2}\ud s_{(1,9)}^2 & = & \sqrt{\frac{2U}{g_sl_s^3N}} \ud x_{(1,6)}^2 +
	\sqrt{\frac{g_sl_s^3N}{2U}} \left( \ud U^2 +
	U^2(d\theta^2 + \sin^2\theta d\varphi^2) \right)\\
e^{\phi} & = & \left( \frac{2Ul_s}{g_sN} \right)^{\frac{3}{4}} \\
A & = & g_sl_s \frac{N}{2} \left( \cos\theta - 1 \right) d\varphi
\end{eqnarray}
which is the near-horizon limit of $N$ coincident D6-branes in type IIA. This is
easily checked by taking the near-horizon limit
\begin{equation}
l_s \rightarrow 0 \;\; , \;\; U = \frac{r}{l_s^2} \;\; \mathrm{fixed} \;\; , \;\;
	g_{YM}^2 = (2\pi)^4 g_s l_s^3 \;\; \mathrm{fixed}
\end{equation}
of the solution for $N$ coincident D6-branes
\begin{eqnarray}
\ud s^2 & = & H_6^{-\frac{1}{2}} \ud x_{(1,6)}^2 +
	H_6^{\frac{1}{2}} (\ud r^2 + r^2 \ud\Omega_2^2) \\
e^{\phi} & = & H_6^{-3/4} \\
\ast \ud A & = & -\ud(H_6^{-1})\wedge\epsilon_{1,6}
\end{eqnarray}
where
\begin{equation}
H_6 = 1 + \frac{g_sl_sN}{2r} \rightarrow \frac{g_sl_s^3N}{2U} \equiv \tilde{H}_6
\end{equation}

Note that the above process only relied upon having an eleven-dimensional
solution which does not depend on the three coordinates of a three-sphere.
Therefore we can repeat the above process of making $\mathbf{Z}_N$ identifications and
then reducing to ten dimensions, starting with non-trivial eleven-dimensional
solutions such as a wave, M2-branes or M5-branes. When reduced to ten
dimensions this will describe the appropriate object (wave, D2-branes or
NS5-branes) along with $N$ D6-branes.
The reason we will always get $N$ D6-branes in the solution is that the
$\mathbf{Z}_N$ identifications along with the Kaluza-Klein reduction will give a
Kaluza-Klein one-form gauge potential with $N$ units of magnetic flux, exactly
as in the Minkowski space example above. Specifically, if we start with an
eleven-dimensional metric
\begin{equation}
\ud s_{(1,10)}^2 = \ud s_{(1,6)}^2 + h\left( d \rho^2 + \rho^2 \left( d\tilde{\theta}^2 +
\cos^2\tilde{\theta} d\tilde{\phi}^2 + \sin^2\tilde{\theta}
d\tilde{\varphi}^2 \right) \right)
\end{equation}
we will end up with the type IIA metric, dilaton and Kaluza-Klein gauge potential
\begin{eqnarray}
\frac{1}{l_s^2}\ud s_{(1,9)}^2 & = & h^{\frac{1}{2}}\tilde{H}_6^{-\frac{1}{2}} \ud s_{(1,6)}^2 +
        h^{\frac{3}{2}}\tilde{H}_6^{\frac{1}{2}} \left( \ud U^2 +
        U^2(d\theta^2 + \sin^2\theta d\varphi^2) \right)\\
e^{\phi} & = & h^{\frac{3}{4}}\tilde{H}_6^{-\frac{3}{4}} \\
A & = & g_sl_s \frac{N}{2} \left( \cos\theta - 1 \right) d\varphi
\end{eqnarray}
For example in the case of $n$ M2-branes we would have
\begin{eqnarray}
\ud s_{(1,6)}^2 & = & H_2^{-\frac{2}{3}} \ud x_{(1,2)}^2 +
	H_2^{\frac{1}{3}} \left(\ud x_3^2 + \cdots + \ud x_6^2 \right) \\
h & = & H_2^{\frac{1}{3}} \\
H_2 & = & 1 + \frac{2^5\pi^2l_P^6n}{(x_3^2 + \cdots + x_6^2 + \rho^2)^3}
\end{eqnarray}
which would give the the type IIA metric for $n$ D2-branes localised within
the near-horizon limit of $N$ D6-branes
\begin{eqnarray}
\frac{1}{l_s^2}\ud s_{(1,9)}^2 & = & H_2^{-\frac{1}{2}}\tilde{H}_6^{-\frac{1}{2}} \ud x_{(1,2)}^2 +
H_2^{\frac{1}{2}}\tilde{H}_6^{-\frac{1}{2}}\left(\ud x_3^2 + \cdots + \ud x_6^2 \right) + \nonumber \\
 & & H_2^{\frac{1}{2}}\tilde{H}_6^{\frac{1}{2}} \left( \ud U^2 + U^2(d\theta^2 + \sin^2\theta d\varphi^2) \right) \\
H_2 & = & 1 + \frac{2^5\pi^2g_s^2l_s^6n}{(x_3^2 + \cdots + x_6^2 + 2g_sl_s^3NU)^3} \\
\tilde{H}_6 & = & \frac{g_sl_s^3N}{2U}
\end{eqnarray}
Note that the metric has the form of the smeared intersecting brane solutions
given in section~\ref{Smeared} and indeed we could smear the D2-branes in the
3456 directions to recover such a solution. However, in the solution here, the
D2-branes are fully localised within the D6-branes which is a significant
improvement on the solutions given by the flat-space harmonic function rules.
Nevertheless, this solution is only valid in the near-horizon limit we have
taken here and it is still an open problem to find any fully localised
intersecting branes solutions, other than those described in
section~\ref{relativelocalised}, without taking such a limit. In particular, it
is not possible to simply replace $\tilde{H}_6$ with $1 + \tilde{H}_6$. Also
note that the harmonic function for the D2-branes, $H_2$, retains the
eleven-dimensional form with an $r^{-6}$ fall-off rather than $r^{-5}$. There
is also the curious linear and quadratic combination of $U$ and
$x_3, \ldots , x_6$ in the effective radial coordinate transverse to the
D2-branes. These seem to be typical features of such near-horizon solutions as
we have seen in section~\ref{nearhorizoncurvedharm}.

The cases of a wave or NS5-branes localised within D6-branes can easily be
constructed using the above expressions. Related solutions can also be
constructed by dualising these solutions. Note also that these solutions
satisfy the curved-space harmonic function rules. Indeed, alternatively we
could have constructed such solutions and others by solving the curved-space
Laplace equation for the smaller brane in the near-core region of the large
brane, similar to the example in section~\ref{nearhorizoncurvedharm}.

Recently the case of D2-branes within D6-branes was considered, without taking
a near-horizon limit. Although the curved-space Laplace equation could not be
completely solved, the solution could be written as a specific (one-dimensional)
integral \cite{Cherkis:2002ir}.

\subsection{Branes intersecting D6-branes}
\label{D4intD6}

It was shown by Hashimoto \cite{Hashimoto:1998ug} that the construction of
branes within D6-branes could easily be generalised to branes intersecting or
ending on D6-branes. The cases where this is possible are those where it is
consistent to embed part of the worldvolume of an M-brane in the orbifold
$\mathbf{C}^2/\mathbf{Z}_N$. For example we could take an M5-brane with two
worldvolume directions spanning the $z^1$-plane at $z^2 = 0$. This embedding is
allowed because it is invariant under the $\mathbf{Z}_N$ action and it
preserves supersymmetry because it is a holomorphic embedding. We then reduce
to type IIA as in section~\ref{D2inD6} to find the supergravity solution for
a D4-brane ending on a D6-brane, in the near horizon limit of the D6-brane.
See also \cite{Cherkis:1999jt} where a series of dualities are used to relate
this to intersecting M5-branes, localised in the four relative transverse
directions but smeared over the three overall transverse directions.

We start with the metric for $n$ M5-branes
\begin{equation}
\ud s^2 = H^{-1/3} \ud x_{(1,3)}^2 +
	H^{2/3}\left(\ud x_4^2 + \cdots + \ud x_6^2 \right) +
	H^{-1/3}\ud z^1 \ud\overline{z}^1 + H^{2/3}\ud z^2 \ud\overline{z}^2
\end{equation}
and perform the same identifications and reduction as in section~\ref{D2inD6}.
It can easily be checked that for these M5-branes filling the $z^1$-plane,
$x^{10}$ is a worldvolume coordinate and so they reduce to D4-branes in type
IIA. The resulting solution can be found in \cite{Hashimoto:1998ug}. It turns
out to be a solution for D4-branes ending on D6-branes rather than intersecting
them. It should also be noted that this solution is not of the form of the
solutions which can be derived from the harmonic function rules.

Similarly we could embed an M2-brane to describe a fundamental string ending
on a D6-brane. We can also consider more general holomorphic embeddings. Since
we must start with the supergravity solution for the M5-branes (or M2-branes,)
only the case of planar embeddings of parallel M5-branes was considered in \cite{Hashimoto:1998ug}. With
the restriction that the embedding should be $\mathbf{Z}_N$-invariant this allows
holomorphic embeddings of the form
\begin{equation}
\left(\cos \alpha z^1 + \sin\alpha z^2 \right)^N = c^N
\end{equation}
However, in such cases there can be problems interpreting the solutions in ten
dimensions. In general the
reduction produces a ten-dimensional solution with some D4-brane and NS5-brane
charge. The problem is essentially that these charges appear not to match with expectations from
considering geometrically how the M5-branes reduce to D4- and NS5-branes \cite{Hashimoto:1998ug}.

Yet another possibility is to have M5-branes spanning all of the orbifold
which will reduce to the near-core limit of D6-branes intersected by D4-branes
with a common two dimensional worldvolume. In this case we start with the
M5-brane metric
\begin{equation}
\ud s^2 = H^{-1/3}\left(-\ud x_0^2 + \ud x_1^2 \right) + H^{2/3}\left(\ud x_2^2 + \ldots + \ud x_6^2 \right) + H^{-1/3}\left(\ud x_7^2 + \ldots + \ud x_{10}^2 \right)
\end{equation}
Since the M5-branes fill the 789(10) space we can actually use the formulae of
section~\ref{D2inD6} with $h = H^{-1/3}$.
Note that this case could equivalently be derived using the
curved-space harmonic function rules. Indeed it is one of the cases with eight
relative transverse dimensions where a fully localised solution can be found,
even without taking the near-core limit of the D6-branes.

\subsection{Intersecting branes from special holonomy manifolds}
\label{IntfromGeom}

We saw in sections \ref{D2inD6} and \ref{D4intD6} how intersecting brane
solutions in type IIA arise from an orbifold in eleven dimensions. These
solutions are related to a large class of intersecting brane solutions which
were previously constructed, starting from pure geometry in eleven dimensions
\cite{Gauntlett:1997pk, Gauntlett:1997cv} (see also \cite{Dasgupta:1998su}.)
The starting point was a supersymmetric eleven-dimensional solution consisting
of a product of $(2+1)$-dimensional Minkowski spacetime with an
eight-dimensional hyper-K\"ahler manifold. The hyper-K\"ahler manifolds
considered had a triholomorphic $T^2$ isometry which means that we can reduce
the solution to nine dimensions while preserving the same amount of
supersymmetry. The idea is to reduce along one isometry direction to type IIA
and then T-dualise along the other isometry direction to generate a type IIB
solution.

Consider the simple example where the eight-dimensional hyper-K\"ahler
manifold is a product to two four-dimensional Taub-NUT spaces. When we reduce
along the isometry direction of one of the Taub-NUT spaces we get a D6-brane
in type IIA while the other Taub-NUT space is unaffected. Then T-duality along
the isometry direction of this Taub-NUT space (or KK5-brane) produces a
NS5-brane in type IIB. The same T-duality is along a worldvolume direction of
the D6-brane so it becomes a D5-brane. So we have a type IIB supergravity
solution describing the orthogonal intersection of a NS5- and D5-brane with a
common $(2+1)$-dimensional worldvolume, smeared over the transverse direction
(along which we T-dualised.)

The general eight-dimensional toric hyper-K\"ahler manifolds can be thought of
as consisting of overlapping Taub-NUT spaces each with isometry direction along
a $(p,q)$-cycle of the $T^2$ (see \cite{Gauntlett:1997pk} for details.) We can
think of this as a configuration of (non-orthogonal) intersecting KK6-branes.
The reduction and T-duality then turns these Taub-NUT spaces into
$(p,q)$5-branes in type IIB with a common $(2+1)$-dimensional worldvolume,
intersecting at angles determined by their $(p,q)$-charges. Again the solution
is smeared over the overall transverse direction.

Alternatively we can reduce the eleven-dimensional solution along one of the
flat directions and then T-dualise along both isometry directions of the
hyper-K\"ahler manifold. This gives a type IIA solution describing NS5-branes
with a common $(1+1)$-dimensional worldvolume smeared over two relative
transverse directions. This solution can be lifted to eleven dimensions, giving
a solution for intersecting M5-branes, additionally smeared over the (totally
transverse) eleventh dimension. Instead of lifting, we could T-dualise along the common
(spatial) worldvolume direction to find the type IIB solution for intersecting
NS5-branes. Then SL(2,$\mathbf{Z}$) transformations map this to an intersection
of $(p,q)$5-branes (all of the same type.) So we have supergravity solutions
for any 5-branes in ten or eleven dimensions intersecting at angles with
a common $(1+1)$-dimensional worldvolume.

Because of the triholomorphic $T^2$ isometry, all the above dimensional
reductions and T-duality transformations relate solutions with the same amount
of supersymmetry. From table~\ref{Holonomy} we see that this is generically
3/16 supersymmetry. However, in the special case where the eight-dimensional
hyper-K\"ahler manifold is a product of two four-dimensional hyper-K\"ahler
manifolds the amount of supersymmetry preserved is 1/4. Note that in this case
the intersections are orthogonal.

The above solutions can also be generalised to include D2-branes, strings or
waves along the common worldvolume directions by starting with the
eleven-dimensional solution for M2-branes with a transverse hyper-K\"ahler
manifold
\begin{eqnarray}
\ud s^2 & = & H^{-2/3}\ud x_{(1,2)}^2 + H^{1/3}\ud s_{HK8}^2 \\
F & = & -\ud(H^{-1}) \wedge \epsilon_{1,2}
\end{eqnarray}
This solution preserves the same amount of supersymmetry (3/16, 1/4 etc.)
provided $H$ is harmonic wrt.\ the eight-dimensional hyper-K\"ahler metric
$\ud s_{HK8}^2$ (see e.g.\
\cite{Becker:1996gj, Brecher:1999xf, Janssen:1999uq, Papadopoulos:1999tw, Ivashchuk:2000ma}.)
We recover the near-core D2-branes within D6-branes example of
section~\ref{D2inD6} when the eight-dimensional hyper-K\"ahler manifold is
$\mathbf{C}^2 \times \mathbf{C}^2/\mathbf{Z}_N$ which allows us to find an
explicit solution for $H$.
Another starting point considered was the solution (using the curved-space
harmonic function rules) for orthogonal M5-branes with a common
$(1+1)$-dimensional worldvolume, with a hyper-K\"ahler metric on the remaining
four worldvolume directions of each M5-brane.

Similar constructions were considered in
\cite{Papadopoulos:1998np, Papadopoulos:1998yx} involving hyper-K\"ahler
manifolds with torsion. These constructions are essentially the same except now
the four-dimensional manifolds describe NS5-branes in ten dimensions rather
than KK6(5)-branes in eleven (ten) dimensions.

Similar intersecting brane configurations can be constructed from other
manifolds of special holonomy (see e.g.\ \cite{Papadopoulos:1999tg}.) In
particular there has been recent interest in manifolds of $G_2$ holonomy
\cite{Acharya:2000gb, Atiyah:2000zz, Acharya:2001hq, Atiyah:2001qf, Acharya:2001gy, Friedmann:2002ct}
partly because compactification from eleven dimensions leads to a
four-dimensional
$\mathcal{N} = 1$ theory, which can be chiral if there are singularities
\cite{Cvetic:2001nr, Cvetic:2001kk}.
Intersecting brane solutions arise from dimensional reduction and T-duality in
a very similar manner to the hyper-K\"ahler manifolds of Sp(2) holonomy
discussed above. For example a $T^2$ isometry is required to relate
eleven-dimensional solutions to type IIB solutions and fixed points of these
isometries correspond to KK6-branes in eleven dimensions
\cite{Acharya:2000gb, Atiyah:2000zz, Gomis:2001vk, Edelstein:2001pu}.
See \cite{Gukov:2001hf, Curio:2001dz, Gukov:2002es, Uranga:2002ag, Lazaroiu:2002jv, Anguelova:2002dd, Blumenhagen:2002wn, Blum:2002aq, Behrndt:2002xm, Gukov:2002zg}
for explicit examples. Note that these constructions are often useful for
understanding properties of the special holonomy manifolds in terms of
intersecting branes. For example changing the relative positions of the branes
corresponds to changing parameters describing the special holonomy manifold and
can give an understanding of processes such as topology change
where the manifold becomes singular but which may be a smooth(er) process in
terms of the intersecting branes.

\section{Hanany-Witten constructions}
\label{HananyWitten}

Hanany-Witten brane configurations provide a very useful method of
describing large classes of supersymmetric gauge theories. One of the
advantages of these constructions is that many features of the gauge theory
can be understood in simple geometric terms. These features include the
moduli space and gauge theory parameters as well as the gauge group and
matter content and even (given enough supersymmetry) the running gauge
coupling. In this section we will review the description of gauge theories in
terms of such brane configurations and some progress towards finding the
corresponding supergravity solutions, at least in the near-horizon limit
appropriate for the AdS/CFT correspondence.

\subsection{Basic Construction}
\label{3dHW}

The basic setup involves taking some parallel D$p$-branes and `compactifying'
them by making them end on some other branes. In the original setup analysed by
Hanany and Witten \cite{Hanany:1997ie}, one of the simplest examples consisted
of $N$ coincident D3-branes with worldvolume directions 0126 ending on two
parallel NS5-branes with worldvolume directions 012345, separated in the $x^6$
direction. In terms of the U($N$) gauge theory on the worldvolume of the
D3-branes, the addition of the NS5-branes has two effects. The theory is
reduced from 3+1 to 2+1 dimensions (at length scales larger than the separation
between the NS5-branes in the 6 direction) and some of the degrees of freedom
are projected out. In particular since the boundary conditions for the D3-branes
to end on the NS5-branes fix the 789 but not the 345 positions, the
three-dimensional theory has only the three massless scalars corresponding to
the transverse 345 directions, rather than all six scalars present in the
four-dimensional $\mathcal{N} = 2$ theory. In fact there are two other scalars
arising from the gauge field. The $A_6$ component of the four-dimensional
gauge potential is a superpartner of the 789 scalars while the three-dimensional
gauge potential is dual to a scalar which is a superpartner of the 345 scalars.

It is also possible to include hypermultiplets in the gauge theory. These can
arise when D3-branes end on an NS5-brane from opposite sides. We can view these
states as corresponding to open strings which end on D3-branes on either side
of the NS5-brane. For example suppose we have three NS5-branes separated in the
6 direction with $N_1$ and $N_2$ D3-branes between each consecutive pair, at
the origin of the 345 space. The
gauge group in this case would be $\mathrm{U}(N_1) \times \mathrm{U}(N_2)$.
There would also be hypermultiplets transforming in the $(N_1, \overline{N}_2)$
representation. All these states are massless. However, if we move say one of
the $N_2$ D3-branes away from the origin of the 345 space then we break the
$\mathrm{U}(N_2)$ gauge group to $\mathrm{U}(N_2-1) \times \mathrm{U}(1)$. The
open strings stretching between this D3-brane and the others have a minimal
length (given by the separation) and so a non-zero minimal mass. This mass is
the mass of the W-boson (multiplet) and a hypermultiplet transforming in the
fundamental representation of $\mathrm{U}(N_1)$. In general, arbitrarily
positioned D3-branes break a U($N$) gauge group to U$(1)^N$. So the Coulomb
branch of the
gauge theory is parameterised by the position of the D3-branes in the 345 space
together with the VEVs of the scalars dual to the U(1) gauge potentials.

It is possible to introduce D5-branes spanning the 012789 directions without
breaking more supersymmetry. Including these D5-branes leads to new
possibilities. For example D3-branes stretched between these D5-branes give
rise to hypermultiplets with a different R-charge to the above
hypermultiplets. The scalars in these multiplets correspond to motions in the
789 directions together with the $A_6$ component of the four dimensional
gauge potentials.

Various properties of these configurations were analysed in
\cite{Hanany:1997ie}. Since S-duality interchanges NS5-branes and D5-branes
while leaving D3-branes unchanged, this effectively exchanges the vector
multiplets and hypermultiplets in the gauge theory. This provides a string
theory realisation of mirror symmetry in three-dimensional gauge theories.
Another type of duality occurs when the configuration consists of $N$ D3-branes
between two NS5-branes. In this case the Coulomb branch of the three-dimensional
gauge theory can equivalently be described in terms of the 5-brane worldvolume
theory. This gives a geometric interpretation of the equivalence between the
moduli space of $N$ SU(2) monopoles and this Coulomb branch. Specifically,
S-duality followed by T-duality in the common 12 directions maps the brane
configuration to that of $N$ D1-branes between two D3-branes which was already
considered from both the D1- and D3-branes' worldvolume theories in
section~\ref{BIons}.

There are many generalisations of these Hanany-Witten constructions to describe
gauge theories in various dimensions, with various gauge groups, matter
content, amounts of supersymmetry etc. See \cite{Giveon:1998sr} for a
comprehensive review with extensive references. In the next section we will describe
some examples of four-dimensional $\mathcal{N} = 2$ gauge theories with the
aim of finding the supergravity duals (in the context of the AdS/CFT
correspondence) in section~\ref{SUGRA_M5SW}.

\subsection{Four-dimensional $\mathcal{N} = 2$ SYM}
\label{4dHW}

Seiberg and Witten's work on $\mathcal{N} = 2$ four-dimensional gauge theories
\cite{Seiberg:1994rs, Seiberg:1994aj} showed how the exact low energy
effective action (including non-perturbative effects) could be described by
a family of Riemann surfaces $\Sigma \subset \mathbf{C}^2$. These Riemann surfaces,
called Seiberg-Witten curves, encode information about the gauge theory
such as the exact mass of BPS states. These families of curves contain
parameters which correspond to VEVs parameterising the Coulomb branch of the
gauge theory.

Using a Hanany-Witten brane construction, Witten \cite{Witten:1997sc} rederived
the description of $\mathcal{N} = 2$ four-dimensional gauge theories in terms
of a Seiberg-Witten curve. Not only did this provide another example of branes
in string theory reproducing field theory results, but it also provided a much
more intuitive geometric interpretation of the previously abstract
Seiberg-Witten curve. In addition, for a large class of field theories there is
a simple prescription for constructing the corresponding Hanany-Witten brane
configuration, from which the Seiberg-Witten curve can easily be described,
thereby solving a difficult field theory problem. We will describe this
construction in the following sections.

\subsubsection{Type IIA brane configuration}

The brane setup in type IIA involves D4-branes with worldvolume directions
01236 and NS5-branes with worldvolume directions 012345. This can be obtained
from the configurations involving D3-branes and NS5-branes in
section~\ref{3dHW} by T-duality in the 3 direction. All the branes are
located at $x^7 = x^8 = x^9 = 0$, the NS5-branes are separated in the $x^6$
direction and the D4-branes can have finite, semi-infinite or infinite extent
in the $x^6$ direction by ending on two, one or no NS5-branes respectively. We
refer to those D4-branes as finite, semi-infinite and infinite. It
is simple to check that the supersymmetry projection operators for the D4-
and NS5-branes are compatible, and so the system is one quarter BPS, preserving
$\mathcal{N} = 2$ supersymmetry in four dimensions.

The simplest configurations involve two NS5-branes with $N_c$ finite D4-branes
between them. This gives gauge group SU$(N_c)$ rather than U$(N_c)$ since a
U(1) is frozen out because the corresponding centre of mass position of the
D4-branes is fixed \cite{Witten:1997sc}. Semi-infinite D4-branes produce
hypermultiplets in the fundamental representation. There are two types of
semi-infinite D4-branes, those ending on the right NS5-brane and extending to
$x^6 = \infty$ and those ending on the left NS5-brane and extending to
$x^6 = -\infty$. They are equivalent with respect to the gauge theory. However,
if there are $N_f$ hypermultiplets then only when we choose all $N_f$
semi-infinite D4-branes to be of the same type is the global SU$(N_f)$ flavour
symmetry manifest in the brane configuration.

So the general configuration we will consider involves $N_L$ semi-infinite
D4-branes to the left of, $N_c$ finite D4-branes between, and $N_R$
semi-infinite D4-branes to the right of the two NS5-branes. The gauge group is
SU$(N_c)$ and there are $N_f = N_L + N_R$ hypermultiplets in the fundamental
representation of SU$(N_c)$. There are obvious further generalisation involving
more NS5-branes and a gauge group which is a product of SU($N_i$) factors, with
hypermultiplets in $(N_i, \overline{N}_{i+1})$ representations. It is also
possible to introduce D6-branes with worldvolume directions 0123789 without
breaking any more supersymmetry. These provide another method of introducing
hypermultiplets in the fundamental representation due to strings stretching
between the D6-branes and the finite D4-branes (between the same NS5-branes.)

\subsubsection{Running gauge coupling}

One of the surprising features of the brane construction is that it gives a
geometric interpretation of the logarithmic running of the gauge coupling. To
see this we first have to identify the gauge coupling in the brane
configuration. It is given by the separation of the NS5-branes, say $L$, in the
$x^6$ direction. This is simply because we view the finite D4-branes as being
compactified on an interval of length $L$, i.e.\
\begin{equation}
S_{D4} \sim \frac{1}{l_sg_s}\int \ud^5x F^2 \sim \frac{L}{l_sg_s}\int \ud^4x F^2
\end{equation}
So we see that the four-dimensional gauge coupling is given by
\begin{equation}
\frac{1}{g_{YM}^2} \sim \frac{L}{l_sg_s}
\end{equation}
Now the point is that since each NS5-brane has D4-branes ending on it, the
tension of the D4-branes distorts the NS5-brane and so the separation between
the NS5-branes depends on the position in the $v = x^4 + ix^5$ plane. The bending
in the $x^6$ direction will be (asymptotically) logarithmic in $|v|$ since the end of the
D4-brane in the NS5-brane is of co-dimension 2. I.e.\ if we ignore the common
worldvolume directions we have the same situation as a three-dimensional system
with a string extended along
$x^6$ (a D4-brane) ending on a membrane (an NS5-brane) extended in the $x^4$ and $x^5$
directions. Obviously strings ending from the left bend the membrane in the
opposite direction to those ending from the right. So if we
consider a single NS5-brane with $n_L$ ($n_R$) D4-branes ending from the left
(right) then the net effect will be a bending of the NS5-brane asymptotically
given by (recalling that $T_{D4} \sim 1/g_s$ whereas $T_{NS5} \sim 1/g_s^2$)
\begin{equation}
x^6 \sim l_sg_s (n_L - n_R) \ln |v|
\end{equation}
A more detailed analysis shows that provided the centre of mass position of the
D4-branes is fixed, giving gauge group SU($N_c$) rather than U($N_c$), moving
the D4-branes to different positions in the $v$-plane does not change the
asymptotic behaviour of the NS5-branes. In the Hanany-Witten construction we
are considering, the logarithmic bending of the NS5-branes means that
\begin{equation}
\frac{1}{g_{YM}^2} \sim (N_c - N_R) \ln |v| - (N_L - N_c) \ln |v|
 = (2N_c - N_f) \ln |v|
\end{equation}
Noticing that the coefficient $2N_c - N_f$ is exactly the same as appears in
the one-loop (perturbatively exact for $\mathcal{N} = 2$ supersymmetry)
beta-function, we see that the brane construction reproduces the correct
running gauge coupling provided we interpret $|v|$ as an energy scale. Note
that this provides an example of a UV/IR correspondence. The origin of this is
the same as in the AdS/CFT correspondence, arising from the identification of
lengths and masses due to the mass of open strings (which are relevant to the
gauge theory) being proportional to their length. In particular, from the
gauge theory point of view,  $w = v/l_s^2$ is a natural variable to use.

So we can understand many details of the field theory from this ten-dimensional
brane configuration. However, we can go even further by considering the lift
to eleven dimensions. The reason for this is that the singular intersections of
the D4-branes and NS5-branes are smoothed out in the eleven-dimensional
configuration. This is related to the question of the shape of the branes
and the preservation of supersymmetry. As we have described the logarithmic
bending of the NS5-branes, it is not obvious why we should still preserve
$\mathcal{N} = 2$ supersymmetry. Indeed the requirement of supersymmetry will
place strong constraints on the exact shape of the branes and we will see that
these constraints can be very easily solved in eleven dimensions.

\subsubsection{Eleven-dimensional description}

We can lift the type IIA configuration to eleven dimensions using the
well-known relation between type IIA branes and M-branes \cite{Townsend:1995kk}.
In particular an NS5-brane is an M5-brane, pointlike in the eleventh dimension,
$x^{10}$, while a D4-brane is also an M5-brane but one which wraps the eleventh
dimension (which is a circle of radius $R$.) We can immediately deduce that the
picture of a D4-brane ending on an NS5-brane is modified for $R \ne 0$ since
the boundary of the D4-brane\footnote{We will continue to refer to D4- and
NS5-branes for the moment in order to distinguish between the orientations of
the M5-branes.}
would span $x^{10}$ and so could not be contained within the worldvolume of
the NS5-brane. So only when we have a genuine intersection (a D4-brane passing
through an NS5-brane rather than ending on it) can we have flat D4-branes
spanning $x^6$ and $x^{10}$ and point-like in $x^4$ and $x^5$. I.e.\ in general
the D4-branes must be deformed, not just the NS5-branes, although this is not
apparent in the singular limit $R \rightarrow 0$. Note that in the special case
where there are $N_c$ infinite D4-branes, neither the D4- nor the NS5-branes
are deformed, and in particular the gauge coupling is constant as expected for
the conformal theory when $N_f = 2N_c$.

To see what happens in general we should consider the conditions for supersymmetry
preservation. We know that in the case of orthogonal intersections,
$\mathcal{N} = 2$ supersymmetry is preserved with the compatible projection
conditions
\begin{equation}
\hat{\Gamma}_{012345} \epsilon = \epsilon = \hat{\Gamma}_{01236(10)} \epsilon
\end{equation}
So we see from the discussion of section~\ref{HolomInt} that any holomorphic
embedding of an M5-brane will preserve $\mathcal{N} = 2$ supersymmetry and
indeed since we know the projection conditions asymptotically, an M5-brane
which preserves $\mathcal{N} = 2$ supersymmetry must be embedded
holomorphically with respect to the given complex coordinates $v = x^4 + ix^5$
and $s = x^6 + ix^{10}$. More precisely, $\exp(s/R)$ is a better coordinate
than $s$ since it is single valued under $x^{10} \rightarrow x^{10} + 2\pi R$
but for simplicity we will continue to use $s$. So the embedding is described by a Riemann surface
$\Sigma \subset \mathbf{C}^2$ which we can determine explicitly up to a finite
number of parameters using the known asymptotic form of the embedding -- i.e.\
how many NS5-branes there are and the positions of the semi-infinite D4-branes.

The Riemann surface $\Sigma$ is in fact the Seiberg-Witten curve for the gauge
theory. BPS states in the field theory correspond to M2-branes ending on the
M5-brane. The mass of the M2-brane gives the mass of the BPS state. This,
together with the conditions for the M2-brane embedding to be supersymmetric
leads to an M-theory derivation of the Seiberg-Witten differential
\cite{Fayyazuddin:1997by, Landsteiner:1997vd, Henningson:1998hy, Mikhailov:1998jv}
(see also \cite{Klemm:1996bj}.) It is also interesting to note that M2-branes
with the topology of a cylinder or disc correspond to vector multiplets or
hypermultiplets respectively \cite{Henningson:1998hy, Mikhailov:1998jv}. It is
also possible to derive the low energy effective action from the M5-brane worldvolume
theory \cite{Howe:1998eu, Witten:1997sc}.

\subsection{Supergravity dual}
\label{SUGRA_M5SW}

If we know how to describe a particular gauge theory in terms of a particular
brane configuration then we can try to describe the gravity dual of the field
theory. We follow essentially the same steps as in the derivation of the
AdS/CFT duality for the case of parallel branes presented in
section~\ref{AdSCFT}. We first identify the field
theory parameters which should be kept fixed while taking a limit to decouple
gravity and string modes. We then take this limit for the appropriate
supergravity solution describing the brane configuration and this should give
a candidate gravity dual of the field theory.

To find the supergravity solution we use the conditions for preservation
of supersymmetry to constrain the metric and four-form field strength. Rather
than start with the most general form of metric we can first impose the
expected symmetries of the solution, namely 3+1 dimensional Poincar{\'e} invariance of the common
worldvolume directions and an SO(3) invariance corresponding to rotations in
the totally transverse directions which is identified with the SU(2) R-symmetry
of the gauge theory. This allows us to write the metric as
\begin{equation}
\ud s^2 = H_1 \eta_{\mu\nu}\ud x^{\mu}\ud x^{\nu} + 2H_1g_{m\overline{n}}\ud z^m\ud z^{\overline{n}} + H_2 \delta_{\alpha\beta}\ud x^{\alpha}\ud x^{\beta}
\end{equation}
where $H_1$, $H_2$ and $g_{m\overline{n}}$ can only depend on the two complex
coordinates $z^m$ and the radial coordinate in the three totally transverse
directions, $r \equiv \sqrt{\delta_{\alpha\beta}x^{\alpha}x^{\beta}}$. We will
use the notation $v = z^1$ and $s = z^2$. It turns
out to be convenient to include the factor $H_1$ with $g_{m\overline{n}}$.
Since the M5-branes are magnetic sources for $F_{(4)}$ we can also deduce that
the only non-vanishing components of $F_{(4)}$ will have at least two indices
in the totally transverse space and no indices in the common worldvolume
directions.

The projection conditions on the 32-component spinor $\epsilon$ are
\begin{equation}
\hat{\Gamma}_{0123m\overline{n}} \epsilon = i \delta_{m\overline{n}} \epsilon
\end{equation}
It is now a straightforward, though rather lengthy, process to write out the
Killing spinor equations~(\ref{11dSUSY}), $\tilde{D}_{\mu}\epsilon = 0$, in
terms of the components of the above metric and
four-form. Then using the above projection conditions we can express this as
a sum of independent antisymmetric combinations of Gamma-matrices acting on
$\epsilon$. The coefficients of these terms must vanish in order to satisfy the
Killing spinor equation for non-vanishing $\epsilon$. The term without any
Gamma-matrices acting on $\epsilon$ is slightly different, it is a first order
differential equation determining the positional dependence of $\epsilon$, with
the result that $\epsilon$ is a specific function times a constant spinor.
The other equations result in a set of first-order differential equations which
reduce to the following relations \cite{Fayyazuddin:1999zu}
\begin{eqnarray}
H_1 & = & H^{-\frac{1}{3}} \\
H_2 & = & H^{\frac{2}{3}} \\
F_{m\overline{n}\alpha\beta} & = & i \epsilon_{\alpha\beta\gamma}\partial_{\gamma}g_{m\overline{n}} \\
F_{m789} & = & -i \partial_{m}H \\
F_{\overline{m}789} & = & i \partial_{\overline{m}}H \\
H = 4g & = & 4\left( g_{v\overline{v}}g_{s\overline{s}} - g_{s\overline{v}}g_{v\overline{s}} \right)
\end{eqnarray}
and the constraint that $g_{m\overline{n}}$ is a K\"ahler metric, with
(square-root) determinant $g$. In deriving
these results it is necessary to fix some integration constants. This has been
done using the condition that asymptotically we recover the usual Minkowski
metric, i.e.\ that asymptotically
\begin{equation}
H_1 \rightarrow 1 \;\; , \;\; H_2 \rightarrow 1 \;\; , \;\; g_{m\overline{n}} \rightarrow \delta_{m\overline{n}} 
\end{equation}

Notice that the metric takes a similar form to what we would expect from the
harmonic function rules but with some extra off-diagonal terms in the
relative transverse space
\begin{equation}
\ud s^2 = H^{-\frac{1}{3}}\ud x_{(1,3)}^2 +
	2H^{-\frac{1}{3}}g_{m\overline{n}}\ud z^m\ud z^{\overline{n}} +
	H^{\frac{2}{3}}\ud x_{(3)}^2
\label{IntM5metric}
\end{equation}

Clearly the four-form satisfies $F \wedge F = 0$ and it is relatively simple to
check that the Bianchi identity is identically satisfied
\begin{equation}
d (\ast F) = 0
\end{equation}
so we are left with the equation of motion for $F$ with a magnetic source $J$
\begin{equation}
\ud F = J = J_{m\overline{n}} \ud z^m \wedge \ud z^{\overline{n}} \wedge \ud x^8 \wedge \ud x^9 \wedge \ud x^{10}
\end{equation}
which results in the equations
\begin{equation}
4\partial_m \partial_{\overline{n}} (2g) +
  \partial_{\gamma}\partial_{\gamma}g_{m\overline{n}} = -iJ_{m\overline{n}}
\label{sourceM5sigma}
\end{equation}
Since the source describes an M5-brane wrapped on a Riemann
surface $\Sigma$, defined by a holomorphic function $f(v, s) = 0$ at $r=0$, we
can write the source terms as
\begin{equation}
\label{holomsource}
J_{m\overline{n}} = -4i\pi^3 l_P^3(\partial_mf)(\overline{\partial_nf})\delta^2(f)\delta^3(r)
\end{equation}
The difficulty lies in solving these source equations~(\ref{sourceM5sigma})
which are non-linear due to the presence of both the metric components and
determinant. We can rewrite these equations as a single (highly non-linear)
partial differential equation for a function $K$, the K\"ahler potential for
the metric, $g_{m\overline{n}} = \partial_m \partial_{\overline{n}} K$. This
results in
\begin{equation}
8g(K) + \partial_{\gamma}\partial_{\gamma} K = -4\pi^3 l_P^3|f|^2\delta^2(f)\delta^3(r)
\end{equation}
which is related to the complex Monge-Amp\`ere equation.

\subsubsection{Special cases}

It can easily be checked that we can reproduce the simple example of parallel
M5-branes. Consider $N$
coincident M5-branes at $r=s=0$ (so $f = s^N$) with the source given by
\begin{equation}
J_{s\overline{s}} = -4i\pi^3l_P^3N\delta^3(r)\delta^2(s)
\end{equation}
Defining $r_{(5)} = \sqrt{r^2 + |s|^2}$ we can easily check that the solution
to the source equations~(\ref{sourceM5sigma}) is
\begin{eqnarray}
g_{v\overline{s}} = g_{s\overline{v}} & = & 0 \\
g_{v\overline{v}} & = & \frac{1}{2} \\
g_{s\overline{s}} & = & \frac{1}{2} + \frac{\pi l_P^3N}{2r_{(5)}^3}
\end{eqnarray}
which is the expected solution for $N$ parallel M5-branes, and provides a check
on the normalisation of the source in equation~(\ref{holomsource}).

A less trivial example to is recover the harmonic function rules. One way
to do this is to try making the simplifying assumption that the metric is
diagonal for orthogonal intersections. I.e.\ if we take the source terms for
$M$ M5-branes at $r=v=0$ intersecting $N$ M5-branes at $r=s=0$ ($f = v^Ms^N$)
so that the sources are
\begin{eqnarray}
J_{v\overline{s}} = J_{s\overline{v}} & = & 0 \\
J_{v\overline{v}} & = & -4i\pi^3l_P^3M\delta^3(r)\delta^2(v) \\
J_{s\overline{s}} & = & -4i\pi^3l_P^3N\delta^3(r)\delta^2(s)
\end{eqnarray}
then we try to find solutions where
\begin{equation}
g_{v\overline{s}} = g_{s\overline{v}} = 0
\end{equation}
Note that the conditions for a K\"ahler metric show that $g_{v\overline{v}}$ is
independent of $s$ and $\overline{s}$, and that $g_{s\overline{s}}$ is
independent of $v$ and $\overline{v}$. For example
\begin{equation}
\partial_v g_{s\overline{s}} = \partial_s g_{v\overline{s}} = 0
\end{equation}
The equations~(\ref{sourceM5sigma}) for the $v\overline{s}$ and $s\overline{v}$
components then reduce to the requirement that $g_{v\overline{v}}$ is also
independent of $v$ and $\overline{v}$, or that $g_{s\overline{s}}$ is
independent of $s$ and $\overline{s}$. Hence we see that the assumption of
having a diagonal metric requires at least one set of M5-branes to be smeared
over the worldvolume directions of the other M5-branes. Without loss of
generality we can require $g_{s\overline{s}}$ to be independent of $s$ and
$\overline{s}$. In this case the remaining equations~(\ref{sourceM5sigma})
become
\begin{eqnarray}
4g_{s\overline{s}} \partial_v \partial_{\overline{v}} g_{v\overline{v}}
	+ \partial_{\gamma}\partial_{\gamma}g_{v\overline{v}} & = &
	-iJ_{v\overline{v}} \nonumber \\
\partial_{\gamma}\partial_{\gamma}g_{s\overline{s}} & = &
        -iJ_{s\overline{s}}
\label{PartLocM5}
\end{eqnarray}
Clearly the second of these equations cannot be satisfied with the fully
localised source term $J_{s\overline{s}}$ since there is no way to produce
the delta function $\delta^2(s)$ on the left hand side. However, if we
compactify the $s$-plane on a 2-torus of volume $V$ we can smear the $N$
M5-branes over the $T^2$ by replacing $\delta^2(s)$ with $1/V$ in the source
term. We then see that $g_{s\overline{s}}$ satisfies the flat-space Laplace
equation in the three totally transverse directions while $g_{v\overline{v}}$
satisfies a curved-space Laplace equation. These are just the expected
conditions of the curved-space harmonic function rules of
section~\ref{GenHarmonicFnRules} which allow partially localised intersecting
branes, with the explicit near-horizon solution given in
section~\ref{nearhorizoncurvedharm}. Again, if we also smear the $M$ M5-branes
over the $v$-plane we recover the flat-space harmonic function rules for this
type of intersection with both branes smeared over the relative transverse
directions.

\subsubsection{General near-horizon solution}
\label{GenNHM5}

Since we are interested in describing the gravity dual of the four-dimensional
gauge theory, we only need to solve the supergravity equations in the
appropriate gauge theory or near-horizon limit. In this limit we keep the
gauge theory masses and coupling constant fixed while taking
$l_P \rightarrow 0$. Specifically we define coordinates which remain fixed in
this limit as
\begin{eqnarray}
w & = & \frac{v}{l_s^2} = \frac{vR}{l_P^3} \\
t^2 & = & \frac{r}{g_sl_s^3} = \frac{r}{l_P^3} \\
y & = & \frac{s}{R}
\end{eqnarray}
Note that $y$ is dimensionless whereas $w$ and $t$ have dimensions of mass. We
can define angular coordinates $\theta$ and $\phi$ so that
\begin{eqnarray}
w & = & \rho \sin\theta e^{i\phi} \\
t & = & \rho \cos\theta
\end{eqnarray}
where $\rho$ has dimensions of mass.
The essential simplification which occurs in such a limit is that we no longer
have a dimensionful constant $l_P$. Therefore the dimension of any quantity
determines its dependence on $\rho$.

In order to find specific solutions we can also use our expectations from the
AdS/CFT correspondence that the dual of a four-dimensional conformal field
theory should involve $AdS_5$. The most general possibility is that the
metric is of the form of a warped product of $AdS_5$ with a six-dimensional
metric
\begin{equation}
\frac{1}{l_P^2}\ud s^2 = \Omega^2 \left( u^2\ud x_{(1,3)}^2 + \frac{1}{u^2}\ud u^2 \right)
	+ \ud s_{(6)}^2
\label{WarpedAdS_5}
\end{equation}
where the warp factor $\Omega$ and the six-dimensional metric $\ud s_6^2$ are
arbitrary functions of the dimensionless coordinates $\theta$, $\phi$ and
$y$. Since $u$ has dimensions of mass we know that $u/\rho$ is also a function
of only $\theta$, $\phi$ and $y$.

Requiring that the metric of equation~(\ref{IntM5metric}) can be written in the
form of equation~(\ref{WarpedAdS_5}) places several constraints on the
components of the K\"ahler metric $g_{m\overline{n}}$ which are not obviously
related to the equations of motion. However, the AdS/CFT correspondence
predicts that they should be compatible and indeed it is possible to find a
solution \cite{Fayyazuddin:2000em}. Furthermore the $w$ and $y$ dependence of
the solution is naturally written in terms of the
holomorphic function defining the Riemann surface -- in this conformal case
with the two NS5-branes separated by $1/g_{YM}^2$ in the $y$-plane, intersected
by (for gauge group SU($N$)) $N$ infinite D4-branes
\begin{equation}
f = \left(y - \frac{1}{2g_{YM}^2}\right) \left(y + \frac{1}{2g_{YM}^2}\right) w^N
\end{equation}
It is then relatively straightforward to check that the solution in
\cite{Fayyazuddin:2000em} generalises to the
case of an arbitrary Riemann surface $\Sigma$, i.e.\ an arbitrary holomorphic
function $f(w, y)$. The supergravity solution can then be determined from the
K\"ahler potential $K$ which is determined by $f$ through two holomorphic functions
$F(w,y)$ and $G(w,y)$ by \cite{Brinne:2000fh}
\begin{eqnarray}
K & = & \frac{\pi N}{2t^2} \ln \left( \frac{\sqrt{t^4 + |F|^4} + t^2}{\sqrt{t^4 + |F|^4} - t^2} \right) + \frac{1}{2}|G|^2 \label{KPot} \\
F & = & f^{1/N}
\end{eqnarray}
where in general $N$ is defined as the degree of $f$ as a polynomial in $w$. To
find explicit solutions we need to solve
\begin{equation}
\left( \partial_y F^2 \right) \left( \partial_w G \right) -
	\left( \partial_w F^2 \right) \left( \partial_y G \right) = 1
\label{PDEforGfromF}
\end{equation}
to find $G$. Whether this can be solved explicitly depends on the choice of $f$.

An interesting observation is that we can interpret the functions $F^2$ and
$G$ as local coordinates transverse and parallel to the M5-brane.
Equation~(\ref{PDEforGfromF}) is simply the condition that the holomorphic
coordinate transformation from $(w,y)$ to $(F^2,G)$ has unit Jacobian. It is
also the necessary condition for the metric
\begin{equation}
g_{m\overline{n}} \equiv 2 \left( \partial_m F^2 \right) \left( \overline{\partial_n F^2} \right) g + \frac{1}{2} \left( \partial_m G \right) ( \overline{\partial_n G} )
\end{equation}
to have determinant $g$. It can then be seen that the source equations
(\ref{sourceM5sigma}) and (\ref{holomsource}) reduce to the condition that
$g$ is a harmonic function in the five-dimensional transverse space with
radial coordinate
\begin{equation}
\tilde{r} \equiv \sqrt{t^4 + |F|^4}
\end{equation}
so that
\begin{equation}
g = \frac{\pi N}{8\tilde{r}^3}
\end{equation}
It can easily be seen that, with $g$ independent of $G$, $g_{m\overline{n}}$
are the components of a K\"ahler metric as required, with K\"ahler potential
$K$ as given in equations~(\ref{KPot}).

\subsubsection{Alternative supergravity constructions}

There are many other methods of constructing brane solutions which should be
dual to $\mathcal{N}=2$ four-dimensional gauge theories. We can consider
wrapped 5-branes in ten dimensions. Such solutions have been constructed using
lower dimensional gauged supergravity theories
\cite{Maldacena:2000mw, Gauntlett:2001ps, Bigazzi:2001aj} or by dimensional reduction of the
analysis presented in the previous section \cite{Ohta:2000gf}.

Alternatively we can use D3-branes in type IIB. The required amount of
supersymmetry is preserved by D3-branes with worldvolume directions 0123
together with D7-branes with worldvolume directions 01236789 and/or a
supersymmetric (SU(2) holonomy) orbifold $\mathbf{C}^2/\mathbf{Z}_n$ in the
6789 directions. This configuration is T-dual (along the 6 direction) to the
type IIA Hanany-Witten configuration described in section~\ref{4dHW}. This has
been shown explicitly for the partially localised solution of
section~\ref{nearhorizoncurvedharm} (in the near-horizon limit) corresponding
to the $N_f = 2N_c$ conformal field theory \cite{Fayyazuddin:1999zu} and for
the usual smeared flat-space harmonic function solution \cite{Dasgupta:1998su}.
The type IIB solution is $AdS_5 \times S^5/\mathbf{Z}_n$ as expected.
In other cases there is no isometry direction
and so T-duality cannot be explicitly performed on the supergravity solution,
although we can still formally relate the descriptions in this way. In
particular the $n$ NS5-branes T-dualise to $n$ coincident KK5-branes which, in
the near-core limit, is simply the orbifold geometry. The D4-branes stretched
between NS5-branes T-dualise to fractional D3-branes \cite{Douglas:1996sw, Douglas:1997xg, Douglas:1997de, Berenstein:1998dw, Diaconescu:1998br, Berenstein:1998ri, Brodie:1998bv, Karch:1998yv, Dasgupta:1999wx} which are fixed at the
orbifold fixed-point. These have the interpretation as D5-branes wrapping the
(zero-size) two-spheres which arise when resolving the singularity (separating
the KK5-branes to produce a smooth multi-centred Taub-NUT metric.) So the
type IIA Hanany-Witten configuration has an equivalent type IIB description
\cite{Karch:1998yv}. However, the gauge couplings which had the geometrical
description in terms of the separation of the NS5-branes are now encoded in
the flux of the NS-NS B-field through the two-spheres \cite{Dasgupta:1999wx}.
Supergravity solutions have been considered for these configurations in
\cite{Klebanov:1999rd, Bertolini:2000dk, Grana:2001xn, Bertolini:2001qa, Bertolini:2001gq}.
See also \cite{Johnson:1999qt} for the closely related case where the orbifold
is replaced by the compact manifold K3. Analysing the supergravity solution with
a probe brane led to the discovery of the enhan{\c c}on mechanism for resolving
singularities. In this case the probe brane becomes tensionless at a finite
radius from the singularity and therefore cannot approach any closer,
effectively cutting-off a finite radius ball around the singularity. I.e.\ the
interpretation is that the physically acceptable geometry is sourced by a
spherical shell of branes rather than branes sitting at a singular point.

\subsubsection{Similar brane configurations}

There are several obvious generalisations of the case of an M5-brane wrapping a
Riemann surface $\Sigma \subset \mathbf{C}^2$. In terms of four-dimensional
gauge theories the most natural case to consider is an M5-brane wrapping
$\Sigma \subset \mathbf{C}^3$, leading to an $\mathcal{N} = 1$ theory. An
analysis of the Killing spinor equations similar to section~\ref{SUGRA_M5SW}
has been performed in \cite{Brinne:2000nf}. Although technically more
complicated, we expect solutions can be found in the near-horizon limit, with
a similar form to those described in section~\ref{GenNHM5}. These solutions are
special cases of more general supersymmetric solutions of eleven-dimensional
supergravity with four-form flux on a (non-compact) seven-dimensional manifold
\cite{Becker:2000rz}. Again, solutions
can be found via gauged supergravity \cite{Maldacena:2000mw, Maldacena:2000yy, Apreda:2001qb}
or in terms of fractional D3-branes in type IIB
\cite{Klebanov:1999rd, Klebanov:2000nc, Bertolini:2001gg}. See also
\cite{Acharya:2000mu, Nieder:2000kc, Gauntlett:2000ng, Nunez:2001pt, Edelstein:2001pu, Schvellinger:2001ib, Maldacena:2001pb, Gauntlett:2001qs, Hernandez:2001bh, Gauntlett:2001jj, Bhattacharyya:2002se}
for other examples of supergravity solutions for wrapped branes.

It is also possible to consider other branes wrapping Riemann surfaces. This
was discussed in \cite{Cho:2000hg} for the cases of M2- and M5-branes in terms
of generalised calibrations, as well as the possibility of the branes wrapping
more general cycles. It was shown in \cite{Gomberoff:1999ps, Ramadevi:1999dy}
that the supergravity solution for a $p$-brane wrapping a Riemann surface
$\Sigma \subset \mathbf{C}^2$ is given by a metric of the form of
equation~(\ref{IntM5metric}), determined by a source equation similar to
equation~(\ref{sourceM5sigma}) with an appropriate number of parallel and
transverse dimensions. The solutions to such  equations were analysed as a
perturbation series around (asymptotic) Minkowski space. The interesting result
was that the perturbative solution does not converge (in ten or eleven
dimensions) for $p$-branes with $p \le 3$. This is consistent with the claim
that no fully localised supergravity solutions exist in such cases. Indeed the
issue of whether localised solutions exist had been considered previously
\cite{Surya:1998dx, Marolf:1999uq}. It was argued using the black hole no-hair
theorem \cite{Surya:1998dx} and gauge theory arguments \cite{Marolf:1999uq}
that there should not be fully localised solutions in some cases. Using the
curved-space harmonic function rules which provide an implicit solution without
smearing for the case of a D$p$-brane parallel to a D$(p+4)$-brane, it was
shown \cite{Marolf:1999uq} that the D$p$-brane would delocalise in the four
relative transverse directions as the separation between the branes was
reduced, in the case where $p \le 1$. T-duality in two of the relative
transverse directions (i.e.\ without changing the number of overall transverse
directions) relates these cases precisely to those cases which were conjectured
not to have localised solutions in \cite{Gomberoff:1999ps}.

There have also been attempts to find localised solutions describing
intersecting D3- and D5-branes \cite{Fayyazuddin:2002bm} relevant to the
Hanany-Witten configurations discussed in section~\ref{3dHW} and intersecting
M2- and M5-branes as well as strings ending on D-branes \cite{Rajaraman:2000ws}.
The Killing spinor equations have been analysed but it is still an open problem
to find fully localised solutions. It would be interesting to find such
solutions, even in the near-horizon limit, to since they are likely to have a
different character from the intersecting brane solutions which are singular
limits of a smoothly wrapped brane.
An additional motivation for the D3/D5 and M2/M5 cases is that it has
been argued \cite{Karch:2001cw, Karch:2000gx} that the intersections will
provide an example of the localisation of gravity with non-compact transverse
dimensions.

\section{Conclusions and outlook}
\label{Conclusions}

We have seen how configurations of intersecting branes have been very useful in
understanding properties of black holes (section~\ref{Smeared}) and gauge
theories (section~\ref{HananyWitten}.) In terms of
supergravity solutions we have seen in section~\ref{Smeared} that the very simple flat-space harmonic
function rules allow us to construct supersymmetric solutions but that the
branes are smeared over some directions. As we saw in section~\ref{semilocalised}, the curved-space harmonic function
rules improve the localisation of the branes, even allowing full localisation
in some cases, but usually it is not possible to find explicit solutions.
However, taking a near-horizon limit simplifies the problem and explicit
solutions can often be found. So, although these harmonic function rules have
proved very useful, especially in relating lower dimensional black holes to
brane configurations, they are only applicable in special cases to the problem
of finding fully localised solutions. We have described some cases of fully
localised solutions in sections \ref{semilocalised} and \ref{HananyWitten} but
it is still unclear what the properties of such solutions are for
general (supersymmetric) configurations of branes, even in the near-horizon
limit where we do know several explicit solutions. For example, while the
smeared solutions given by the harmonic function rules all have the same
general form, it is not clear whether different fully localised solutions will
have such a similar description. It is hoped that progress can be made in this
direction. A particular motivation is the relation to gauge theories via
Hanany-Witten constructions and the AdS/CFT correspondence, as described in section~\ref{HananyWitten}.

We have also seen in sections \ref{IntandEnd}, \ref{semilocalised} and \ref{HananyWitten} the close relation between intersecting branes, wrapped
branes and solutions which don't involve any branes. Viewing intersecting
branes as singular limits of a smoothly wrapped brane proved particularly
useful in describing four-dimensional $\mathcal{N} = 2$ gauge theories using
Hanany-Witten configurations. This relation was probably also an important
factor enabling a fully localised supergravity solution, also described in section~\ref{HananyWitten}, to be found in the
near-horizon limit. As we saw in section~\ref{semilocalised}, a less obvious connection between intersecting branes and
smooth geometry arises via the connection between various branes and
Kaluza-Klein monopoles. One application of these relations is that some
properties of particular special holonomy manifolds can be described in terms
of intersecting branes. This particularly illustrates the fact that
(intersecting) branes is not an isolated subject. So we can expect that
finding new supergravity solutions for intersecting branes will contribute to
our understanding of supersymmetric solutions in general.

\vspace{1cm}
{\bf \large Acknowledgements}
\vspace{0.4cm}
\newline
I would like to thank
Dominic Brecher, Chong-Sun Chu, Ruth Gregory, Emily Hackett-Jones, Bert Janssen, Clifford Johnson, Ken Lovis, Jos{\'e} Figueroa-O'Farrill and Paul Saffin
for various useful discussions and comments.
I am very grateful to
Simon Ross
for also providing helpful comments on a draft version of this paper.
I would especially like to thank Ansar Fayyazuddin and David Page for
collaborations on and discussions of several topics covered in this review.

\newpage

\bibliographystyle{utphys}

\bibliography{references}

\providecommand{\href}[2]{#2}\begingroup\raggedright\begin{thebibliography}{10%
0}

\bibitem{Duff:1995an}
M.~J. Duff, R.~R. Khuri, and J.~X. Lu, ``String solitons,'' Phys. Rept. {\bf
  259} (1995) 213--326,
hep-th/9412184.

\bibitem{Polchinski:1996fm}
J.~Polchinski, S.~Chaudhuri, and C.~V. Johnson, ``Notes on D-Branes,''
hep-th/9602052.

\bibitem{Polchinski:1996na}
J.~Polchinski, ``TASI lectures on D-branes,''
hep-th/9611050.

\bibitem{Duff:1996zn}
M.~J. Duff, ``Supermembranes,''
hep-th/9611203.

\bibitem{Stelle:1996tz}
K.~S. Stelle, ``Lectures on supergravity p-branes,''
hep-th/9701088.

\bibitem{Lu:1998hb}
H.~Lu, C.~N. Pope, T.~A. Tran, and K.~W. Xu, ``Classification of p-branes,
  NUTs, waves and intersections,'' Nucl. Phys. {\bf B511} (1998) 98--154,
hep-th/9708055.

\bibitem{Townsend:1997wg}
P.~K. Townsend, ``M-theory from its superalgebra,''
hep-th/9712004.

\bibitem{Stelle:1998xg}
K.~S. Stelle, ``BPS branes in supergravity,''
hep-th/9803116.

\bibitem{Bachas:1998rg}
C.~P. Bachas, ``Lectures on D-branes,''
hep-th/9806199.

\bibitem{Johnson:2000ch}
C.~V. Johnson, ``D-brane primer,''
hep-th/0007170.

\bibitem{Green:1987sp}
M.~B. Green, J.~H. Schwarz, and E.~Witten, ``Superstring theory. Vol. 1:
  Introduction,''. Cambridge, UK: Univ. Pr. (1987) 469 p. (Cambridge Monographs
  On Mathematical Physics).

\bibitem{Green:1987mn}
M.~B. Green, J.~H. Schwarz, and E.~Witten, ``Superstring theory. Vol. 2: Loop
  amplitudes, anomalies and phenomenology,''. Cambridge, UK: Univ. Pr. (1987)
  596 p. (Cambridge Monographs On Mathematical Physics).

\bibitem{Polchinski:1998rq}
J.~Polchinski, ``String theory. Vol. 1: An introduction to the bosonic
  string,''. Cambridge, UK: Univ. Pr. (1998) 402 p. (Cambridge Monographs On
  Mathematical Physics).

\bibitem{Polchinski:1998rr}
J.~Polchinski, ``String theory. Vol. 2: Superstring theory and beyond,''.
  Cambridge, UK: Univ. Pr. (1998) 531 p. (Cambridge Monographs On Mathematical
  Physics).

\bibitem{Johnson:2003gi}
C.~V. Johnson, ``D-branes,''. Cambridge, UK: Univ. Pr. (2003) 548 p. (Cambridge
  Monographs On Mathematical Physics).

\bibitem{Gauntlett:1997cv}
J.~P. Gauntlett, ``Intersecting branes,''
hep-th/9705011.

\bibitem{Duff:1991xz}
M.~J. Duff and K.~S. Stelle, ``Multi-membrane solutions of D = 11
  supergravity,'' Phys. Lett. {\bf B253} (1991)
113--118.

\bibitem{Nepomechie:1985wu}
R.~I. Nepomechie, ``Magnetic monopoles from antisymmetric tensor gauge
  fields,'' Phys. Rev. {\bf D31} (1985)
1921.

\bibitem{Teitelboim:1986yc}
C.~Teitelboim, ``Monopoles of higher rank,'' Phys. Lett. {\bf B167} (1986)
69.

\bibitem{Gueven:1992hh}
R.~Gueven, ``Black p-brane solutions of D = 11 supergravity theory,'' Phys.
  Lett. {\bf B276} (1992)
49--55.

\bibitem{Duff:1995wd}
M.~J. Duff, J.~T. Liu, and R.~Minasian, ``Eleven-dimensional origin of string /
  string duality: A one-loop test,'' Nucl. Phys. {\bf B452} (1995) 261--282,
hep-th/9506126.

\bibitem{Bergshoeff:1995vh}
E.~Bergshoeff, M.~B. Green, G.~Papadopoulos, and P.~K. Townsend, ``The IIA
  super-eightbrane,''
hep-th/9511079.

\bibitem{Polchinski:1995mt}
J.~Polchinski, ``Dirichlet-branes and Ramond-Ramond charges,'' Phys. Rev. Lett.
  {\bf 75} (1995) 4724--4727,
hep-th/9510017.

\bibitem{Townsend:1995kk}
P.~K. Townsend, ``The eleven-dimensional supermembrane revisited,'' Phys. Lett.
  {\bf B350} (1995) 184--187,
hep-th/9501068.

\bibitem{Ooguri:1996wj}
H.~Ooguri and C.~Vafa, ``Two-Dimensional Black Hole and Singularities of CY
  Manifolds,'' Nucl. Phys. {\bf B463} (1996) 55--72,
hep-th/9511164.

\bibitem{Gregory:1997te}
R.~Gregory, J.~A. Harvey, and G.~W. Moore, ``Unwinding strings and T-duality of
  Kaluza-Klein and H-monopoles,'' Adv. Theor. Math. Phys. {\bf 1} (1997)
  283--297,
hep-th/9708086.

\bibitem{Buscher:1987sk}
T.~H. Buscher, ``A symmetry of the string background field equations,'' Phys.
  Lett. {\bf B194} (1987)
59.

\bibitem{Buscher:1988qj}
T.~H. Buscher, ``Path integral derivation of quantum duality in nonlinear sigma
  models,'' Phys. Lett. {\bf B201} (1988)
466.

\bibitem{Bergshoeff:1995as}
E.~Bergshoeff, C.~M. Hull, and T.~Ortin, ``Duality in the type II superstring
  effective action,'' Nucl. Phys. {\bf B451} (1995) 547--578,
hep-th/9504081.

\bibitem{Dabholkar:1990yf}
A.~Dabholkar, G.~W. Gibbons, J.~A. Harvey, and F.~Ruiz~Ruiz, ``Superstrings and
  solitons,'' Nucl. Phys. {\bf B340} (1990)
33--55.

\bibitem{Callan:1991dj}
C.~G. Callan, J.~A. Harvey, and A.~Strominger, ``World sheet approach to
  heterotic instantons and solitons,'' Nucl. Phys. {\bf B359} (1991)
611--634.

\bibitem{Cremmer:1978km}
E.~Cremmer, B.~Julia, and J.~Scherk, ``Supergravity theory in 11 dimensions,''
  Phys. Lett. {\bf B76} (1978)
409--412.

\bibitem{Sorokin:1999jx}
D.~P. Sorokin, ``Superbranes and superembeddings,'' Phys. Rept. {\bf 329}
  (2000) 1--101,
hep-th/9906142.

\bibitem{Bergshoeff:1997kr}
E.~Bergshoeff, R.~Kallosh, T.~Ortin, and G.~Papadopoulos, ``Kappa-symmetry,
  supersymmetry and intersecting branes,'' Nucl. Phys. {\bf B502} (1997)
  149--169,
hep-th/9705040.

\bibitem{Bergshoeff:1987cm}
E.~Bergshoeff, E.~Sezgin, and P.~K. Townsend, ``Supermembranes and
  eleven-dimensional supergravity,'' Phys. Lett. {\bf B189} (1987)
75--78.

\bibitem{Bandos:1997ui}
I.~Bandos {\em et.~al.}, ``Covariant action for the super-five-brane of
  M-theory,'' Phys. Rev. Lett. {\bf 78} (1997) 4332--4334,
hep-th/9701149.

\bibitem{Aganagic:1997zq}
M.~Aganagic, J.~Park, C.~Popescu, and J.~H. Schwarz, ``World-volume action of
  the M-theory five-brane,'' Nucl. Phys. {\bf B496} (1997) 191--214,
hep-th/9701166.

\bibitem{Aganagic:1997pe}
M.~Aganagic, C.~Popescu, and J.~H. Schwarz, ``D-brane actions with local kappa
  symmetry,'' Phys. Lett. {\bf B393} (1997) 311--315,
hep-th/9610249.

\bibitem{Cederwall:1997ri}
M.~Cederwall, A.~von Gussich, B.~E.~W. Nilsson, P.~Sundell, and A.~Westerberg,
  ``The Dirichlet super-p-branes in ten-dimensional type IIA and IIB
  supergravity,'' Nucl. Phys. {\bf B490} (1997) 179--201,
hep-th/9611159.

\bibitem{Bergshoeff:1997tu}
E.~Bergshoeff and P.~K. Townsend, ``Super D-branes,'' Nucl. Phys. {\bf B490}
  (1997) 145--162,
hep-th/9611173.

\bibitem{Cederwall:1997pv}
M.~Cederwall, A.~von Gussich, B.~E.~W. Nilsson, and A.~Westerberg, ``The
  Dirichlet super-three-brane in ten-dimensional type IIB supergravity,'' Nucl.
  Phys. {\bf B490} (1997) 163--178,
hep-th/9610148.

\bibitem{Sen:1998xi}
A.~Sen, ``String network,'' JHEP {\bf 03} (1998) 005,
hep-th/9711130.

\bibitem{Ortin:1995su}
T.~Ortin, ``SL(2,R) duality covariance of Killing spinors in axion - dilaton
  black holes,'' Phys. Rev. {\bf D51} (1995) 790--794,
hep-th/9404035.

\bibitem{Kaya:1999mm}
A.~Kaya, ``A note on a relation between the Killing spinor and Einstein
  equations,'' Phys. Lett. {\bf B458} (1999) 267--273,
hep-th/9902010.

\bibitem{Kaya:2000zs}
A.~Kaya, ``New brane solutions from Killing spinor equations,'' Nucl. Phys.
  {\bf B583} (2000) 411--430,
hep-th/0004199.

\bibitem{Douglas:1998uz}
M.~R. Douglas, ``Gauge Fields and D-branes,'' J. Geom. Phys. {\bf 28} (1998)
  255--262,
hep-th/9604198.

\bibitem{Banks:1996nj}
T.~Banks, M.~R. Douglas, and N.~Seiberg, ``Probing F-theory with branes,''
  Phys. Lett. {\bf B387} (1996) 278--281,
hep-th/9605199.

\bibitem{Sen:1997sk}
A.~Sen, ``BPS states on a three brane probe,'' Phys. Rev. {\bf D55} (1997)
  2501--2503,
hep-th/9608005.

\bibitem{Douglas:1997yp}
M.~R. Douglas, D.~Kabat, P.~Pouliot, and S.~H. Shenker, ``D-branes and short
  distances in string theory,'' Nucl. Phys. {\bf B485} (1997) 85--127,
hep-th/9608024.

\bibitem{Tseytlin:1997hi}
A.~A. Tseytlin, ``No-force condition and BPS combinations of p-branes in 11 and
  10 dimensions,'' Nucl. Phys. {\bf B487} (1997) 141--154,
hep-th/9609212.

\bibitem{Leigh:1989jq}
R.~G. Leigh, ``Dirac-Born-Infeld action from Dirichlet sigma model,'' Mod.
  Phys. Lett. {\bf A4} (1989)
2767.

\bibitem{Dai:1989ua}
J.~Dai, R.~G. Leigh, and J.~Polchinski, ``New connections between string
  theories,'' Mod. Phys. Lett. {\bf A4} (1989)
2073--2083.

\bibitem{Douglas:1995bn}
M.~R. Douglas, ``Branes within branes,''
hep-th/9512077.

\bibitem{Green:1997dd}
M.~B. Green, J.~A. Harvey, and G.~W. Moore, ``I-brane inflow and anomalous
  couplings on D-branes,'' Class. Quant. Grav. {\bf 14} (1997) 47--52,
hep-th/9605033.

\bibitem{Connes:1998cr}
A.~Connes, M.~R. Douglas, and A.~Schwarz, ``Noncommutative geometry and matrix
  theory: Compactification on tori,'' JHEP {\bf 02} (1998) 003,
hep-th/9711162.

\bibitem{Douglas:1998fm}
M.~R. Douglas and C.~M. Hull, ``D-branes and the noncommutative torus,'' JHEP
  {\bf 02} (1998) 008,
hep-th/9711165.

\bibitem{Hofman:1998iy}
C.~Hofman and E.~Verlinde, ``U-duality of Born-Infeld on the noncommutative
  two-torus,'' JHEP {\bf 12} (1998) 010,
hep-th/9810116.

\bibitem{Chu:1998qz}
C.-S. Chu and P.-M. Ho, ``Noncommutative open string and D-brane,'' Nucl. Phys.
  {\bf B550} (1999) 151--168,
hep-th/9812219.

\bibitem{Chu:1999gi}
C.-S. Chu and P.-M. Ho, ``Constrained quantization of open string in background
  B field and noncommutative D-brane,'' Nucl. Phys. {\bf B568} (2000) 447--456,
hep-th/9906192.

\bibitem{Seiberg:1999vs}
N.~Seiberg and E.~Witten, ``String theory and noncommutative geometry,'' JHEP
  {\bf 09} (1999) 032,
hep-th/9908142.

\bibitem{Moyal:1949sk}
J.~E. Moyal, ``Quantum mechanics as a statistical theory,'' Proc. Cambridge
  Phil. Soc. {\bf 45} (1949)
99--124.

\bibitem{Tseytlin:1997cs1}
A.~A. Tseytlin, ``On non-abelian generalisation of the Born-Infeld action in
  string theory,'' Nucl. Phys. {\bf B501} (1997) 41--52,
hep-th/9701125.

\bibitem{Tseytlin:1999dj}
A.~A. Tseytlin, ``Born-Infeld action, supersymmetry and string theory,''
hep-th/9908105.

\bibitem{Myers:1999ps}
R.~C. Myers, ``Dielectric-branes,'' JHEP {\bf 12} (1999) 022,
hep-th/9910053.

\bibitem{Maldacena:1998re}
J.~Maldacena, ``The large $N$ limit of superconformal field theories and
  supergravity,'' Adv. Theor. Math. Phys. {\bf 2} (1998) 231--252,
hep-th/9711200.

\bibitem{Witten:1998qj}
E.~Witten, ``Anti-de Sitter space and holography,'' Adv. Theor. Math. Phys.
  {\bf 2} (1998) 253--291,
hep-th/9802150.

\bibitem{Gubser:1998bc}
S.~S. Gubser, I.~R. Klebanov, and A.~M. Polyakov, ``Gauge theory correlators
  from non-critical string theory,'' Phys. Lett. {\bf B428} (1998) 105--114,
hep-th/9802109.

\bibitem{Aharony:1999ti}
O.~Aharony, S.~S. Gubser, J.~M. Maldacena, H.~Ooguri, and Y.~Oz, ``Large N
  field theories, string theory and gravity,'' Phys. Rept. {\bf 323} (2000)
  183--386,
hep-th/9905111.

\bibitem{Strominger:1996ac}
A.~Strominger, ``Open p-branes,'' Phys. Lett. {\bf B383} (1996) 44--47,
hep-th/9512059.

\bibitem{Townsend:1997em}
P.~K. Townsend, ``Brane surgery,'' Nucl. Phys. Proc. Suppl. {\bf 58} (1997)
  163--175,
hep-th/9609217.

\bibitem{Argurio:1997qv}
R.~Argurio, F.~Englert, L.~Houart, and P.~Windey, ``On the opening of branes,''
  Phys. Lett. {\bf B408} (1997) 151--156,
hep-th/9704190.

\bibitem{Gauntlett:2001ur}
J.~P. Gauntlett, N.~Kim, D.~Martelli, and D.~Waldram, ``Fivebranes wrapped on
  SLAG three-cycles and related geometry,'' JHEP {\bf 11} (2001) 018,
hep-th/0110034.

\bibitem{Gauntlett:2002sc}
J.~P. Gauntlett, D.~Martelli, S.~Pakis, and D.~Waldram, ``G-structures and
  wrapped NS5-branes,''
hep-th/0205050.

\bibitem{Gauntlett:2002nw}
J.~P. Gauntlett, J.~B. Gutowski, C.~M. Hull, S.~Pakis, and H.~S. Reall, ``All
  supersymmetric solutions of minimal supergravity in five dimensions,''
hep-th/0209114.

\bibitem{Becker:1995kb}
K.~Becker, M.~Becker, and A.~Strominger, ``Five-branes, membranes and
  nonperturbative string theory,'' Nucl. Phys. {\bf B456} (1995) 130--152,
hep-th/9507158.

\bibitem{Townsend:1998mk}
P.~K. Townsend, ``M-branes at angles,'' Nucl. Phys. Proc. Suppl. {\bf 67}
  (1998) 88--92,
hep-th/9708074.

\bibitem{SheikhJabbari:1998cv}
M.~M. Sheikh~Jabbari, ``Classification of different branes at angles,'' Phys.
  Lett. {\bf B420} (1998) 279--284,
hep-th/9710121.

\bibitem{Cvetic:1998yf}
M.~Cvetic and C.~M. Hull, ``Wrapped branes and supersymmetry,'' Nucl. Phys.
  {\bf B519} (1998) 141--158,
hep-th/9709033.

\bibitem{Ooguri:1996ck}
H.~Ooguri, Y.~Oz, and Z.~Yin, ``D-branes on Calabi-Yau spaces and their
  mirrors,'' Nucl. Phys. {\bf B477} (1996) 407--430,
hep-th/9606112.

\bibitem{Berkooz:1996km}
M.~Berkooz, M.~R. Douglas, and R.~G. Leigh, ``Branes intersecting at angles,''
  Nucl. Phys. {\bf B480} (1996) 265--278,
hep-th/9606139.

\bibitem{HarveyLawson}
F.~Harvey and H.~Lawson, ``Calibrated geometries,'' Acta. Math. {\bf 148}
  (1982) 47--157.

\bibitem{Harvey:Book}
F.~Harvey, {\em Spinors and calibrations}.
\newblock Academic Press, New York, 1990.

\bibitem{Becker:1996ay}
K.~Becker {\em et.~al.}, ``Supersymmetric cycles in exceptional holonomy
  manifolds and Calabi-Yau 4-folds,'' Nucl. Phys. {\bf B480} (1996) 225--238,
hep-th/9608116.

\bibitem{Bershadsky:1996qy}
M.~Bershadsky, C.~Vafa, and V.~Sadov, ``D-Branes and Topological Field
  Theories,'' Nucl. Phys. {\bf B463} (1996) 420--434,
hep-th/9511222.

\bibitem{Gauntlett:1998vk}
J.~P. Gauntlett, N.~D. Lambert, and P.~C. West, ``Branes and calibrated
  geometries,'' Commun. Math. Phys. {\bf 202} (1999) 571--592,
hep-th/9803216.

\bibitem{Acharya:1998en}
B.~S. Acharya, J.~M. Figueroa-O'Farrill, and B.~Spence, ``Branes at angles and
  calibrated geometry,'' JHEP {\bf 04} (1998) 012,
hep-th/9803260.

\bibitem{Figueroa-O'Farrill:1998su}
J.~M. Figueroa-O'Farrill, ``Intersecting brane geometries,'' J. Geom. Phys.
  {\bf 35} (2000) 99--125,
hep-th/9806040.

\bibitem{Joyce:2001xt}
D.~Joyce, ``Lectures on Calabi-Yau and special Lagrangian geometry,''
math.dg/0108088.

\bibitem{Karch:1998sj}
A.~Karch, D.~Lust, and A.~Miemiec, ``N = 1 supersymmetric gauge theories and
  supersymmetric 3- cycles,'' Nucl. Phys. {\bf B553} (1999) 483--510,
hep-th/9810254.

\bibitem{Stanciu:1998sk}
S.~Stanciu, ``D-branes in Kazama-Suzuki models,'' Nucl. Phys. {\bf B526} (1998)
  295--310,
hep-th/9708166.

\bibitem{Gauntlett:1998wb}
J.~P. Gauntlett, N.~D. Lambert, and P.~C. West, ``Supersymmetric fivebrane
  solitons,'' Adv. Theor. Math. Phys. {\bf 3} (1999) 91--118,
hep-th/9811024.

\bibitem{Gauntlett:1999aw}
J.~P. Gauntlett, ``Membranes on fivebranes,'' Adv. Theor. Math. Phys. {\bf 3}
  (1999) 775--790,
hep-th/9906162.

\bibitem{Lust:1999pq}
D.~Lust and A.~Miemiec, ``Supersymmetric M5-branes with H-field,'' Phys. Lett.
  {\bf B476} (2000) 395--401,
hep-th/9912065.

\bibitem{Gutowski:1999iu}
J.~Gutowski and G.~Papadopoulos, ``AdS calibrations,'' Phys. Lett. {\bf B462}
  (1999) 81--88,
hep-th/9902034.

\bibitem{Gutowski:1999tu}
J.~Gutowski, G.~Papadopoulos, and P.~K. Townsend, ``Supersymmetry and
  generalized calibrations,'' Phys. Rev. {\bf D60} (1999) 106006,
hep-th/9905156.

\bibitem{Barwald:1999ux}
O.~Barwald, N.~D. Lambert, and P.~C. West, ``A calibration bound for the
  M-theory fivebrane,'' Phys. Lett. {\bf B463} (1999) 33--40,
hep-th/9907170.

\bibitem{Gutowski:2002bc}
J.~Gutowski, S.~Ivanov, and G.~Papadopoulos, ``Deformations of generalized
  calibrations and compact non- Kahler manifolds with vanishing first Chern
  class,''
math.dg/0205012.

\bibitem{Townsend:1999nf}
P.~K. Townsend, ``PhreMology: Calibrating M-branes,'' Class. Quant. Grav. {\bf
  17} (2000) 1267--1276,
hep-th/9911154.

\bibitem{Acharya:1998st}
B.~S. Acharya, J.~M. Figueroa-O'Farrill, B.~Spence, and S.~Stanciu, ``Planes,
  branes and automorphisms. II: Branes in motion,'' JHEP {\bf 07} (1998) 005,
hep-th/9805176.

\bibitem{Strominger:1986uh}
A.~Strominger, ``Superstrings with torsion,'' Nucl. Phys. {\bf B274} (1986)
253.

\bibitem{Papadopoulos:2000iv}
G.~Papadopoulos, ``Brane solitons and hypercomplex structures,''
math.dg/0003024.

\bibitem{Friedrich:2001nh}
T.~Friedrich and S.~Ivanov, ``Parallel spinors and connections with
  skew-symmetric torsion in string theory,''
math.dg/0102142.

\bibitem{Ivanov:2001ma}
S.~Ivanov, ``Connection with torsion, parallel spinors and geometry of Spin(7)
  manifolds,''
math.dg/0111216.

\bibitem{Friedrich:2001yp}
T.~Friedrich and S.~Ivanov, ``Killing spinor equations in dimension 7 and
  geometry of integrable G$_2$-manifolds,''
math.dg/0112201.

\bibitem{Tod:1983pm}
K.~P. Tod, ``All metrics admitting supercovariantly constant spinors,'' Phys.
  Lett. {\bf B121} (1983)
241--244.

\bibitem{Gibbons:1982fy}
G.~W. Gibbons and C.~M. Hull, ``A Bogomolny bound for general relativity and
  solitons in N=2 supergravity,'' Phys. Lett. {\bf B109} (1982)
190.

\bibitem{Tod:1995jf}
K.~P. Tod, ``More on supercovariantly constant spinors,'' Class. Quant. Grav.
  {\bf 12} (1995)
1801--1820.

\bibitem{Figueroa-O'Farrill:2001tw}
J.~Figueroa-O'Farrill, ``Maximal supersymmetry in ten and eleven dimensions,''
math.dg/0109162.

\bibitem{Figueroa-O'Farrill:2001nz}
J.~Figueroa-O'Farrill and G.~Papadopoulos, ``Homogeneous fluxes, branes and a
  maximally supersymmetric solution of M-theory,'' JHEP {\bf 08} (2001) 036,
hep-th/0105308.

\bibitem{Blau:2001ne}
M.~Blau, J.~Figueroa-O'Farrill, C.~Hull, and G.~Papadopoulos, ``A new maximally
  supersymmetric background of IIB superstring theory,'' JHEP {\bf 01} (2002)
  047,
hep-th/0110242.

\bibitem{Vancea:2000zu}
I.~V. Vancea, ``Note on four Dp-branes at angles,'' JHEP {\bf 04} (2001) 020,
hep-th/0011251.

\bibitem{Matusis:1997qp}
A.~Matusis, ``Interaction of non-parallel D1-branes,'' Int. J. Mod. Phys. {\bf
  A14} (1999) 1153--1162,
hep-th/9707135.

\bibitem{Bachas:1996kx}
C.~Bachas, ``D-brane dynamics,'' Phys. Lett. {\bf B374} (1996) 37--42,
hep-th/9511043.

\bibitem{Hashimoto:1997gm}
A.~Hashimoto and W.~Taylor~IV, ``Fluctuation spectra of tilted and intersecting
  D-branes from the Born-Infeld action,'' Nucl. Phys. {\bf B503} (1997)
  193--219,
hep-th/9703217.

\bibitem{Ohta:1998fr}
N.~Ohta and P.~K. Townsend, ``Supersymmetry of M-branes at angles,'' Phys.
  Lett. {\bf B418} (1998) 77--84,
hep-th/9710129.

\bibitem{Gibbons:1998hm}
G.~W. Gibbons and G.~Papadopoulos, ``Calibrations and intersecting branes,''
  Commun. Math. Phys. {\bf 202} (1999) 593--619,
hep-th/9803163.

\bibitem{Acharya:1998yv}
B.~S. Acharya, J.~M. Figueroa-O'Farrill, and B.~Spence, ``Planes, branes and
  automorphisms. I: Static branes,'' JHEP {\bf 07} (1998) 004,
hep-th/9805073.

\bibitem{Biswas:2002yz}
A.~Biswas, A.~Kumar, and K.~L. Panigrahi, ``p - p' branes in pp-wave
  background,'' Phys. Rev. {\bf D66} (2002) 126002,
hep-th/0208042.

\bibitem{Nayak:2002ty}
R.~R. Nayak, ``D-branes at angle in pp-wave background,''
hep-th/0210230.

\bibitem{Aharony:1997ju}
O.~Aharony and A.~Hanany, ``Branes, superpotentials and superconformal fixed
  points,'' Nucl. Phys. {\bf B504} (1997) 239--271,
hep-th/9704170.

\bibitem{Aharony:1998bh}
O.~Aharony, A.~Hanany, and B.~Kol, ``Webs of (p,q) 5-branes, five dimensional
  field theories and grid diagrams,'' JHEP {\bf 01} (1998) 002,
hep-th/9710116.

\bibitem{Aharony:1996xr}
O.~Aharony, J.~Sonnenschein, and S.~Yankielowicz, ``Interactions of strings and
  D-branes from M theory,'' Nucl. Phys. {\bf B474} (1996) 309--322,
hep-th/9603009.

\bibitem{Schwarz:1997bh}
J.~H. Schwarz, ``Lectures on superstring and M theory dualities,'' Nucl. Phys.
  Proc. Suppl. {\bf 55B} (1997) 1--32,
hep-th/9607201.

\bibitem{Gaberdiel:1998ud}
M.~R. Gaberdiel and B.~Zwiebach, ``Exceptional groups from open strings,''
  Nucl. Phys. {\bf B518} (1998) 151--172,
hep-th/9709013.

\bibitem{Dasgupta:1998pu}
K.~Dasgupta and S.~Mukhi, ``BPS nature of 3-string junctions,'' Phys. Lett.
  {\bf B423} (1998) 261--264,
hep-th/9711094.

\bibitem{Krogh:1998dx}
M.~Krogh and S.-M. Lee, ``String network from M-theory,'' Nucl. Phys. {\bf
  B516} (1998) 241--254,
hep-th/9712050.

\bibitem{Matsuo:1998jw}
Y.~Matsuo and K.~Okuyama, ``BPS condition of string junction from M theory,''
  Phys. Lett. {\bf B426} (1998) 294--296,
hep-th/9712070.

\bibitem{Bhattacharyya:1998vr}
S.~Bhattacharyya, A.~Kumar, and S.~Mukhopadhyay, ``String network and
  U-duality,'' Phys. Rev. Lett. {\bf 81} (1998) 754--757,
hep-th/9801141.

\bibitem{Kol:1998cf}
B.~Kol and J.~Rahmfeld, ``BPS spectrum of 5 dimensional field theories, (p,q)
  webs and curve counting,'' JHEP {\bf 08} (1998) 006,
hep-th/9801067.

\bibitem{Bergman:1998yw}
O.~Bergman, ``Three-pronged strings and 1/4 BPS states in N = 4 super-
  Yang-Mills theory,'' Nucl. Phys. {\bf B525} (1998) 104--116,
hep-th/9712211.

\bibitem{Bergman:1998br}
O.~Bergman and A.~Fayyazuddin, ``String junctions and BPS states in
  Seiberg-Witten theory,'' Nucl. Phys. {\bf B531} (1998) 108--124,
hep-th/9802033.

\bibitem{Hashimoto:1998zs}
K.~Hashimoto, H.~Hata, and N.~Sasakura, ``3-string junction and BPS saturated
  solutions in SU(3) supersymmetric Yang-Mills theory,'' Phys. Lett. {\bf B431}
  (1998) 303--310,
hep-th/9803127.

\bibitem{Bergman:1998gs}
O.~Bergman and B.~Kol, ``String webs and 1/4 BPS monopoles,'' Nucl. Phys. {\bf
  B536} (1998) 149--174,
hep-th/9804160.

\bibitem{Bergman:1998ej}
O.~Bergman and A.~Fayyazuddin, ``String junction transitions in the moduli
  space of N = 2 SYM,'' Nucl. Phys. {\bf B535} (1998) 139--151,
hep-th/9806011.

\bibitem{Ohtake:1998sd}
Y.~Ohtake, ``String junctions and the BPS spectrum of N = 2 SU(2) theory with
  massive matter fields,'' Prog. Theor. Phys. {\bf 102} (1999) 671--684,
hep-th/9812227.

\bibitem{DeWolfe:1998zf}
O.~DeWolfe and B.~Zwiebach, ``String junctions for arbitrary Lie algebra
  representations,'' Nucl. Phys. {\bf B541} (1999) 509--565,
hep-th/9804210.

\bibitem{Iqbal:1998xb}
A.~Iqbal, ``Self-intersection number of BPS junctions in backgrounds of three
  and seven-branes,'' JHEP {\bf 10} (1999) 032,
hep-th/9807117.

\bibitem{DeWolfe:1998yf}
O.~DeWolfe, ``Affine Lie algebras, string junctions and 7-branes,'' Nucl. Phys.
  {\bf B550} (1999) 622--637,
hep-th/9809026.

\bibitem{DeWolfe:1998pr}
O.~DeWolfe, T.~Hauer, A.~Iqbal, and B.~Zwiebach, ``Uncovering infinite
  symmetries on (p,q) 7-branes: Kac-Moody algebras and beyond,'' Adv. Theor.
  Math. Phys. {\bf 3} (1999) 1835--1891,
hep-th/9812209.

\bibitem{Hauer:1999pt}
T.~Hauer and A.~Iqbal, ``Del Pezzo surfaces and affine 7-brane backgrounds,''
  JHEP {\bf 01} (2000) 043,
hep-th/9910054.

\bibitem{Mohri:2000wu}
K.~Mohri, Y.~Ohtake, and S.-K. Yang, ``Duality between string junctions and
  D-branes on del Pezzo surfaces,'' Nucl. Phys. {\bf B595} (2001) 138--164,
hep-th/0007243.

\bibitem{Ohtake:2001fv}
Y.~Ohtake, ``String webs and curves of marginal stability in five- dimensional
  E(N) theories on S(1),''
hep-th/0105291.

\bibitem{Callan:1998kz}
C.~G. Callan and J.~M. Maldacena, ``Brane dynamics from the Born-Infeld
  action,'' Nucl. Phys. {\bf B513} (1998) 198--212,
hep-th/9708147.

\bibitem{Howe:1998ue}
P.~S. Howe, N.~D. Lambert, and P.~C. West, ``The self-dual string soliton,''
  Nucl. Phys. {\bf B515} (1998) 203--216,
hep-th/9709014.

\bibitem{Gibbons:1998xz}
G.~W. Gibbons, ``Born-Infeld particles and Dirichlet p-branes,'' Nucl. Phys.
  {\bf B514} (1998) 603--639,
hep-th/9709027.

\bibitem{Lee:1998xh}
S.-M. Lee, A.~Peet, and L.~Thorlacius, ``Brane-waves and strings,'' Nucl. Phys.
  {\bf B514} (1998) 161--176,
hep-th/9710097.

\bibitem{Constable:1999ac}
N.~R. Constable, R.~C. Myers, and O.~Tafjord, ``The noncommutative bion core,''
  Phys. Rev. {\bf D61} (2000) 106009,
hep-th/9911136.

\bibitem{Diaconescu:1997rk}
D.-E. Diaconescu, ``D-branes, monopoles and Nahm equations,'' Nucl. Phys. {\bf
  B503} (1997) 220--238,
hep-th/9608163.

\bibitem{Constable:2001kv}
N.~R. Constable, R.~C. Myers, and O.~Tafjord, ``Fuzzy funnels: Non-abelian
  brane intersections,''
hep-th/0105035.

\bibitem{Howe:1998et}
P.~S. Howe, N.~D. Lambert, and P.~C. West, ``The threebrane soliton of the
  M-fivebrane,'' Phys. Lett. {\bf B419} (1998) 79--83,
hep-th/9710033.

\bibitem{Gauntlett:1998ss}
J.~P. Gauntlett, J.~Gomis, and P.~K. Townsend, ``BPS bounds for worldvolume
  branes,'' JHEP {\bf 01} (1998) 003,
hep-th/9711205.

\bibitem{Lambert:2002ms}
N.~D. Lambert, ``Moduli and brane intersections,'' Phys. Rev. {\bf D67} (2003)
  026006,
hep-th/0209141.

\bibitem{Gutowski:1998xt}
J.~Gutowski and G.~Papadopoulos, ``The dynamics of D-3-brane dyons and toric
  hyper-Kaehler manifolds,'' Nucl. Phys. {\bf B551} (1999) 650--666,
hep-th/9811207.

\bibitem{Savvidy:1999wx}
K.~G. Savvidy and G.~K. Savvidy, ``Von Neumann boundary conditions from
  Born-Infeld dynamics,'' Nucl. Phys. {\bf B561} (1999) 117--124,
hep-th/9902023.

\bibitem{Kastor:1999ag}
D.~Kastor and J.~Traschen, ``Dynamics of the DBI spike soliton,'' Phys. Rev.
  {\bf D61} (2000) 024034,
hep-th/9906237.

\bibitem{Townsend:1999hi}
P.~K. Townsend, ``Brane theory solitons,''
hep-th/0004039.

\bibitem{Gauntlett:2000de}
J.~P. Gauntlett, R.~Portugues, D.~Tong, and P.~K. Townsend, ``D-brane solitons
  in supersymmetric sigma-models,'' Phys. Rev. {\bf D63} (2001) 085002,
hep-th/0008221.

\bibitem{Portugues:2002ih}
R.~Portugues and P.~K. Townsend, ``Sigma-model soliton intersections from
  exceptional calibrations,'' JHEP {\bf 04} (2002) 039,
hep-th/0203181.

\bibitem{Gauntlett:1999xz}
J.~P. Gauntlett, C.~Koehl, D.~Mateos, P.~K. Townsend, and M.~Zamaklar, ``Finite
  energy Dirac-Born-Infeld monopoles and string junctions,'' Phys. Rev. {\bf
  D60} (1999) 045004,
hep-th/9903156.

\bibitem{Michishita:2000hu}
Y.~Michishita, ``The M2-brane soliton on the M5-brane with constant 3-form,''
  JHEP {\bf 09} (2000) 036,
hep-th/0008247.

\bibitem{Youm:2000kr}
D.~Youm, ``BPS solitons in M5-brane worldvolume theory with constant three-form
  field,'' Phys. Rev. {\bf D63} (2001) 045004,
hep-th/0009082.

\bibitem{Karczmarek:2001pn}
J.~L. Karczmarek and C.~G. Callan, ``Tilting the noncommutative bion,'' JHEP
  {\bf 05} (2002) 038,
hep-th/0111133.

\bibitem{Bak:1998xp}
D.~Bak, J.-H. Lee, and H.~Min, ``Dynamics of BPS states in the
  Dirac-Born-Infeld theory,'' Phys. Rev. {\bf D59} (1999) 045011,
hep-th/9806149.

\bibitem{Brecher:1998tv}
D.~Brecher, ``BPS states of the non-Abelian Born-Infeld action,'' Phys. Lett.
  {\bf B442} (1998) 117--124,
hep-th/9804180.

\bibitem{Constable:2001ag}
N.~R. Constable, R.~C. Myers, and O.~Tafjord, ``Non-Abelian brane
  intersections,'' JHEP {\bf 06} (2001) 023,
hep-th/0102080.

\bibitem{Savvidy:1998xx}
K.~G. Savvidy, ``Brane death via Born-Infeld string,''
hep-th/9810163.

\bibitem{Gorsky:1999gk}
A.~Gorsky and K.~Selivanov, ``Junctions and the fate of branes in external
  fields,'' Nucl. Phys. {\bf B571} (2000) 120--134,
hep-th/9904041.

\bibitem{Hashimoto:1998qh}
K.~Hashimoto, ``String junction from worldsheet gauge theory,'' Prog. Theor.
  Phys. {\bf 101} (1999) 1353--1370,
hep-th/9808185.

\bibitem{Bain:1999hu}
P.~Bain, ``On the non-Abelian Born-Infeld action,''
hep-th/9909154.

\bibitem{Thorlacius:1998zd}
L.~Thorlacius, ``Born-Infeld string as a boundary conformal field theory,''
  Phys. Rev. Lett. {\bf 80} (1998) 1588--1590,
hep-th/9710181.

\bibitem{Hashimoto:1998px}
A.~Hashimoto, ``The shape of branes pulled by strings,'' Phys. Rev. {\bf D57}
  (1998) 6441--6451,
hep-th/9711097.

\bibitem{Nahm:1980yw}
W.~Nahm, ``A simple formalism for the BPS monopole,'' Phys. Lett. {\bf B90}
  (1980)
413.

\bibitem{Constable:2002yn}
N.~R. Constable and N.~D. Lambert, ``Calibrations, monopoles and fuzzy
  funnels,'' Phys. Rev. {\bf D66} (2002) 065016,
hep-th/0206243.

\bibitem{Tseytlin:1996bh}
A.~A. Tseytlin, ``Harmonic superpositions of M-branes,'' Nucl. Phys. {\bf B475}
  (1996) 149--163,
hep-th/9604035.

\bibitem{Gauntlett:1996pb}
J.~P. Gauntlett, D.~A. Kastor, and J.~Traschen, ``Overlapping Branes in
  M-Theory,'' Nucl. Phys. {\bf B478} (1996) 544--560,
hep-th/9604179.

\bibitem{Tseytlin:1997cs}
A.~A. Tseytlin, ``Composite BPS configurations of p-branes in 10 and 11
  dimensions,'' Class. Quant. Grav. {\bf 14} (1997) 2085--2105,
hep-th/9702163.

\bibitem{Bergshoeff:1997tt}
E.~Bergshoeff, M.~de~Roo, E.~Eyras, B.~Janssen, and J.~P. van~der Schaar,
  ``Intersections involving monopoles and waves in eleven dimensions,'' Class.
  Quant. Grav. {\bf 14} (1997) 2757--2769,
hep-th/9704120.

\bibitem{Lu:1998mi}
H.~Lu and C.~N. Pope, ``Interacting intersections,'' Int. J. Mod. Phys. {\bf
  A13} (1998) 4425--4443,
hep-th/9710155.

\bibitem{Balasubramanian:1997uc}
V.~Balasubramanian and R.~G. Leigh, ``D-branes, moduli and supersymmetry,''
  Phys. Rev. {\bf D55} (1997) 6415--6422,
hep-th/9611165.

\bibitem{Behrndt:1997ph}
K.~Behrndt and M.~Cvetic, ``BPS-saturated bound states of tilted p-branes in
  type II string theory,'' Phys. Rev. {\bf D56} (1997) 1188--1193,
hep-th/9702205.

\bibitem{Costa:1997dt}
M.~S. Costa and M.~Cvetic, ``Non-threshold D-brane bound states and black holes
  with non-zero entropy,'' Phys. Rev. {\bf D56} (1997) 4834--4843,
hep-th/9703204.

\bibitem{Breckenridge:1997ar}
J.~C. Breckenridge, G.~Michaud, and R.~C. Myers, ``New angles on D-branes,''
  Phys. Rev. {\bf D56} (1997) 5172--5178,
hep-th/9703041.

\bibitem{Hambli:1997uq}
N.~Hambli, ``Comments on Dirichlet branes at angles,'' Phys. Rev. {\bf D56}
  (1997) 2369--2377,
hep-th/9703179.

\bibitem{Balasubramanian:1998az}
V.~Balasubramanian, F.~Larsen, and R.~G. Leigh, ``Branes at angles and black
  holes,'' Phys. Rev. {\bf D57} (1998) 3509--3528,
hep-th/9704143.

\bibitem{Michaud:1997tk}
G.~Michaud and R.~C. Myers, ``Hermitian D-brane solutions,'' Phys. Rev. {\bf
  D56} (1997) 3698--3705,
hep-th/9705079.

\bibitem{Biswas:2002sa}
A.~Biswas and K.~L. Panigrahi, ``Intersecting membranes from charged
  macroscopic strings,'' Mod. Phys. Lett. {\bf A17} (2002) 1297--1304,
hep-th/0205138.

\bibitem{Duff:1995yh}
M.~J. Duff, S.~Ferrara, R.~R. Khuri, and J.~Rahmfeld, ``Supersymmetry and dual
  string solitons,'' Phys. Lett. {\bf B356} (1995) 479--486,
hep-th/9506057.

\bibitem{Bergshoeff:1996sq}
E.~Bergshoeff, H.~J. Boonstra, and T.~Ortin, ``S duality and dyonic p-brane
  solutions in type II string theory,'' Phys. Rev. {\bf D53} (1996) 7206--7212,
hep-th/9508091.

\bibitem{Izquierdo:1996ms}
J.~M. Izquierdo, N.~D. Lambert, G.~Papadopoulos, and P.~K. Townsend, ``Dyonic
  Membranes,'' Nucl. Phys. {\bf B460} (1996) 560--578,
hep-th/9508177.

\bibitem{Russo:1997if}
J.~G. Russo and A.~A. Tseytlin, ``Waves, boosted branes and BPS states in
  M-theory,'' Nucl. Phys. {\bf B490} (1997) 121--144,
hep-th/9611047.

\bibitem{Breckenridge:1997tt}
J.~C. Breckenridge, G.~Michaud, and R.~C. Myers, ``More D-brane bound states,''
  Phys. Rev. {\bf D55} (1997) 6438--6446,
hep-th/9611174.

\bibitem{Costa:1998zd}
M.~S. Costa and G.~Papadopoulos, ``Superstring dualities and p-brane bound
  states,'' Nucl. Phys. {\bf B510} (1998) 217--231,
hep-th/9612204.

\bibitem{Sorokin:1997ps}
D.~P. Sorokin and P.~K. Townsend, ``M-theory superalgebra from the M-5-brane,''
  Phys. Lett. {\bf B412} (1997) 265--273,
hep-th/9708003.

\bibitem{Kumar:2001ag}
A.~Kumar, R.~R. Nayak, and K.~L. Panigrahi, ``Bound states of string networks
  and D-branes,'' Phys. Rev. Lett. {\bf 88} (2002) 121601,
hep-th/0108174.

\bibitem{Papadopoulos:1996uq}
G.~Papadopoulos and P.~K. Townsend, ``Intersecting M-branes,'' Phys. Lett. {\bf
  B380} (1996) 273--279,
hep-th/9603087.

\bibitem{Yang:1999ze}
H.~Yang, ``Localized intersecting brane solutions of D = 11 supergravity,''
hep-th/9902128.

\bibitem{Strominger:1996sh}
A.~Strominger and C.~Vafa, ``Microscopic Origin of the Bekenstein-Hawking
  Entropy,'' Phys. Lett. {\bf B379} (1996) 99--104,
hep-th/9601029.

\bibitem{Bekenstein:1972tm}
J.~D. Bekenstein, ``Black holes and the second law,'' Nuovo Cim. Lett. {\bf 4}
  (1972)
737--740.

\bibitem{Bekenstein:1973ur}
J.~D. Bekenstein, ``Black holes and entropy,'' Phys. Rev. {\bf D7} (1973)
2333--2346.

\bibitem{Bardeen:1973gs}
J.~M. Bardeen, B.~Carter, and S.~W. Hawking, ``The Four laws of black hole
  mechanics,'' Commun. Math. Phys. {\bf 31} (1973)
161--170.

\bibitem{Hawking:1974rv}
S.~W. Hawking, ``Black hole explosions,'' Nature {\bf 248} (1974)
30--31.

\bibitem{Bekenstein:1974ax}
J.~D. Bekenstein, ``Generalized second law of thermodynamics in black hole
  physics,'' Phys. Rev. {\bf D9} (1974)
3292--3300.

\bibitem{Hawking:1975sw}
S.~W. Hawking, ``Particle creation by black holes,'' Commun. Math. Phys. {\bf
  43} (1975)
199--220.

\bibitem{Maldacena:1996ky}
J.~M. Maldacena, ``Black holes in string theory,''
hep-th/9607235.

\bibitem{Youm:1997hw}
D.~Youm, ``Black holes and solitons in string theory,'' Phys. Rept. {\bf 316}
  (1999) 1--232,
hep-th/9710046.

\bibitem{Peet:1998es}
A.~W. Peet, ``The Bekenstein formula and string theory (N-brane theory),''
  Class. Quant. Grav. {\bf 15} (1998) 3291--3338,
hep-th/9712253.

\bibitem{Mohaupt:2000mj}
T.~Mohaupt, ``Black hole entropy, special geometry and strings,'' Fortsch.
  Phys. {\bf 49} (2001) 3--161,
hep-th/0007195.

\bibitem{Callan:1996dv}
C.~G. Callan and J.~M. Maldacena, ``D-brane Approach to Black Hole Quantum
  Mechanics,'' Nucl. Phys. {\bf B472} (1996) 591--610,
hep-th/9602043.

\bibitem{Maldacena:1996ds}
J.~M. Maldacena and L.~Susskind, ``D-branes and Fat Black Holes,'' Nucl. Phys.
  {\bf B475} (1996) 679--690,
hep-th/9604042.

\bibitem{Breckenridge:1997is}
J.~C. Breckenridge, R.~C. Myers, A.~W. Peet, and C.~Vafa, ``D-branes and
  spinning black holes,'' Phys. Lett. {\bf B391} (1997) 93--98,
hep-th/9602065.

\bibitem{Breckenridge:1996sn}
J.~C. Breckenridge {\em et.~al.}, ``Macroscopic and Microscopic Entropy of
  Near-Extremal Spinning Black Holes,'' Phys. Lett. {\bf B381} (1996) 423--426,
hep-th/9603078.

\bibitem{Gubser:1996de}
S.~S. Gubser, I.~R. Klebanov, and A.~W. Peet, ``Entropy and Temperature of
  Black 3-Branes,'' Phys. Rev. {\bf D54} (1996) 3915--3919,
hep-th/9602135.

\bibitem{Horowitz:1996rn}
G.~T. Horowitz, ``Quantum states of black holes,''
gr-qc/9704072.

\bibitem{Cvetic:1996uj}
M.~Cvetic and D.~Youm, ``Dyonic BPS saturated black holes of heterotic string
  on a six torus,'' Phys. Rev. {\bf D53} (1996) 584--588,
hep-th/9507090.

\bibitem{Cvetic:1996yq}
M.~Cvetic and A.~A. Tseytlin, ``General class of BPS saturated dyonic black
  holes as exact superstring solutions,'' Phys. Lett. {\bf B366} (1996)
  95--103,
hep-th/9510097.

\bibitem{Cvetic:1996bj}
M.~Cvetic and A.~A. Tseytlin, ``Solitonic strings and BPS saturated dyonic
  black holes,'' Phys. Rev. {\bf D53} (1996) 5619--5633,
hep-th/9512031.

\bibitem{Cvetic:1996gq}
M.~Cvetic and A.~A. Tseytlin, ``Non-extreme black holes from non-extreme
  intersecting M- branes,'' Nucl. Phys. {\bf B478} (1996) 181--198,
hep-th/9606033.

\bibitem{Klebanov:1996mh}
I.~R. Klebanov and A.~A. Tseytlin, ``Intersecting M-branes as four-dimensional
  black holes,'' Nucl. Phys. {\bf B475} (1996) 179--192,
hep-th/9604166.

\bibitem{Balasubramanian:1996rx}
V.~Balasubramanian and F.~Larsen, ``On D-Branes and Black Holes in Four
  Dimensions,'' Nucl. Phys. {\bf B478} (1996) 199--208,
hep-th/9604189.

\bibitem{Behrndt:1997mm}
K.~Behrndt and T.~Mohaupt, ``Entropy of N = 2 black holes and their M-brane
  description,'' Phys. Rev. {\bf D56} (1997) 2206--2211,
hep-th/9611140.

\bibitem{Bertolini:2000ei}
M.~Bertolini and M.~Trigiante, ``Regular BPS black holes: Macroscopic and
  microscopic description of the generating solution,'' Nucl. Phys. {\bf B582}
  (2000) 393--406,
hep-th/0002191.

\bibitem{Bertolini:2000ya}
M.~Bertolini and M.~Trigiante, ``Microscopic entropy of the most general
  four-dimensional BPS black hole,'' JHEP {\bf 10} (2000) 002,
hep-th/0008201.

\bibitem{Maldacena:1997qk}
J.~M. Maldacena, ``N = 2 extremal black holes and intersecting branes,'' Phys.
  Lett. {\bf B403} (1997) 20--22,
hep-th/9611163.

\bibitem{Maldacena:1997de}
J.~M. Maldacena, A.~Strominger, and E.~Witten, ``Black hole entropy in
  M-theory,'' JHEP {\bf 12} (1997) 002,
hep-th/9711053.

\bibitem{Tseytlin:1996as}
A.~A. Tseytlin, ``Extreme dyonic black holes in string theory,'' Mod. Phys.
  Lett. {\bf A11} (1996) 689--714,
hep-th/9601177.

\bibitem{Tseytlin:1996qg}
A.~A. Tseytlin, ``Extremal black hole entropy from conformal string sigma
  model,'' Nucl. Phys. {\bf B477} (1996) 431--448,
hep-th/9605091.

\bibitem{Vafa:1998gr}
C.~Vafa, ``Black holes and Calabi-Yau threefolds,'' Adv. Theor. Math. Phys.
  {\bf 2} (1998) 207--218,
hep-th/9711067.

\bibitem{Harvey:1998bx}
J.~A. Harvey, R.~Minasian, and G.~W. Moore, ``Non-abelian tensor-multiplet
  anomalies,'' JHEP {\bf 09} (1998) 004,
hep-th/9808060.

\bibitem{LopesCardoso:1999ur}
G.~Lopes~Cardoso, B.~de~Wit, and T.~Mohaupt, ``Macroscopic entropy formulae and
  non-holomorphic corrections for supersymmetric black holes,'' Nucl. Phys.
  {\bf B567} (2000) 87--110,
hep-th/9906094.

\bibitem{Horowitz:1996ac}
G.~T. Horowitz, D.~A. Lowe, and J.~M. Maldacena, ``Statistical Entropy of
  Nonextremal Four-Dimensional Black Holes and U-Duality,'' Phys. Rev. Lett.
  {\bf 77} (1996) 430--433,
hep-th/9603195.

\bibitem{Ortin:1998bz}
T.~Ortin, ``Extremality versus supersymmetry in stringy black holes,'' Phys.
  Lett. {\bf B422} (1998) 93--100,
hep-th/9612142.

\bibitem{Dabholkar:1997rk}
A.~Dabholkar, ``Microstates of non-supersymmetric black holes,'' Phys. Lett.
  {\bf B402} (1997) 53--58,
hep-th/9702050.

\bibitem{Sheinblatt:1998nt}
H.~J. Sheinblatt, ``Statistical entropy of an extremal black hole with 0- and
  6-brane charge,'' Phys. Rev. {\bf D57} (1998) 2421--2426,
hep-th/9705054.

\bibitem{Itzhaki:1997fm}
N.~Itzhaki, ``Stringy corrections to Kaluza-Klein black holes,'' Nucl. Phys.
  {\bf B508} (1997) 700--714,
hep-th/9704096.

\bibitem{Ortin:1997yn}
T.~Ortin, ``Non-supersymmetric (but) extreme black holes, scalar hair and other
  open problems,''
hep-th/9705095.

\bibitem{Dabholkar:1998fc}
A.~Dabholkar, G.~Mandal, and P.~Ramadevi, ``Nonrenormalization of mass of some
  nonsupersymmetric string states,'' Nucl. Phys. {\bf B520} (1998) 117--131,
hep-th/9705239.

\bibitem{Pierre:1997zd}
J.~M. Pierre, ``Comparing D-branes and black holes with 0- and 6-brane
  charge,'' Phys. Rev. {\bf D56} (1997) 6710--6713,
hep-th/9707102.

\bibitem{Dhar:1998ip}
A.~Dhar and G.~Mandal, ``Probing 4-dimensional nonsupersymmetric black holes
  carrying D0- and D6-brane charges,'' Nucl. Phys. {\bf B531} (1998) 256--274,
hep-th/9803004.

\bibitem{Barbon:1998nx}
J.~L.~F. Barbon, J.~L. Manes, and M.~A. Vazquez-Mozo, ``Large N limit of
  extremal non-supersymmetric black holes,'' Nucl. Phys. {\bf B536} (1998)
  279--300,
hep-th/9805154.

\bibitem{Youm:1999zs}
D.~Youm, ``Localized intersecting BPS branes,''
hep-th/9902208.

\bibitem{Fayyazuddin:1999zu}
A.~Fayyazuddin and D.~J. Smith, ``Localized intersections of M5-branes and
  four-dimensional superconformal field theories,'' JHEP {\bf 04} (1999) 030,
hep-th/9902210.

\bibitem{Loewy:1999mn}
A.~Loewy, ``Semi localized brane intersections in SUGRA,'' Phys. Lett. {\bf
  B463} (1999) 41--47,
hep-th/9903038.

\bibitem{Youm:1999tt}
D.~Youm, ``Supergravity solutions for BI dyons,'' Phys. Rev. {\bf D60} (1999)
  105006,
hep-th/9905155.

\bibitem{Brecher:2000pa}
D.~Brecher, A.~Chamblin, and H.~S. Reall, ``AdS/CFT in the infinite momentum
  frame,'' Nucl. Phys. {\bf B607} (2001) 155--190,
hep-th/0012076.

\bibitem{Gibbons:1996vg}
G.~W. Gibbons, M.~B. Green, and M.~J. Perry, ``Instantons and Seven-Branes in
  Type IIB Superstring Theory,'' Phys. Lett. {\bf B370} (1996) 37--44,
hep-th/9511080.

\bibitem{Chu:1998in}
C.-S. Chu, P.-M. Ho, and Y.-Y. Wu, ``D-instanton in AdS(5) and instanton in
  SYM(4),'' Nucl. Phys. {\bf B541} (1999) 179--194,
hep-th/9806103.

\bibitem{Kogan:1998re}
I.~I. Kogan and G.~Luzon, ``D-instantons on the boundary,'' Nucl. Phys. {\bf
  B539} (1999) 121--134,
hep-th/9806197.

\bibitem{Bianchi:1998nk}
M.~Bianchi, M.~B. Green, S.~Kovacs, and G.~Rossi, ``Instantons in
  supersymmetric Yang-Mills and D-instantons in IIB superstring theory,'' JHEP
  {\bf 08} (1998) 013,
hep-th/9807033.

\bibitem{Park:1998uv}
C.~Park and S.-J. Sin, ``Large N limit and instanton effects in the AdS/CFT
  correspondence,'' Phys. Lett. {\bf B444} (1998) 156--162,
hep-th/9807156.

\bibitem{Cvetic:2000cj}
M.~Cvetic, H.~Lu, C.~N. Pope, and J.~F. Vazquez-Poritz, ``AdS in warped
  spacetimes,'' Phys. Rev. {\bf D62} (2000) 122003,
hep-th/0005246.

\bibitem{Khuri:1993cs}
R.~R. Khuri, ``Remark on string solitons,'' Phys. Rev. {\bf D48} (1993)
2947--2948.

\bibitem{Khuri:1993ii}
R.~R. Khuri, ``A Comment on string solitons,''
hep-th/9305143.

\bibitem{Edelstein:1998vs}
J.~D. Edelstein, L.~Tataru, and R.~Tatar, ``Rules for localized overlappings
  and intersections of p- branes,'' JHEP {\bf 06} (1998) 003,
hep-th/9801049.

\bibitem{Gauntlett:1998kc}
J.~P. Gauntlett, R.~C. Myers, and P.~K. Townsend, ``Supersymmetry of rotating
  branes,'' Phys. Rev. {\bf D59} (1999) 025001,
hep-th/9809065.

\bibitem{Ohta:1997gw}
N.~Ohta, ``Intersection rules for non-extreme p-branes,'' Phys. Lett. {\bf
  B403} (1997) 218--224,
hep-th/9702164.

\bibitem{Aref'eva:1998uh}
I.~Y. Aref'eva, M.~G. Ivanov, O.~A. Rytchkov, and I.~V. Volovich,
  ``Non-extremal localized branes and vacuum solutions in M- theory,'' Class.
  Quant. Grav. {\bf 15} (1998) 2923--2936,
hep-th/9802163.

\bibitem{Itzhaki:1998uz}
N.~Itzhaki, A.~A. Tseytlin, and S.~Yankielowicz, ``Supergravity solutions for
  branes localized within branes,'' Phys. Lett. {\bf B432} (1998) 298--304,
hep-th/9803103.

\bibitem{Cherkis:2002ir}
S.~A. Cherkis and A.~Hashimoto, ``Supergravity Solution of Intersecting Branes
  and AdS/CFT with Flavor,'' JHEP {\bf 11} (2002) 036,
hep-th/0210105.

\bibitem{Hashimoto:1998ug}
A.~Hashimoto, ``Supergravity solutions for localized intersections of branes,''
  JHEP {\bf 01} (1999) 018,
hep-th/9812159.

\bibitem{Cherkis:1999jt}
S.~A. Cherkis, ``Supergravity solution for M5-brane intersection,''
hep-th/9906203.

\bibitem{Gauntlett:1997pk}
J.~P. Gauntlett, G.~W. Gibbons, G.~Papadopoulos, and P.~K. Townsend,
  ``Hyper-Kaehler manifolds and multiply intersecting branes,'' Nucl. Phys.
  {\bf B500} (1997) 133--162,
hep-th/9702202.

\bibitem{Dasgupta:1998su}
K.~Dasgupta and S.~Mukhi, ``Brane constructions, conifolds and M-theory,''
  Nucl. Phys. {\bf B551} (1999) 204--228,
hep-th/9811139.

\bibitem{Becker:1996gj}
K.~Becker and M.~Becker, ``M-Theory on Eight-Manifolds,'' Nucl. Phys. {\bf
  B477} (1996) 155--167,
hep-th/9605053.

\bibitem{Brecher:1999xf}
D.~Brecher and M.~J. Perry, ``Ricci-flat branes,'' Nucl. Phys. {\bf B566}
  (2000) 151--172,
hep-th/9908018.

\bibitem{Janssen:1999uq}
B.~Janssen, ``Curved branes and cosmological (a,b)-models,'' JHEP {\bf 01}
  (2000) 044,
hep-th/9910077.

\bibitem{Papadopoulos:1999tw}
G.~Papadopoulos, J.~G. Russo, and A.~A. Tseytlin, ``Curved branes from string
  dualities,'' Class. Quant. Grav. {\bf 17} (2000) 1713--1728,
hep-th/9911253.

\bibitem{Ivashchuk:2000ma}
V.~D. Ivashchuk, ``On supersymmetric solutions in D = 11 supergravity with
  Ricci-flat internal spaces,'' Grav. Cosmol. {\bf 6} (2000) 344,
hep-th/0012263.

\bibitem{Papadopoulos:1998np}
G.~Papadopoulos and A.~Teschendorff, ``Multi-angle five-brane intersections,''
  Phys. Lett. {\bf B443} (1998) 159--166,
hep-th/9806191.

\bibitem{Papadopoulos:1998yx}
G.~Papadopoulos and A.~Teschendorff, ``Grassmannians, calibrations and
  five-brane intersections,'' Class. Quant. Grav. {\bf 17} (2000) 2641--2662,
hep-th/9811034.

\bibitem{Papadopoulos:1999tg}
G.~Papadopoulos, ``Rotating rotated branes,'' JHEP {\bf 04} (1999) 014,
hep-th/9902166.

\bibitem{Acharya:2000gb}
B.~S. Acharya, ``On realising N = 1 super Yang-Mills in M theory,''
hep-th/0011089.

\bibitem{Atiyah:2000zz}
M.~Atiyah, J.~M. Maldacena, and C.~Vafa, ``An M-theory flop as a large N
  duality,'' J. Math. Phys. {\bf 42} (2001) 3209--3220,
hep-th/0011256.

\bibitem{Acharya:2001hq}
B.~S. Acharya, ``Confining strings from G(2)-holonomy spacetimes,''
hep-th/0101206.

\bibitem{Atiyah:2001qf}
M.~Atiyah and E.~Witten, ``M-theory dynamics on a manifold of G(2) holonomy,''
  Adv. Theor. Math. Phys. {\bf 6} (2003) 1--106,
hep-th/0107177.

\bibitem{Acharya:2001gy}
B.~Acharya and E.~Witten, ``Chiral fermions from manifolds of G(2) holonomy,''
hep-th/0109152.

\bibitem{Friedmann:2002ct}
T.~Friedmann, ``On the quantum moduli space of M theory compactifications,''
  Nucl. Phys. {\bf B635} (2002) 384--394,
hep-th/0203256.

\bibitem{Cvetic:2001nr}
M.~Cvetic, G.~Shiu, and A.~M. Uranga, ``Chiral four-dimensional N = 1
  supersymmetric type IIA orientifolds from intersecting D6-branes,'' Nucl.
  Phys. {\bf B615} (2001) 3--32,
hep-th/0107166.

\bibitem{Cvetic:2001kk}
M.~Cvetic, G.~Shiu, and A.~M. Uranga, ``Chiral type II orientifold
  constructions as M theory on G(2) holonomy spaces,''
hep-th/0111179.

\bibitem{Gomis:2001vk}
J.~Gomis, ``D-branes, holonomy and M-theory,'' Nucl. Phys. {\bf B606} (2001)
  3--17,
hep-th/0103115.

\bibitem{Edelstein:2001pu}
J.~D. Edelstein and C.~Nunez, ``D6 branes and M-theory geometrical transitions
  from gauged supergravity,'' JHEP {\bf 04} (2001) 028,
hep-th/0103167.

\bibitem{Gukov:2001hf}
S.~Gukov and J.~Sparks, ``M-theory on Spin(7) manifolds. I,'' Nucl. Phys. {\bf
  B625} (2002) 3--69,
hep-th/0109025.

\bibitem{Curio:2001dz}
G.~Curio, B.~Kors, and D.~Lust, ``Fluxes and branes in type II vacua and
  M-theory geometry with G(2) and Spin(7) holonomy,'' Nucl. Phys. {\bf B636}
  (2002) 197--224,
hep-th/0111165.

\bibitem{Gukov:2002es}
S.~Gukov and D.~Tong, ``D-brane probes of special holonomy manifolds, and
  dynamics of N = 1 three-dimensional gauge theories,'' JHEP {\bf 04} (2002)
  050,
hep-th/0202126.

\bibitem{Uranga:2002ag}
A.~M. Uranga, ``Localized instabilities at conifolds,''
hep-th/0204079.

\bibitem{Lazaroiu:2002jv}
C.~I. Lazaroiu and L.~Anguelova, ``M-theory compactifications on certain
  'toric' cones of G(2) holonomy,'' JHEP {\bf 01} (2003) 066,
hep-th/0204249.

\bibitem{Anguelova:2002dd}
L.~Anguelova and C.~I. Lazaroiu, ``M-theory on 'toric' G(2) cones and its type
  II reduction,'' JHEP {\bf 10} (2002) 038,
hep-th/0205070.

\bibitem{Blumenhagen:2002wn}
R.~Blumenhagen, V.~Braun, B.~Kors, and D.~Lust, ``Orientifolds of K3 and
  Calabi-Yau manifolds with intersecting D-branes,'' JHEP {\bf 07} (2002) 026,
hep-th/0206038.

\bibitem{Blum:2002aq}
J.~D. Blum, ``Triple intersections and geometric transitions,'' Nucl. Phys.
  {\bf B646} (2002) 524--539,
hep-th/0207016.

\bibitem{Behrndt:2002xm}
K.~Behrndt, G.~Dall'Agata, D.~Lust, and S.~Mahapatra, ``Intersecting 6-branes
  from new 7-manifolds with G(2) holonomy,'' JHEP {\bf 08} (2002) 027,
hep-th/0207117.

\bibitem{Gukov:2002zg}
S.~Gukov, J.~Sparks, and D.~Tong, ``Conifold Transitions and Five-Brane
  Condensation in M- Theory on Spin(7) Manifolds,'' Class. Quant. Grav. {\bf
  20} (2003) 665--706,
hep-th/0207244.

\bibitem{Hanany:1997ie}
A.~Hanany and E.~Witten, ``Type IIB superstrings, BPS monopoles, and
  three-dimensional gauge dynamics,'' Nucl. Phys. {\bf B492} (1997) 152--190,
hep-th/9611230.

\bibitem{Giveon:1998sr}
A.~Giveon and D.~Kutasov, ``Brane dynamics and gauge theory,'' Rev. Mod. Phys.
  {\bf 71} (1999) 983--1084,
hep-th/9802067.

\bibitem{Seiberg:1994rs}
N.~Seiberg and E.~Witten, ``Electric - magnetic duality, monopole condensation,
  and confinement in N=2 supersymmetric Yang-Mills theory,'' Nucl. Phys. {\bf
  B426} (1994) 19--52,
hep-th/9407087.

\bibitem{Seiberg:1994aj}
N.~Seiberg and E.~Witten, ``Monopoles, duality and chiral symmetry breaking in
  N=2 supersymmetric QCD,'' Nucl. Phys. {\bf B431} (1994) 484--550,
hep-th/9408099.

\bibitem{Witten:1997sc}
E.~Witten, ``Solutions of four-dimensional field theories via M-theory,'' Nucl.
  Phys. {\bf B500} (1997) 3--42,
hep-th/9703166.

\bibitem{Fayyazuddin:1997by}
A.~Fayyazuddin and M.~Spalinski, ``The Seiberg-Witten differential from
  M-theory,'' Nucl. Phys. {\bf B508} (1997) 219--228,
hep-th/9706087.

\bibitem{Landsteiner:1997vd}
K.~Landsteiner, E.~Lopez, and D.~A. Lowe, ``N = 2 supersymmetric gauge
  theories, branes and orientifolds,'' Nucl. Phys. {\bf B507} (1997) 197--226,
hep-th/9705199.

\bibitem{Henningson:1998hy}
M.~Henningson and P.~Yi, ``Four-dimensional BPS-spectra via M-theory,'' Phys.
  Rev. {\bf D57} (1998) 1291--1298,
hep-th/9707251.

\bibitem{Mikhailov:1998jv}
A.~Mikhailov, ``BPS states and minimal surfaces,'' Nucl. Phys. {\bf B533}
  (1998) 243--274,
hep-th/9708068.

\bibitem{Klemm:1996bj}
A.~Klemm, W.~Lerche, P.~Mayr, C.~Vafa, and N.~P. Warner, ``Self-Dual Strings
  and N=2 Supersymmetric Field Theory,'' Nucl. Phys. {\bf B477} (1996)
  746--766,
hep-th/9604034.

\bibitem{Howe:1998eu}
P.~S. Howe, N.~D. Lambert, and P.~C. West, ``Classical M-fivebrane dynamics and
  quantum N = 2 Yang- Mills,'' Phys. Lett. {\bf B418} (1998) 85--90,
hep-th/9710034.

\bibitem{Fayyazuddin:2000em}
A.~Fayyazuddin and D.~J. Smith, ``Warped AdS near-horizon geometry of
  completely localized intersections of M5-branes,'' JHEP {\bf 10} (2000) 023,
hep-th/0006060.

\bibitem{Brinne:2000fh}
B.~Brinne, A.~Fayyazuddin, S.~Mukhopadhyay, and D.~J. Smith, ``Supergravity
  M5-branes wrapped on Riemann surfaces and their QFT duals,'' JHEP {\bf 12}
  (2000) 013,
hep-th/0009047.

\bibitem{Maldacena:2000mw}
J.~M. Maldacena and C.~Nunez, ``Supergravity description of field theories on
  curved manifolds and a no go theorem,'' Int. J. Mod. Phys. {\bf A16} (2001)
  822--855,
hep-th/0007018.

\bibitem{Gauntlett:2001ps}
J.~P. Gauntlett, N.~Kim, D.~Martelli, and D.~Waldram, ``Wrapped fivebranes and
  N = 2 super Yang-Mills theory,'' Phys. Rev. {\bf D64} (2001) 106008,
hep-th/0106117.

\bibitem{Bigazzi:2001aj}
F.~Bigazzi, A.~L. Cotrone, and A.~Zaffaroni, ``N = 2 gauge theories from
  wrapped five-branes,'' Phys. Lett. {\bf B519} (2001) 269--276,
hep-th/0106160.

\bibitem{Ohta:2000gf}
K.~Ohta and T.~Yokono, ``Linear dilaton background and fully localized
  intersecting five-branes,'' Phys. Rev. {\bf D63} (2001) 105011,
hep-th/0012030.

\bibitem{Douglas:1996sw}
M.~R. Douglas and G.~W. Moore, ``D-branes, Quivers, and ALE Instantons,''
hep-th/9603167.

\bibitem{Douglas:1997xg}
M.~R. Douglas, ``Enhanced gauge symmetry in M(atrix) theory,'' JHEP {\bf 07}
  (1997) 004,
hep-th/9612126.

\bibitem{Douglas:1997de}
M.~R. Douglas, B.~R. Greene, and D.~R. Morrison, ``Orbifold resolution by
  D-branes,'' Nucl. Phys. {\bf B506} (1997) 84--106,
hep-th/9704151.

\bibitem{Berenstein:1998dw}
D.~Berenstein, R.~Corrado, and J.~Distler, ``Aspects of ALE matrix models and
  twisted matrix strings,'' Phys. Rev. {\bf D58} (1998) 026005,
hep-th/9712049.

\bibitem{Diaconescu:1998br}
D.-E. Diaconescu, M.~R. Douglas, and J.~Gomis, ``Fractional branes and wrapped
  branes,'' JHEP {\bf 02} (1998) 013,
hep-th/9712230.

\bibitem{Berenstein:1998ri}
D.~Berenstein and R.~Corrado, ``Matrix theory on ALE spaces and wrapped
  membranes,'' Nucl. Phys. {\bf B529} (1998) 225--245,
hep-th/9803048.

\bibitem{Brodie:1998bv}
J.~H. Brodie, ``Fractional branes, confinement, and dynamically generated
  superpotentials,'' Nucl. Phys. {\bf B532} (1998) 137--152,
hep-th/9803140.

\bibitem{Karch:1998yv}
A.~Karch, D.~Lust, and D.~J. Smith, ``Equivalence of geometric engineering and
  Hanany-Witten via fractional branes,'' Nucl. Phys. {\bf B533} (1998)
  348--372,
hep-th/9803232.

\bibitem{Dasgupta:1999wx}
K.~Dasgupta and S.~Mukhi, ``Brane constructions, fractional branes and anti-de
  Sitter domain walls,'' JHEP {\bf 07} (1999) 008,
hep-th/9904131.

\bibitem{Klebanov:1999rd}
I.~R. Klebanov and N.~A. Nekrasov, ``Gravity duals of fractional branes and
  logarithmic RG flow,'' Nucl. Phys. {\bf B574} (2000) 263--274,
hep-th/9911096.

\bibitem{Bertolini:2000dk}
M.~Bertolini {\em et.~al.}, ``Fractional D-branes and their gauge duals,'' JHEP
  {\bf 02} (2001) 014,
hep-th/0011077.

\bibitem{Grana:2001xn}
M.~Grana and J.~Polchinski, ``Gauge / gravity duals with holomorphic dilaton,''
  Phys. Rev. {\bf D65} (2002) 126005,
hep-th/0106014.

\bibitem{Bertolini:2001qa}
M.~Bertolini, P.~Di~Vecchia, M.~Frau, A.~Lerda, and R.~Marotta, ``N = 2 gauge
  theories on systems of fractional D3/D7 branes,'' Nucl. Phys. {\bf B621}
  (2002) 157--178,
hep-th/0107057.

\bibitem{Bertolini:2001gq}
M.~Bertolini, P.~Di~Vecchia, and R.~Marotta, ``N = 2 four-dimensional gauge
  theories from fractional branes,''
hep-th/0112195.

\bibitem{Johnson:1999qt}
C.~V. Johnson, A.~W. Peet, and J.~Polchinski, ``Gauge theory and the excision
  of repulson singularities,'' Phys. Rev. {\bf D61} (2000) 086001,
hep-th/9911161.

\bibitem{Brinne:2000nf}
B.~Brinne, A.~Fayyazuddin, T.~Z. Husain, and D.~J. Smith, ``N = 1 M5-brane
  geometries,'' JHEP {\bf 03} (2001) 052,
hep-th/0012194.

\bibitem{Becker:2000rz}
K.~Becker and M.~Becker, ``Compactifying M-theory to four dimensions,'' JHEP
  {\bf 11} (2000) 029,
hep-th/0010282.

\bibitem{Maldacena:2000yy}
J.~M. Maldacena and C.~Nunez, ``Towards the large N limit of pure N = 1 super
  Yang Mills,'' Phys. Rev. Lett. {\bf 86} (2001) 588--591,
hep-th/0008001.

\bibitem{Apreda:2001qb}
R.~Apreda, F.~Bigazzi, A.~L. Cotrone, M.~Petrini, and A.~Zaffaroni, ``Some
  comments on N = 1 gauge theories from wrapped branes,'' Phys. Lett. {\bf
  B536} (2002) 161--168,
hep-th/0112236.

\bibitem{Klebanov:2000nc}
I.~R. Klebanov and A.~A. Tseytlin, ``Gravity duals of supersymmetric SU(N) x
  SU(N+M) gauge theories,'' Nucl. Phys. {\bf B578} (2000) 123--138,
hep-th/0002159.

\bibitem{Bertolini:2001gg}
M.~Bertolini, P.~Di~Vecchia, G.~Ferretti, and R.~Marotta, ``Fractional branes
  and N = 1 gauge theories,'' Nucl. Phys. {\bf B630} (2002) 222--240,
hep-th/0112187.

\bibitem{Acharya:2000mu}
B.~S. Acharya, J.~P. Gauntlett, and N.~Kim, ``Fivebranes wrapped on associative
  three-cycles,'' Phys. Rev. {\bf D63} (2001) 106003,
hep-th/0011190.

\bibitem{Nieder:2000kc}
H.~Nieder and Y.~Oz, ``Supergravity and D-branes wrapping special Lagrangian
  cycles,'' JHEP {\bf 03} (2001) 008,
hep-th/0011288.

\bibitem{Gauntlett:2000ng}
J.~P. Gauntlett, N.~Kim, and D.~Waldram, ``M-fivebranes wrapped on
  supersymmetric cycles,'' Phys. Rev. {\bf D63} (2001) 126001,
hep-th/0012195.

\bibitem{Nunez:2001pt}
C.~Nunez, I.~Y. Park, M.~Schvellinger, and T.~A. Tran, ``Supergravity duals of
  gauge theories from F(4) gauged supergravity in six dimensions,'' JHEP {\bf
  04} (2001) 025,
hep-th/0103080.

\bibitem{Schvellinger:2001ib}
M.~Schvellinger and T.~A. Tran, ``Supergravity duals of gauge field theories
  from SU(2) x U(1) gauged supergravity in five dimensions,'' JHEP {\bf 06}
  (2001) 025,
hep-th/0105019.

\bibitem{Maldacena:2001pb}
J.~M. Maldacena and H.~Nastase, ``The supergravity dual of a theory with
  dynamical supersymmetry breaking,'' JHEP {\bf 09} (2001) 024,
hep-th/0105049.

\bibitem{Gauntlett:2001qs}
J.~P. Gauntlett, N.~Kim, S.~Pakis, and D.~Waldram, ``Membranes wrapped on
  holomorphic curves,'' Phys. Rev. {\bf D65} (2002) 026003,
hep-th/0105250.

\bibitem{Hernandez:2001bh}
R.~Hernandez, ``Branes wrapped on coassociative cycles,'' Phys. Lett. {\bf
  B521} (2001) 371--375,
hep-th/0106055.

\bibitem{Gauntlett:2001jj}
J.~P. Gauntlett and N.~Kim, ``M-fivebranes wrapped on supersymmetric cycles.
  II,'' Phys. Rev. {\bf D65} (2002) 086003,
hep-th/0109039.

\bibitem{Bhattacharyya:2002se}
S.~Bhattacharyya, A.~Kumar, and S.~Mukhopadhyay, ``Curved membrane solutions in
  D = 11 supergravity,'' Int. J. Mod. Phys. {\bf A17} (2002) 4647--4660,
hep-th/0206032.

\bibitem{Cho:2000hg}
H.~Cho, M.~Emam, D.~Kastor, and J.~Traschen, ``Calibrations and
  Fayyazuddin-Smith spacetimes,'' Phys. Rev. {\bf D63} (2001) 064003,
hep-th/0009062.

\bibitem{Gomberoff:1999ps}
A.~Gomberoff, D.~Kastor, D.~Marolf, and J.~Traschen, ``Fully localized brane
  intersections: The plot thickens,'' Phys. Rev. {\bf D61} (2000) 024012,
hep-th/9905094.

\bibitem{Ramadevi:1999dy}
P.~Ramadevi, ``Supergravity solution for three-string junction in M- theory,''
  JHEP {\bf 06} (2000) 005,
hep-th/9906247.

\bibitem{Surya:1998dx}
S.~Surya and D.~Marolf, ``Localized branes and black holes,'' Phys. Rev. {\bf
  D58} (1998) 124013,
hep-th/9805121.

\bibitem{Marolf:1999uq}
D.~Marolf and A.~Peet, ``Brane baldness vs. superselection sectors,'' Phys.
  Rev. {\bf D60} (1999) 105007,
hep-th/9903213.

\bibitem{Fayyazuddin:2002bm}
A.~Fayyazuddin, ``Supersymmetric webs of D3/D5-branes in supergravity,''
hep-th/0207129.

\bibitem{Rajaraman:2000ws}
A.~Rajaraman, ``Supergravity solutions for localised brane intersections,''
  JHEP {\bf 09} (2001) 018,
hep-th/0007241.

\bibitem{Karch:2001cw}
A.~Karch and L.~Randall, ``Localized gravity in string theory,'' Phys. Rev.
  Lett. {\bf 87} (2001) 061601,
hep-th/0105108.

\bibitem{Karch:2000gx}
A.~Karch and L.~Randall, ``Open and closed string interpretation of SUSY CFT's
  on branes with boundaries,'' JHEP {\bf 06} (2001) 063,
hep-th/0105132.

\end{thebibliography}\endgroup

\end{document}